\documentclass[11pt]{article}

\usepackage{authblk}  
\usepackage{graphicx} 
\usepackage{epsfig}
\usepackage{dcolumn}  
\usepackage{bm}       
\usepackage{amsmath}
\usepackage{amssymb}

\usepackage{multirow}
\usepackage{booktabs}

\newcommand{\newsection}{ \setcounter{equation}{0} \section}

\newcommand{\beq}{\begin{equation}} \newcommand{\eeq}{\end{equation}}
\newcommand{\bea}{\begin{eqnarray}} \newcommand{\eea}{\end{eqnarray}}

  \newcommand
{\Romannumeral}[1]{\uppercase\expandafter{\romannumeral#1}}

\newcommand{\be}{\begin{enumerate}} \newcommand{\ee}{\end{enumerate}}
\newcommand{\bi}{\begin{itemize}} \newcommand{\ei}{\end{itemize}}
\newcommand{\ba}{\begin{array}} \newcommand{\ea}{\end{array}}
\newcommand{\bc}{\begin{center}} \newcommand{\ec}{\end{center}}
\newcommand{\bt}{\begin{tabular}} \newcommand{\et}{\end{tabular}}

\def\lsim{\mathrel{\rlap{\lower4pt\hbox{\hskip1pt$\sim$}}
    \raise1pt\hbox{$<$}}}           
\def\gsim{\mathrel{\rlap{\lower4pt\hbox{\hskip1pt$\sim$}}
    \raise1pt\hbox{$>$}}}           

\newcommand{\tr}{\mathop{\rm tr}}           
\newcommand{\half}{\textstyle {1\over2} \displaystyle}    
\newcommand{\third}{\textstyle {1\over3} \displaystyle}   
\newcommand{\Dslash}{{\hbox{D}\kern-0.6em\raise0.15ex\hbox{/}}} 

\renewcommand{\et}{\eta}

\begin{document}

\setlength{\oddsidemargin}{0cm} \setlength{\baselineskip}{7mm}

\begin{normalsize}\begin{flushright}

June 2015 \\

\end{flushright}\end{normalsize}

\begin{center}
  
\vspace{5pt}

{\Large \bf Scaling Exponents for Lattice Quantum Gravity in Four Dimensions }

\vspace{30pt}

{\sl Herbert W. Hamber}
$^{}$\footnote{e-mail address : HHamber@uci.edu} 
\\
Department of Physics and Astronomy \\
University of California \\
Irvine, CA 92697-4575 \\

\vspace{10pt}

\end{center}

\begin{center} {\bf ABSTRACT } \end{center}

\noindent

In this work nonperturbative aspects of quantum gravity are investigated 
using the lattice formulation, and some new results are presented for
critical exponents, amplitudes and invariant correlation functions. 
Values for the universal scaling dimensions are compared 
with other nonperturbative approaches to gravity in 
four dimensions, and specifically to the conjectured value for the
universal critical exponent $\nu =1 /3$. 
It is found that the lattice results are generally consistent with gravitational 
anti-screening, which would imply a slow increase in the strength
of the gravitational coupling with distance,
and here detailed estimates for exponents and amplitudes characterizing 
this slow rise are presented.
Furthermore, it is shown that in the lattice approach (as for gauge
theories) the quantum theory is highly constrained, and eventually by
virtue of scaling depends on a rather small set of physical parameters. 
Arguments are given in support of the statement that the fundamental
reference scale for the growth of the gravitational coupling $G$
with distance is represented by the observed scaled cosmological
constant $\lambda$, which in gravity acts as an effective nonperturbative infrared cutoff.
In this nonperturbative vacuum condensate picture
a fundamental relationship emerges between the scale characterizing 
the running of $G$ at large distances, the macroscopic scale for the 
curvature as described by the observed cosmological constant, 
and the behavior of invariant gravitational correlation 
functions at large distances.
Overall, the lattice results suggest that the slow infrared growth of $G$ with distance 
should become observable only on very large distance scales, comparable
to $\lambda$.
It is hoped that future high precision satellite experiments will possibly 
come within reach of this small quantum correction, as suggested by
a vacuum condensate picture of quantum gravity.



\vfill

\pagestyle{empty}

\newpage

\pagestyle{plain}


\section{Introduction}

\label{sec:intro}

\vskip 20pt

In this work nonperturbative aspects of the ground state for
quantum gravity will be discussed, based on the lattice theory.
So far the lattice formulation represents the only known
first-principle method for reproducing correctly the low 
energy properties of non-Abelian gauge theories, including
confinement and chiral symmetry breaking.
It is hoped therefore that a lattice approach to quantum gravity will
shed useful light on the low energy properties of quantum gravity as
well.
A key aspect of this method is a recognition of the importance of
Wilson's modern interpretation of the renormalization group
\cite{wil72,wil75} as it applies to perturbatively non-renormalizable 
theories \cite{par73,par76}, including gravity.
In previous work the elegant lattice formulation of gravity of Regge and 
Wheeler \cite{reg61,whe64} was used to compute a number of observable quantities 
expected to be relevant for the ground state properties of
quantum gravity (for a detailed discussion of the Feynman path integral
approach to quantum gravity the reader is referred to \cite{book}).
The lattice formulation implies the existence of an ultraviolet
cutoff, nevertheless in the real world such a cutoff would presumably
arise from short distance details derived from an underlying more fundamental
theory, such as higher derivative gravity, supergravity or string theory.
Nevertheless, it is expected that a softer cutoff would lead
to significant short distance modifications of gravity, while leaving 
the quantum infrared behavior largely unchanged.
It is these universal long distance effects that form the subject 
of the present paper.
As in QED and QCD, it will turn out that these quantum infrared modifications
to gravity are intrinsically non-local.

In four dimensions (and for the Euclidean theory) it was found that
for gravity two phases are possible, a pathological gravitational screening phase
for $G < G_c$, and an anti-screening phase for $G > G_c$.
It has been known for some time that the screening phase corresponds 
to a branched polymer, with no physically acceptable continuum limit,
and the lattice results were therefore interpreted as suggesting that 
ultimately the only physical acceptable phase is the strong coupling phase 
for $G > G_c$.
Furthermore, it was found that in this phase the average local 
curvature approaches zero towards the critical point at $G_c$, 
indicating that in this phase the recovery of the semiclassical 
limit for gravity appears to be possible. 
In previous work detailed estimates were given for the location of the 
critical point at $G_c$, for the critical exponents and scaling
dimensions characterizing the growth of the gravitational coupling
with distance, and for the scaling behavior of gravitational correlation functions.
Of central importance in these results is the value for the 
universal critical exponent $\nu$, related to the derivative 
of the beta function for $G$ at the fixed point.
Here more refined estimates for the critical point and scaling 
dimensions will be provided; 
the analysis will later be extended to correlation 
functions of invariant operators at fixed geodesic distances.
In this context the present discussion includes both local operators, as 
well as extended ones such as the Wilson loop and the correlation between
gravitational Wilson loops. 
In previous work it was argued that the gravitational 
Wilson loop provides information about the macroscopic
curvature, and therefore about the hoped-for recovery of the 
semiclassical limit. 
This last result is quite different from what is found in gauge theories,
since in gravity the gravitational Wilson loop has no significance
for the static potential. 
Also, it was shown earlier that the
the area law for the gravitational Wilson loop provides a connection 
between the nonperturbative scale $\xi$ that arises in 
nonperturbative gravity and the macroscopic large-scale curvature,
and thus the observed effective cosmological constant $\lambda$. 
Here it will be argued that the
numerical results so far are consistent with many of the 
previous answers, including the conjecture that the exponent 
$\nu$ is exactly equal to one third in four dimensions.
The latter part of the paper will therefore deal with a detailed
discussion of the possible physical significance
of having an exponent $\nu$ exactly equal to one third in four
dimensions, as this relates to a number of physical consequences,
such as the scale dependence of $G$ and the behavior of 
invariant gravitational correlation functions at large separation.

The structure of the paper is as follows. 
In Section 2 the form of the discretized lattice gravitational 
Feynman path integral will be recalled, and basic notation will be established. 
Here the basic definitions for local averages and their fluctuations
will be laid out as well.
Section 3 will introduce basic diffeomorphism invariant 
correlation functions, and show how these can 
be transcribed to the lattice theory.
Section 4 will extend the previous discussion to correlation 
functions involving operators that are not necessarily local, 
such as the correlation between smeared operators, the definition 
of the gravitational Wilson loop, and the form and basic
expected properties of correlations between these loop operators.
Section 5 will recall the basic fundamental
scaling assumptions for the gravitational path integral, 
and how those assumptions and definitions affect the critical behavior 
of various averages, fluctuations and correlations defined 
in the previous sections.
Section 6 summarize how the quantum continuum limit 
in lattice quantum gravity should be taken, in accordance 
with the general principles of the renormalization group. 
A discussion is provided to show how the interplay between the 
bare coupling constants, the critical point 
and the correlation length $\xi$ lead to a definite expression 
for the running of Newton's $G$. 
In addition, it is shown how the prediction for 
a running of $G$ can be described in terms of universal quantities,
to leading order in the vicinity of that fixed point, and
specifically in terms of universal exponents and amplitudes. 
Sections 7,8,9 and 10 later provide details 
on the how the numerical calculations are performed, and 
on the methods by which the universal critical exponents and amplitudes 
are extracted from the numerical results.
A discussion is given to show the overall consistency of the results, 
based on a variety of different observables and methods of analysis.
At the end of Section 10 the results obtained 
from a variety of different observables are compared in 
a comprehensive summary table. 
Two additional tables later provide a comparison between 
the lattice results for the universal exponents and the values
obtained by other nonperturbative methods.
In Section 10 a separate comparison table is provided for four
dimensions, 
and a second table is added later for the case of three dimensions.
It is argued that the numerical results so far are consistent with the 
expectation that the universal critical exponent $\nu$
for quantum gravity in four dimensions is equal to one third.
Section 11 then discusses the physical implications of having an exponent
$\nu$ exactly equal to one third as it applies to various 
local averages, fluctuations and correlation functions 
introduced earlier in the paper. 
It is shown that many lattice results become particularly 
simple and perhaps more transparent for this choice of exponent. 
The numerical calculation also supply a value for several
critical amplitudes which appear in the running of $G$ in 
the vicinity of the fixed point at $G_c$. 
Section 12 at the end of the paper is devoted to a discussion 
of the curvature correlation function, and how this correlation
can, in suitable cases via the field equations, be related to the 
analogous correlation function for matter density fluctuations. 
The final section presents some conclusions, and
elaborates further on a suggestive analogy between the vacuum
condensate picture derived from lattice quantum gravity and the
well understood nonperturbative properties
of non-Abelian gauge theories, and specifically the case of 
lattice $QCD$.



\vskip 40pt

\section{Path Integral, Invariant Local Gravitational Averages and their Fluctuations}

\label{sec:ave}

\vskip 20pt

In this section the basic definitions for diffeomorphism invariant
gravitational averages and correlations will be
recalled briefly, in a form suitable for later discussions.
Here the starting point for a nonperturbative formulation of quantum gravity
is the discretized form for the Feynman path integral for pure gravity
\cite{fey}, written originally as
\beq
Z_C \; = \; \int d \mu [ g_{\mu\nu} ] \;
\exp \left \{ - I [ g_{\mu\nu} ] \right \}  \;\; ,
\label{eq:z_cont}
\eeq
with invariant gravitational action
\beq
I [ g_{\mu\nu} ] \; = \; 
\int d^4 x \, \sqrt g \, \Bigl ( \lambda_0 - { k \over 2 } \, R
+ { a_0 \over 4 } \, R_{\mu\nu\rho\sigma} R^{\mu\nu\rho\sigma}
+ \cdots \Bigr )
\label{eq:ac_cont}
\eeq
and DeWitt invariant functional measure \cite{dew67}
\beq
\int d \mu [ g_{\mu\nu} ] \; = \; \int \prod_x \;
\left ( {\textstyle \sqrt{g(x)} \displaystyle} \right )^{\sigma}
\; \prod_{ \mu \ge \nu } \, d g_{ \mu \nu } (x) \; .
\label{eq:meas_cont}
\eeq
In the above expression $k^{-1} \equiv 8 \pi G $ with $G$ the bare
Newton's constant,
$\lambda_0$ the bare cosmological constant, and $a_0$ a possible
higher derivative coupling \cite{hdqg}.
In the absence of matter fields the DeWitt invariant measure for pure
gravity in four dimensions corresponds to the simple choice $\sigma=0$.
In the following we will only consider the case $a_0=0$, i.e. no
higher derivative $R^2$-type terms.
\footnote{
A well known problem of the Euclidean path integral formulation
is the conformal instability of the classical gravitational action \cite{haw79,gib77}.
The latter is seen by considering conformal transformations
$ \tilde g_{\mu \nu} = \Omega^2 \, g_{\mu \nu} $ with $\Omega$ a positive function.
Then the Einstein-Hilbert action transforms into
\beq
I [ \tilde g ] = - { 1 \over 16 \pi G }  \int  d ^ 4 x  \sqrt g \;
( \Omega^2 R + 6 \, g^{\mu \nu} \partial_\mu \Omega \, \partial_\nu \Omega ) \;\; .
\eeq
which can be made arbitrarily negative by choosing a rapidly varying conformal
factor $\Omega$.
The wrong sign for the kinetic term of the $\Omega$ fields then
implies that the Euclidean gravitational functional integral is possibly 
badly divergent, depending on the detailed nature of the gravitational 
measure contribution $ d \mu [ g_{\mu\nu} ]$ (the ``entropy'' or phase space part), and 
specifically its behavior in the regime of strong fields and
rapidly varying conformal factors.}

The continuum Feynman path integral given above is generally
ill-defined, and has to be formulated more precisely by 
introducing a suitable discretization \cite{fey65}.
The last step is particularly important for nonperturbative calculations,
where the nontrivial invariant measure over the $g_{\mu\nu}$'s plays a key
role.
Regge and Wheeler proposed an elegant discretization of
the classical gravitational action \cite{reg61,whe64},
which forms the basis for the lattice formulation of
quantum gravity discussed in this paper.
Once the measure and the path integral have been discretized,
the ultimate goal then becomes to recover the original 
continuum theory of Eq.~(\ref{eq:z_cont}) in the limit of a small 
lattice spacing (this limit is rather subtle, and involves in
a nontrivial way fundamental aspects of the renormalization group).
This approach then leads, as a starting point, to the following discrete
form for the Euclidean Feynman path integral for pure gravity
\beq
Z_L \; = \; \int d \mu [ l^2 ] 
\exp \left \{  - I [l^2 ] \right \}  \;\; ,
\label{eq:z_latt} 
\eeq
with lattice gravitational action
\beq
I [ l^2 ] \; = \; \sum_h \, \Bigl ( \lambda_0 \, V_h - k \, \delta_h A_h 
+ a \, \delta_h^2 A_h^2 / V_h  + \cdots \Bigr )
\label{eq:ac_latt} 
\eeq
and lattice functional measure
\beq
\int d \mu [ l^2 ] \; = \;
\int_0^\infty \; \prod_s \; \left ( V_d (s) \right )^{\sigma} \;
\prod_{ ij } \, dl_{ij}^2 \; \Theta [l_{ij}^2]  \; .
\label{eq:meas_latt} 
\eeq
Here the sum over hinges $h$ in four dimensions corresponds to a sum 
over all lattice triangles with area $A_h$, with deficit angles 
$\delta_h$ describing the curvature around them.
\footnote{
In the following we will deal almost exclusively, as is customary
in lattice field theories, with dimensionless
quantities. Thus the couplings $\lambda_0$ and $G$ appearing in
the continuum theory will be expressed from the start
in units of the fundamental lattice ultraviolet cutoff $\Lambda = 1/a$ \cite{lesh84}.
As is standard procedure in ordinary lattice field theories and
lattice gauge theories \cite{par81,zin02}, the 
latter is then later set equal to one, which means that from then on all observable
quantities, correlations and couplings are measured in units of this
fundamental ultraviolet cutoff.
The actual value for the ultraviolet cutoff (in $MeV$ or $cm^{-1}$) is later determined
by comparing suitable physical quantities, see Eqs.~(\ref{eq:gc_phys})
and (\ref{eq:a_phys}) towards the end of the paper.}

In the discrete formulation a functional integration over metric
is replaced by an integration over squared edge lengths,
which are taken as fundamental variables in the discrete theory.
The basis for this step is a rather direct correspondence between the squared
edge lengths in a four-simplex and the induced metric within
that same simplex.
Within each $n$-simplex $s$ one can define a metric in terms
of unit vectors $e_i$ pointing along the edges
\beq
g_{ij} (s) \; = \; e_i \cdot e_j \;\; , 
\eeq
with $1 \leq i,j \leq n $, and a positive definite quantity in the Euclidean case.
In terms of the edge lengths $l_{ij} \, = \, | e_i - e_ j | $, or conversely
\beq
g_{ij} (s) \; = \; \half \, 
\left ( l_{0i}^2 + l_{0j}^2 - l_{ij}^2 \right ) \;\;
\label{eq:latmet}
\eeq
for a simplex based at $0$.
This last result then provides a key connection between the metric
$g_{\mu\nu} (x) $ in the continuum and the lattice degrees of freedom $l^2_i$,
which is essential in establishing a fairly unambiguous relationship between lattice and 
continuum operators, just as is the case in ordinary lattice gauge theories.
It is also known that the lattice action in Eq.~(\ref{eq:z_latt}
generally reduces to the continuum one of Eq.~(\ref{eq:z_cont} for smooth enough field
configurations \cite{cms82}, and that it contains the correct
physical degrees of freedom for gravity in the weak field limit, namely
transverse-traceless (massless spin two) modes \cite{rowi81}.

The general aim of the calculations presented later will be to evaluate
the lattice path integral exactly by numerical means, by performing a
(correctly weighted) sum over all fields configurations, without relying on the weak
field expansion, or an expansion around suitable saddle points or some other
approximate scheme, which generally tends to involve a number of assumptions on
what configurations (smooth or otherwise) might or might not play a dominant role
in the path integral (indeed the general expectation for such path
integrals is that smooth field configurations tend to have measure zero).
Here the functional integration over edge lengths is highly nontrivial, due to the 
constraint coming from the generalized triangle inequalities
[expressed in the function $\Theta [l_{ij}^2]$ in Eq.~(\ref{eq:meas_latt})],
which is placed there in the Euclidean formulation to insure that all
edge lengths, triangle areas, tetrahedra and simplex volumes are strictly positive.
The discrete gravitational measure in $Z_L$ of Eq.~(\ref{eq:z_latt})
can then be regarded as a regularized version of the DeWitt continuum 
functional measure \cite{dew67}.
Also, a bare cosmological constant term is essential for the convergence of the path
integral, while curvature squared terms allow one to further control the
fluctuations in the curvature \cite{lesh84,hw84}.
It is generally understood that these last terms are generated by radiative
corrections within a perturbative diagrammatic treatment in the continuum.
In practice, and for obvious phenomenological reasons, one is
nevertheless only interested eventually in a limit where effective 
higher derivative contributions are negligible compared to the rest 
of the action, $a_0 \rightarrow 0$.

In this limit the theory depends, in the absence of matter and after a 
suitable rescaling of the metric (in the continuum) or the edge
lengths (on the lattice), only on {\it one} bare parameter, the 
dimensionless coupling $ k / \sqrt{\lambda_0} $.
Indeed already in the continuum one finds in $d$ dimensions under a rescaling of the metric
\beq
g_{\mu\nu} = \omega \, g_{\mu\nu}' \; ,
\label{eq:metric_scale}
\eeq
with $\omega$ a constant, that the cosmological constant term $ \lambda_0 \sqrt{g} $
turns into $ \lambda_0 \, \omega^{d/2} \, \sqrt{g'} $ so that a
subsequent rescaling
\beq
G \rightarrow \omega^{-d/2+1} G \; , \;\;\;\; 
\lambda_0 \rightarrow \lambda_0 \, \omega^{d/2}
\eeq
leaves only the dimensionless combination $G^d \lambda_0^{d-2}$ unchanged.
Clearly only the latter combination has physical meaning in pure
gravity, and in particular one can always choose the scale $\omega = \lambda_0^{-2/d}$
so as to adjust the volume term to have a unit coefficient.
Equivalently, this shows that it seems physically meaningless to discuss separately the 
renormalization properties of $G$ and $\lambda_0$.
Without any loss of generality one can therefore set the bare cosmological
constant $\lambda_0 = 1$ in units of the ultraviolet cutoff \cite{lesh84}.
The latter contribution controls the overall scale for the edge
lengths, contains (like a mass term) no derivatives in the 
continuum, and does not affect the construction of a 
suitable lattice continuum limit, which is determined by the relative interplay
between the curvature and volume terms.
It seems therefore redundant to vary $\lambda_0$, as this will
only change the overall length scale, without any discernible effect on the
quantum lattice continuum limit.
In the continuum a similar result can be derived; there on can show
that the renormalization of $\lambda_0$ is gauge- and
scheme-dependent, and that only the renormalization of $G$ is
independent of the choice of gauge condition \cite{eps,aid97,lambda}.


Some partial information about the behavior of physical correlations
can be obtained indirectly from averages of suitable local invariant operators.
In \cite{hw84} a set of diffeomorphism invariant gravitational
observables, such as the average curvature and its fluctuation, were
introduced.
Appropriate lattice analogs of these quantities are easily written
down, making use of the following well understood correspondences
\bea
\sqrt{g} \, (x) \; & \to & \; 
\sum_{{\rm hinges} \, h \, \supset \, x } \; V_h 
\nonumber \\
\sqrt{g} \, R (x) \; & \to & \; 
2 \sum_{{\rm hinges} \, h \, \supset \, x } \; \delta_h A_h
\nonumber \\
\sqrt{g} \, R_{\mu\nu\lambda\sigma} \, R^{\mu\nu\lambda\sigma} (x) \; & \to & \;
4 \sum_{{\rm hinges} \, h \, \supset \, x } \; ( \delta_h A_h )^2 / V_h  \;\; .
\label{eq:ops}
\eea
An overall numerical normalization coefficient has been omitted on the r.h.s., since it
will depend on how many hinges are actually included in the summation.
In the following we will not consider any further higher derivative
terms, which means that the subsequent discussion will
be limited almost exclusively to the first and second type of operators.

First consider the average local curvature, defined as
\beq
{\cal R} (k) \; \sim \;
{ < \int d^4 x \, \sqrt{ g } \, R(x) >
\over < \int d^4 x \, \sqrt{ g } > } \;\;\; .
\label{eq:avr_cont}
\eeq
The above quantity is relevant for parallel transports of
vectors around an elementary, infinitesimal parallel transport loop,
and is by construction manifestly diffeomorphism invariant.
On the lattice one first notes that it is preferable to define quantities in such a way that
variations in the average lattice spacing $l_0 \sim \sqrt{< \! l^2 \! >}$ are compensated by
a suitable multiplicative factor, determined entirely from dimensional
considerations.
It would be possible to adjust $\lambda_0$ in Eq.~(\ref{eq:z_latt}) to
achieve $l_0 =1$, but here we choose to have simply $\lambda_0 =1 $ in
units of the ultraviolet cutoff $\Lambda=1/a$.
As stated previously, in the following all quantities will be
expressed in units of this
fundamental cutoff $a$, whose value, as is customary in lattice gauge
theories, is set initially equal to one, $a =1$.
In the case of the average local curvature a useful lattice definition is
therefore \cite{hw84,lesh84}
\beq
{\cal R} (k) \; \equiv \; 
< \! l^2 \! > { < 2 \; \sum_h \delta_h A_h > \over < \sum_h V_h > } \;\;\; .
\label{eq:avr_latt} 
\eeq
Note that by construction this quantity is dimensionless, and consequently
if all edge lengths are rescaled by a common factor it remains unchanged.
Again, this choice factors out an entirely irrelevant overall length scale (a
phenomenon peculiar to gravity, which does not arise in ordinary 
lattice gauge theories).

A second quantity of interest is the local curvature fluctuation
\beq
\chi_{\cal R}  (k) \; \sim \;
{ < ( \int d^4 x \, \sqrt{g} \, R )^2 > - < \int d^4 x \, \sqrt{g} \, R >^2
\over < \int d^4 x \, \sqrt{g} > } \;\; .
\label{eq:chi_cont}
\eeq
A suitable lattice transcription of this last quantity is
\beq
\chi_{\cal R}  (k) \; \equiv \; 
{ < (\sum_h 2 \, \delta_h A_h)^2 >
- < \sum_h 2 \, \delta_h A_h >^2 \over < \sum_h V_h > } \;\;\; .
\label{eq:chi_latt}
\eeq
Note that in the functional integral formulation of 
Eqs.~(\ref{eq:z_cont}) and (\ref{eq:z_latt})
both the average curvature ${\cal R} (k) $ and its fluctuation
$\chi_{\cal R} (k) $ can be obtained by taking derivatives of
the functional $Z_L$ in Eq.~(\ref{eq:z_latt}) with respect to $k$.
Therefore on the lattice one has
\beq
{\cal R} (k) \, \sim \,
\frac{1}{<\!V\!>} \, \frac{\partial}{\partial k} \ln Z_L \;
\label{eq:avr_z} 
\eeq
and 
\beq
\chi_{\cal R}  (k) \, \sim \,
\frac{1}{<\!V\!>} \, \frac{\partial^2}{\partial k^2} \ln Z_L \; ,
\label{eq:chir_z} 
\eeq
just as the analogous continuum quantities in Eqs.~(\ref{eq:avr_cont})
and (\ref{eq:chi_cont}) can be obtained as derivatives of 
the expression in Eq.~(\ref{eq:z_cont}).
In a similar way, the average volume per site is defined as
\beq
< \! V \! > \; \equiv \; { 1 \over N_0 } < \sum_h V_h > \;\;\; ,
\label{eq:avv} 
\eeq
and again one has
\beq
< \! V \! > \; = \; - \, \frac{\partial}{\partial \lambda_0} \, 
{ 1 \over  N_0} \, \ln Z_L \;\;\; .
\label{eq:avv_z} 
\eeq
Furthermore, its fluctuation $\chi_V$ can also be obtained as a 
second derivative of $Z_L$ with respect to the bare cosmological constant $\lambda_0$.
A simple scaling argument, based on neglecting the effects
of curvature terms entirely (which vanish in the vicinity of the critical
point), is found to give a rather accurate estimate for the average 
volume per edge
\beq
< \! V_l \! > \; \sim \; { 2 \, ( 1 + \sigma d ) \over \lambda_0 \, d }
\;\; \mathrel{\mathop\rightarrow_{ d=4, \; \sigma=0 }} \;\;
{ 1 \over 2 \, \lambda_0 } \;\; .
\eeq
In four dimensions numerical simulations agree quite well with this simple formula.
Finally, a set of exact sume rules can be derived from the scaling properties 
of the action and measure in Eq.~(\ref{eq:z_latt}).
As an example, for the case of the $dl^2$ measure one finds the
following exact lattice Ward identity
\beq
2 \lambda_0 < \sum_h V_h > \; - \; 
k < \sum_h \delta_h A_h > \; - \;  N_1  \; = \; 0 \;\;\; ,
\label{eq:sum_rule} 
\eeq
which is easily derived from Eq.~(\ref{eq:z_latt}) and the definitions
in Eqs.~(\ref{eq:avr_z}) and (\ref{eq:avv_z}).
Here $N_0$ represents the number of sites in the lattice, and the
averages are defined per site.
For the hypercubic lattices used in this paper, 
$N_1 = 15 \, N_0$, $N_2 = 50 \, N_0$, $N_3 = 36 \, N_0$ and $N_4 = 24 \, N_0$.
The above exact identity can be a useful tool
in establishing the numerical convergence of the integration method used
for the lattice path integral.
\footnote{
As an example, in practice one can achieve that the l.h.s. of Eq.~(\ref{eq:sum_rule})
is zero to about one part in $10^5$ for $200k$ individual lattice edge length
configurations containing around 25 million simplices.}



\vskip 40pt

\section{Diffeomorphism Invariant Gravitational Correlation Functions}

\label{sec:corr}

\vskip 20pt

Generally in a quantum theory of gravity the physical distance between any two
points $x$ and $y$ in a fixed background geometry is determined by the metric
\beq
d(x,y \, \vert \, g) \; = \; \min_{\xi} \; \int_{\tau(x)}^{\tau(y)} d \tau 
\sqrt{ \textstyle g_{\mu\nu} ( \xi )
{d \xi^{\mu} \over d \tau} {d \xi^{\nu} \over d \tau} \displaystyle } \;\; .
\eeq
Because of quantum fluctuations the latter depends on the metric
or, equivalently, in the lattice case on the edge length configuration considered.
Correlation functions of local operators need to account for this
fluctuating distance, and as a result these correlations have
to be computed at some fixed geodesic distance between 
a set of given spacetime points \cite{lesh84,cor94}.
In addition, in gravity one generally requires that the local
operators involved should be coordinate scalars.
In principle one could also smear such operators over a small region
of spacetime, an option which will be discussed later.
It is also possible to compute nonlocal gravitational observables in
analogy to what is done in Yang-Mills theories, by defining objects
such as the gravitational Wilson loop (which carries information about
the parallel transport of vectors around large loops, and therefore
about large scale curvature) \cite{modacorr,modaloop,npb400,loops},
or the correlation between Wilson lines closed by the lattice
periodicity (which can be used for extracting the static potential
in quantum gravity) \cite{modapot,lines}. 
One more different type of gravitational correlation was studied in \cite{smit}.

A fundamental correlation function in te quantum theory of gravity is the one 
associated with the scalar curvature, with physical points $x$ and $y$ separated 
by a fixed geodesic distance $d$
\beq
G_R (d) \; \sim \; < \sqrt{g} \; R(x) \; \sqrt{g} \; R(y) \;
\delta ( | x - y | -d ) >_c \; .
\label{eq:corr_cont}
\eeq
It is then straightforward to define the same type of object on the lattice.
If the lattice deficit angles are averaged over a number of contiguous hinges
which share a common vertex, one is lead to consider the
connected lattice correlation function at fixed geodesic distance $d$
\beq
G_R (d) \; \equiv \; < \sum_{ h \supset x } 2 \, \delta_h A_h \;
\sum_{ h' \supset y } 2 \, \delta_{h'} A_{h'} \;
\delta ( | x - y | -d ) >_c \; .
\label{eq:corr_latt}
\eeq
The need to compute physical distances between points for any given 
metric (or edge length) field configuration
complicates the problem considerably, as compared for example
to ordinary gauge theories, where the distance between points
is assigned a priori based on a fixed immutable underlying lattice structure.
\footnote{Practical useful methods for calculating such diffeomorphism 
invariant correlations were described in detail in \cite{cor94}. 
For each given metric configuration which, properly weighted, contributes to the path integral
one needs to compute both the geodesic distance between any two points,
as well as the correlation between a set of given invariant operators 
centered at those points. By far the most time consuming part
of the calculation is the determination of the actual physical 
distance between any two given points for an assigned background metric configuration.
The latter part can be done by generating a large number of
random walks that start at one of the two points, and then obtaining
the physical distance from the shortest walk.
Alternatively, the geodesic distance can be determined directly from 
the propagator, and specifically the 
exponential decay in distance of a covariantly
coupled lattice scalar field propagator with a given mass.
Either way, the calculation is then later repeated for every metric configuration
contributing to the chosen ensemble, resulting eventually in the
sought-after final average.}

For the curvature correlation at fixed geodesic distance 
one expects at short distances (i.e. distances much shorter
than the gravitational correlation length $\xi$) a power law decay
\beq
< \sqrt{g} \; R(x) \; \sqrt{g} \; R(y) \; \delta ( | x - y | -d ) >_c
\;\; \mathrel{\mathop\sim_{d \; \ll \; \xi }} \;\; 
d^{- 2 n}  \;\;\;\; ,
\label{eq:corr_pow}
\eeq
with the power characterized by a universal exponent $n$;
how $n$ is related to another calculable universal critical
exponent (in particular $\nu$) will be discussed further below.

One notes on the other hand that for sufficiently strong coupling (large $G$, or small
$k$) fluctuations in different spacetime regions largely decouple (the
kinetic or derivative term in Eqs.~(\ref{eq:z_cont}) or
(\ref{eq:z_latt}) is responsible for coupling fluctuations in different
regions, and it comes with a coefficient $1/G$).
In this regime one then expects a faster, exponential decay, 
controlled by the correlation length $\xi$
\beq
< \sqrt{g} \; R(x) \; \sqrt{g} \; R(y) \; \delta ( | x - y | -d ) >_c
\;\; \mathrel{\mathop\sim_{d \; \gg \; \xi }} \;\;
e^{ - d / \xi } \;\;\;\; .
\label{eq:corr_exp}
\eeq
This last result shows that the fundamental gravitational correlation
length $\xi$, if nonzero, can be defined through the long-distance
decay of the connected invariant correlations at fixed geodesic distance $d$.
\footnote{
This rather general result can be proven easily by using the same type of arguments
used in ordinary field theories (and lattice gauge
theories) to show that Euclidean correlation functions generally decay 
exponentially at strong coupling. 
There one shows that it takes $n$ actions of the kinetic (or hopping) term to
connect, via the shortest possible lattice path, two points that 
are $n$ lattice sites apart.
A similar result holds for lattice gravity, where the relevant kinetic or hopping
term is the curvature ($R$) contribution, proportional
to $1/G$ \cite{larged}. 
Of course in the extreme limit of infinite $G$, due to
the absence of a kinetic term, fluctuations
in the fields at different spacetime locations completely decouple,
and in this limit the correlation length shrinks to zero (or more
precisely, to one lattice spacing).}
Note also that the behavior in Eq.~(\ref{eq:corr_pow}) is expected to hold at
short distances, i.e. distances much larger than the fundamental
lattice spacing but significantly shorter than the correlation
length, $ l_0 \ll d \ll \xi $, whereas
the behavior in Eq.~(\ref{eq:corr_exp}) is expected to hold
at much larger distances, $ d \gg \xi \gg l_0 $.
In either case, in order to reach the lattice continuum limit
the distances considered need to be much larger than the
fundamental lattice spacing, $ d, \xi \gg l_0 $.
This last constraint is referred to as the scaling limit, where short distance lattice
artifacts are presumably washed out, and the true (and physically relevant)
continuum limit is expected to emerge.
Later it will be shown, from rather elementary scaling considerations,
that the exponent $n$ in Eq.~(\ref{eq:corr_pow}) is related to the
so-called correlation length exponent $\nu$ in four dimensions by $n=4-1/\nu$.

Another key result of relevance here lies in the fact that the local curvature fluctuation 
defined in Eqs.~(\ref{eq:chi_cont}) and (\ref{eq:chi_latt}) is directly 
related to the connected curvature correlation of
Eqs.~(\ref{eq:corr_cont}) and (\ref{eq:corr_latt}) at zero momentum
\beq
\chi_{\cal R} \; 
\sim \; { \int d^4 x \int d^4 y < \sqrt{g(x)} \, R(x) \; \sqrt{g(y)} \, R(y) >_c
\over < \int d^4 x \sqrt{g(x)} > } \;\;\; ,
\label{eq:corr_chi}
\eeq
a well-known and rather useful result already in ordinary field theories.
This connection will be used extensively further below, and
its relevance will lie in the fact that it allows one to relate the
exponent $\nu$, obtained for example from the curvature fluctuation
of Eq.~(\ref{eq:chi_cont}), to the physical correlation in
Eqs.~(\ref{eq:corr_cont}) and (\ref{eq:corr_pow}).
The latter can in turn be related to other physical correlations,
such as the matter density correlation, by the use for example of the
effective, long distance gravitational field equations.


\vskip 40pt

\section{Correlations Between Smeared and Nonlocal Operators}

\label{sec:smear}

\vskip 20pt


The discussion up to this point has dealt with local operators
and their correlations, i.e. operators defined at a point $x$ in spacetime, or 
on the lattice at a single lattice point.
Some rather mild nonlocality does in fact appear, due to the circumstance that
the gravitational action involves, via the affine connection and the
Riemann tensor, the parallel transport of a test vector around an 
infinitesimally small loop.
The latter is encoded on the lattice by the deficit angles, which
describe the parallel transport of a vector around a loop whose
size is comparable to the lattice spacing, or in physical terms
of size comparable to the ultraviolet cutoff or the Planck length.
It is nevertheless possible to define smeared operators, which
involve a new length scale: the linear size of the smearing
volume $r_s$.
On the basis of rather general renormalization group arguments one expects
correlations for these operators to have milder short distance 
divergences.

Consider first the average of an operators over a spherically shaped
smearing region $\Omega (x,r_s)$, centered at the point $x$ and of
linear size $r_s$.
In other words, all points within a physical distance $r_s$ from
the point in question are considered; 
on a lattice of course the number of points within a given physical neighborhood
of the point $x$ with linear size $r_s$ will in general be finite.
Then define the smeared operator $ {\cal O}_S (x) $ by the spacetime average
\beq
{\cal O}_S (x) \; \equiv \; \int_{\Omega (x, r_s )} d^4 z \, \sqrt{g} \; {\cal O} (z) \;\; .
\label{eq:sme_op}
\eeq
A natural candidate operator for smearing is of course the scalar
curvature, but various curvature squared terms would also be viable,
for example.
A suitable invariant correlation function is then defined as
\beq
G_{r_s} (d) \; = \; 
< {\cal O}_S (x) \; {\cal O}_S (y) \; \delta ( | x - y | -d ) >_c \; ,
\label{eq:sme_corr}
\eeq
where again the correlation between the two operators  $ {\cal O}_S (x) $
is taken at a fixed geodesic distance $d$.
The general expectation is that the short distance ($ d \gsim r_s $) behavior
for this correlation function is less singular than for 
correlations of operators defined at
a single point.
Nevertheless at larger distances $ d \gg r_s $ the asymptotic decay
of the correlation function should be the same as for the one in
Eq.~(\ref{eq:corr_pow}), provided the operators in question have 
the same quantum numbers.


\begin{figure}
\begin{center}
\includegraphics[width=0.7\textwidth]{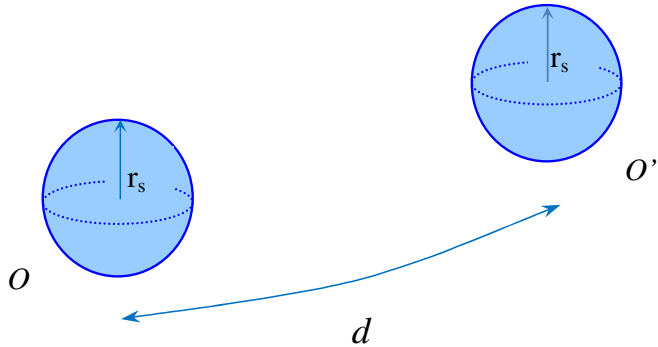}
\end{center}
\caption{
Correlation function for two smeared operators of size $r_s$, separated by a geodesic
distance $d$.
}
\end{figure}

A second class of invariant correlation functions for smeared operators
involves the correlation of parallel transport loops \cite{npb400,modacorr}.
First note that infinitesimal transport loops appear already in
the definition of the correlation function for
the local scalar curvature, as in Eqs.~(\ref{eq:corr_cont}) and (\ref{eq:corr_latt}).
Next consider the parallel transport of a vector around a loop $C$
which is {\it not} infinitesimal; 
in the following this loop will be assumed to be close to planar,
a well defined geometric construction described in detail in \cite{loops}.
First define the total rotation matrix ${\bf U}(C)$ along the path $C$ via a 
path-ordered (${\cal P}$) exponential of the integral of the
affine connection $ \Gamma^{\lambda}_{\mu \nu}$
\beq
U^\mu_{\;\; \nu} (C) \; = \; \Bigl [ \; {\cal P} \, \exp
\left \{ \oint_{C}
\Gamma^{\cdot}_{\lambda \, \cdot} d x^\lambda
\right \}
\, \Bigr ]^\mu_{\;\; \nu}  \;\; .
\label{eq:rot_cont}
\eeq
The lattice action itself already contains contributions from infinitesimal
loops, but more generally one might want to consider near-planar, 
but noninfinitesimal, lattice closed loops $C$.
Along such a closed loop the overall rotation matrix is given by a
product of elementary rotations defined along the lattice path
\beq
U^{\mu}_{\;\; \nu} (C) \; = \;
\Bigl [ \prod_{s \, \subset C}  U_{s,s+1} \Bigr ]^{\mu}_{\;\; \nu}  \; .
\label{eq:latt_wloop_a}
\eeq
In analogy with the infinitesimal loop case, one expects for the 
overall rotation matrix
\beq
U^{\mu}_{\;\; \nu} (C) \; \approx \; 
\Bigl [ \, e^{\delta (C) \omega (C))} \Bigr ]^{\mu}_{\;\; \nu}  \;\; ,
\eeq
where $\omega_{\mu\nu} (C)$ is an area bivector perpendicular to the
loop and $\delta (C)$ the corresponding deficit angle.
This will work if the loop is close to planar, so
that $\omega_{\mu\nu}$ can be taken to be approximately constant
along the path $C$, or defined by some suitable average
over the loop.
Here by a near-planar loop around the point $P$ what is meant is
a loop that is constructed by drawing outgoing geodesics 
on a plane through $P$, so that this unit bivector plays 
the role of a normal to the loop.
A coordinate scalar can be defined 
by contracting the above rotation matrix ${\bf U}(C)$ 
with the appropriate unit length bivector, namely
\beq
W_C \; = \; \omega_{\mu\nu}(C) \; U^{\mu\nu} (C) 
\label{eq:loop}
\eeq
where the bivector $\omega_{\alpha\beta} (C )$ is taken to be representative 
of the overall geometric features of the loop.
Now if the parallel transport loop in question is centered at the point
$x$, then one can define the operator $W_C (x)$ by
\beq
W_C (x) \; = \; \omega_{\mu\nu}(C,x) \; U^{\mu\nu} (C,x) 
\label{eq:loop_x}
\eeq
with the near-planar loop centered at $x$ and of linear size $r_C$.
A suitable invariant correlation two-point function for these 
operators is then defined as
\beq
G_{C} (d) \; = \; < W_C (x) \; W_C (y) \; \delta ( | x - y | - d ) >_c
\;\; ,
\label{eq:sme_loop}
\eeq
where again the correlation between the loop operators  $ W_C (x) $
is taken at some given fixed geodesic distance $d$.
Of course for {\it infinitesimal} loops one recovers the expressions given
earlier in Eqs.~(\ref{eq:corr_cont}) and (\ref{eq:corr_latt}).


\begin{figure}
\begin{center}
\includegraphics[width=0.7\textwidth]{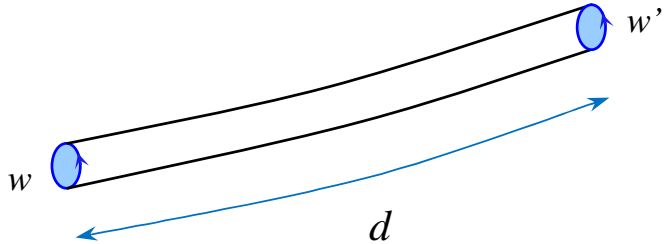}
\end{center}
\caption{
Correlation function of two infinitesimal parallel transport loops, 
separated by a geodesic distance $d$.
}
\end{figure}

In general one needs to specify the relative orientation of the two loops.
So, for example, one can take the first loop in a plane perpendicular to the
direction associated with the geodesic connecting the two points, and
the same for the second loop;
the parallel transport of a vector along this geodesic will then be
sufficient to establish the relative orientation of the two loops.
Nevertheless if one is interested in the analog (for large loops) of the scalar
curvature, then it will be adequate to perform a weighted sum over all
possible loop orientations at both ends.
This is in fact precisely what is done for infinitesimal loops of size $ r_C \sim
a $, if one looks carefully at the way the Regge lattice action is originally defined.
Again here the expectation is that the short distance $ d \gsim r_C $ behavior of this
correlation function for extended loop objects 
is less singular than for correlations of operators defined at
a point; nevertheless at larger distances $d$ such that 
$ r_C \ll d \ll \xi $ the decay
of these correlation function should be the same as the local ones in
Eq.~(\ref{eq:corr_pow}).

It is possible to give a more quantitative description for the behavior of the loop-loop
correlation function given in Eq.~(\ref{eq:sme_loop}), at least in
the strong coupling limit.
The following estimate is based on the previous results and definitions, and
the important analogy and correspondence of lattice gravity
to non-Abelian gauge theories outlined in \cite{larged,loops}.
First it will be assumed here that the two (near planar) loops are of comparable shape
and size, with overall linear sizes $r_C \sim L$ and perimeter $P \simeq 2 \pi L$.
In addition, the two loops will be separated by a distance $ d \gg L $, 
and for both loops it will be assumed that this separation is much
larger than the lattice spacing, $ d \gg a$ and $L \gg a$.
Then to get a nonvanishing correlation in the strong coupling, large $G$
limit it will be necessary to completely tile a tube connecting the two
loops, due to the area law arising from the use of the Haar
measure for the local rotation matrices at strong coupling, again as discussed 
in detail in \cite{loops}.
In this last paper extensive use is made of a modified first order
formalism for the Regge lattice theory, based on the work
of \cite{cas89}, which then allows the separation of metric degrees of
freedom into local Lorentz rotations and tetrads, as is done in the continuum.
Consequently in this limit one obtains an area law
\beq
G_{C} (d) \; \simeq \; 
\exp \left \{ - \, { 2 \pi L \cdot d \over \xi \cdot \xi_0 (L) } \right \} \; 
= \; \exp \left \{ - \, { A (L,d) \over \xi \cdot \xi_0 (L) } \right \} \; .
\label{eq:sme_loop_asy}
\eeq
Consistency of the above expression with the result for small
(infinitesimal) loops given in
Eqs.~(\ref{eq:corr_pow}) and (\ref{eq:corr_exp}), and the area law for
large loops requires the
following limits for the quantity $ \xi_0 (L) $
\beq
\xi_0 (L) \; \mathrel{\mathop\sim_{ L \; \simeq \; a}} \; a 
\;\;\;\;\;\;\;\;
{\rm and}
\;\;\;\;\;\;\;\;
\xi_0 (L) \; \mathrel{\mathop\sim_{ L \; \gg \; a}} \; \xi \;\;\; .
\label{eq:xi_l}
\eeq
From these results one concludes that the asymptotic decay of correlations for
large loops is fundamentally different in form as compared to the decay of
correlations for infinitesimal loops, with an additional factor of $\xi$ appearing
for large loops.
In other words, the results of Eqs.~(\ref{eq:corr_pow}) and
(\ref{eq:corr_exp}) only apply to infinitesimal loops which probe the
parallel transport on infinitesimal (cutoff) scales, and these
results will have to be suitably amended when much larger loops, 
of semiclassical significance, are considered.


\begin{figure}
\begin{center}
\includegraphics[width=0.7\textwidth]{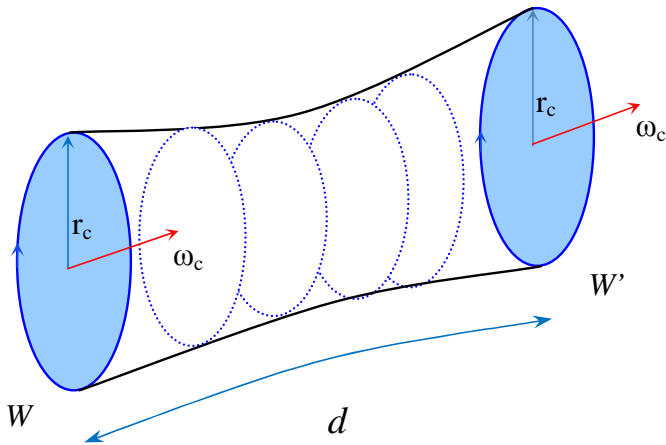}
\end{center}
\caption{
Correlation function for two large parallel transport loops of size
$r_c$ and orientation $\omega_c$, separated by a geodesic distance $d$.
}
\end{figure}



\vskip 40pt

\section{Renormalization Group Scaling Relations for Gravity}

\label{sec:scale}

\vskip 20pt

In this section some of the basic scaling relations for quantum
gravity will be summarized.
It is by now established wisdom, at least in most field theories besides
gravity, that standard scaling arguments allow one to determine the scaling
behavior of local averages, correlation functions and even suitable
nonlocal observables such as the Wilson loop from the knowledge of 
the basic renormalization group behavior, and specifically from
the universal critical exponents. 
The latter generally characterize the singular behavior of local 
averages in the vicinity of the critical point, a nontrivial fixed
point of the renormalization group (RG) in field theory language.
For extensive reviews on the subject see for example \cite{par81,itz91,zin02,car96,bre10}.
There is by now a rather well established body of knowledge in quantum field
theory and statistical field theory on this subject, and there is no 
apparent reason why its basic tenets should not apply to gravity as
well, with quantum gravity describing the unique theory of a massless 
spin two particle coupled to a covariantly conserved energy momentum
tensor \cite{fey}.

It is also understood that in the vicinity of a critical point (seen as equivalent to a
nontrivial fixed point of the renormalization group) long range
correlations arise due to the appearance of a massless particle.
In statistical field theory language, the presence of
a massless particle is reflected in a divergent correlation length
$\xi = 1/m $, or equivalently a power law in the relevant correlation functions.
Let us summarize here the basis for the scaling assumptions for
local averages, fluctuations and their correlations.
\footnote{
It is well established that for theories with a nontrivial ultraviolet
fixed point \cite{wil72,wil75}, 
the long distance (and thus infrared) universal scaling 
properties are uniquely determined, up to subleading corrections
to exponents and scaling amplitudes, by the (generally nontrivial) scaling 
dimensions obtained via renormalization group methods in the vicinity
of an ultraviolet fixed point \cite{par81,zin02,itz91,car96,bre10}. 
These sets of results form the basis of universal
predictions for, as an example, the (perturbatively nonrenormalizable)
nonlinear sigma model \cite{bre76,bre76a}.
The latter provides today one of the
most accurate test of quantum field theory \cite{gui98}, after the $g-2$ prediction
for $QED$
(for a comprehensive set of references, see \cite{zin02,book}, and the
references therein). 
It is also a well established fact of modern renormalization group
theory that in lattice $QCD$
the scaling behavior of the theory in the vicinity of 
the asymptotic freedom ultraviolet fixed point unambiguously determines 
the universal nonperturbative scaling properties of
the theory \cite{gro75}, as quantified by physical
observables such as hadron masses, vacuum condensates,
decay amplitudes, the QCD string tension etc. \cite{hag10,fod12}.}
In brief, since $\log Z$ in either Eq.~(\ref{eq:z_cont}) or (\ref{eq:z_latt})
is both dimensionless and extensive, for a volume $V \sim L^d $ it
has to have the form
\beq
\log \, Z (k) \; = \; 
f_a \; \left ( { L \over l_0 } \right )^d \; + \; 
f_s \; \left ( { L \over \xi } \right )^d
\eeq
where here $f_a$ and $f_s$ are nonsingular functions of dimensionless
parameters, $\xi$ is the fundamental correlation length (the distance
over which quantum fluctuations are strongly correlated),
and $ l_0 \sim a $ the fundamental lattice spacing or ultraviolet cutoff.
Then the free energy or generating function, defined as
\beq
F (k) \; = \; - { 1 \over V} \; \log \, Z (k) \; ,
\eeq
is expected, based on purely dimensional grounds, to aquire 
a singular part $F_{sing} (k) $ such that \cite{scaling}
\beq
F_{sing} (k) \; \sim \; \xi^{-d} \; \; .
\label{eq:z_sing}
\eeq
If one sets for the nonperturbative correlation length $\xi$
\beq
\xi (k) \;\; \mathrel{\mathop\sim_{ k \rightarrow k_c}} \;\; A_\xi \;
( k_c - k ) ^{ - \nu } \; ,
\label{eq:xi_k}
\eeq
where  $ A_\xi $ is the correlation length amplitude, $k_c$ the
critical point and $\nu$ the correlation length exponent 
characterizing the divergence of $\xi$ at the critical point, then one
obtains for the singular part of the free energy
\beq
F_{sing} \;\; \mathrel{\mathop\sim_{ k \rightarrow k_c}} \;\; 
( k_c - k )^{ d \, \nu } \; .
\label{eq:f_sing}
\eeq
One concludes that a divergent correlation length signals the presence of a phase
transition, and this in turn leads to the appearance of 
nonanaliticities in thermodynamic quantities such as $Z(k)$
and the free energy $F(k)$.
The origin of these nonanaliticities in $Z(k)$ lies therefore in the 
divergence of $\xi$ in the vicinity of the critical point at $k_c$,
where the theory becomes scale invariant.

The following results then follow more or less immediately from the definitions in
Eqs.~(\ref{eq:z_cont}) or (\ref{eq:z_latt}), and in Eq.~(\ref{eq:avr_z}).
Near the singularity the average curvature behaves as
\beq
{\cal R} (k) \;\; \mathrel{\mathop\sim_{ k \rightarrow k_c}} \;\;
- \, A_{\cal R} \, ( k_c - k )^\delta \;\; ,
\label{eq:r_sing}
\eeq
with a curvature exponent $\delta$ related to the exponent $\nu$
introduced earlier in Eq.~(\ref{eq:xi_k}) by the scaling relation
$ \delta \, = \, d \, \nu \, - \, 1 $.
Consequently the presence of a phase transition can already
be inferred directly from the appearance of nonanalytic terms in 
invariant local averages, such as the average curvature.
\footnote{
An additive constant could be present in Eq.~(\ref{eq:r_sing}), 
but the evidence so far points to this constant being 
consistent with zero for the Regge lattice gravity theory.}
Similarly, one has for the curvature fluctuation defined in
Eq.~(\ref{eq:chi_latt}), using Eqs.~(\ref{eq:chir_z}) and (\ref{eq:xi_k}),
\beq
\chi_{\cal R} (k) \;\; \mathrel{\mathop\sim_{ k \rightarrow k_c}} \;\;
\delta \, A_{\cal R} \; ( k_c - k ) ^{ -(1-\delta) } \;\;\;\; .
\label{eq:chi_sing}
\eeq
Again, scaling [Eqs.~(\ref{eq:xi_k}) and (\ref{eq:f_sing})] relates 
the exponent $\delta$ appearing in the curvature fluctuation
to $\nu$, so that the exponent in Eq.~(\ref{eq:chi_sing}) is simply
$ 1 \, - \, \delta \; = \; 2 \, - \, d \, \nu $.
These results show that from suitable averages one can extract 
the correlation length exponent $\nu$ [defined in Eq.~(\ref{eq:xi_k})],
without even a need to compute an invariant two-point function, such as the 
ones in Eqs.~(\ref{eq:corr_cont}),
(\ref{eq:corr_latt}), (\ref{eq:corr_pow}) and (\ref{eq:corr_exp}),
provided scaling holds.
Furthermore, in the vicinity of the critical point $k_c$
one can trade the distance from the critical point for the
correlation length $\xi$, and obtain the following
equivalent result relating the quantum expectation value
of the local curvature to the physical correlation length $\xi$,
\beq
{\cal R} ( \xi ) \;\; \mathrel{\mathop\sim_{ k \rightarrow k_c}} \;\;
\xi^{ 1 / \nu - d } \;\; .
\label{eq:r_xi}
\eeq
This last expression is obtained from Eqs.~(\ref{eq:xi_k}) and
(\ref{eq:r_sing}), using $ \delta \, = \, d \, \nu \, - \, 1 $.
Matching of dimensionalities here can always
be achieved by supplying appropriate powers of the lattice spacing
$l_0 \sim a $ or, equivalently, the Planck length $l_P=\sqrt{G}$.

In addition, the above results allow one to relate the fundamental
scaling exponent $\nu$ of Eq.~(\ref{eq:xi_k}) to the 
scaling behavior of some correlation functions at large distances.
Thus, for example, the curvature fluctuation of Eq.~(\ref{eq:chi_cont}) 
is related to the connected scalar curvature 
correlation of Eq.~(\ref{eq:corr_pow}) evaluated at zero momentum 
\beq
\chi_{\cal R} (k) 
\sim { \int d^4 x \int d^4 y < \sqrt{g} R (x) \, \sqrt{g} R (y) >_c
\over < \int d^4 x \sqrt{g} > } 
\;\; \mathrel{\mathop\sim_{ k \rightarrow k_c}} \;\;
A_{\chi} \, ( k_c - k )^{\delta -1} \; .
\label{eq:chi_corr}
\eeq
It follows that a divergence in the curvature fluctuation is
indicative of long range correlations, corresponding to the 
presence of a massless particle, the graviton.
Close to the critical point one expects, in the scaling limit, 
i.e. for physical distances much larger than the fundamental lattice
spacing, a power law decay in the geodesic distance $d$, as in
Eq.~(\ref{eq:corr_pow}),
\beq
< \sqrt{g} R (x) \sqrt{g} R (y) >_c
\;\; \mathrel{\mathop\sim_{ \vert x - y \vert \rightarrow \infty}} \;\;
\frac{1}{ \vert x-y \vert^{2n} } \;\;\;\; .
\label{eq:corr_nu}
\eeq
After inserting the correlation function from Eq.~(\ref{eq:corr_nu}) 
in the expression of Eq.~(\ref{eq:chi_corr}) and then integrating 
over a region of size $\xi$  one obtains immediately for the power 
in Eq.~(\ref{eq:corr_nu}) 
\footnote{
Note that in weak field perturbation theory \cite{modacorr}
$ < \sqrt{g} R (x) \sqrt{g} R (y) >_c 
\; \sim \; < \partial^2 h (x) \, \partial^2 h (y) > 
\; \sim \; 1/ \vert x-y \vert^{d+2}$, 
which is quite different from the result in Eq.~(\ref{eq:corr_nu})
unless $\nu = 2/(d-2) $, which is only correct for $d$ close to two,
where Einstein gravity becomes perturbatively renormalizable and the
corrections to free field behavior become small.
}
\beq
n \; = \; d \, - \, 1 / \nu \;\; .
\label{eq:n_pow}
\eeq
Thus knowledge of the scaling exponent $\nu$ uniquely 
determines the power in Eqs.~(\ref{eq:corr_nu}) and
(\ref{eq:corr_pow}).

So, the scaling theory for gravity just outlined implies a universal relationship
between various quantities and the exponents that appear in them.
Of course the nonperturbative scaling exponents can be determined, in
principle, separately for each individual observable.
Nevertheless scaling theory, based on the assumption of the existence
of a massless particle in the vicinity of an ultraviolet fixed
point, immediately implies a direct (and testable) relationship between the
scaling behavior of various quantities, such as the ones in
Eqs.~(\ref{eq:avr_cont}),(\ref{eq:chi_cont}), 
(\ref{eq:corr_pow}), (\ref{eq:corr_exp}) and (\ref{eq:xi_k}).
Ultimately, the physical relevance of the above results is that 
Eq.~(\ref{eq:xi_k}) gives, when solved for $k$ or $G$, the running
of $G$ with scale in the vicinity of the fixed point at $G_c$.
Thus, again, Eqs.(\ref{eq:r_sing}) and (\ref{eq:chi_sing}) are useful 
for an accurate determination of the universal scaling exponent $\nu$,
and this quantity in turn determines the scaling behavior
of invariant curvature correlation functions in 
Eqs.~(\ref{eq:corr_pow}) and (\ref{eq:corr_nu}) as a function
of geodesic distance.



\vskip 40pt

\section{Continuum Limit of Lattice Quantum Gravity}

\vskip 20pt

The long distance behavior of quantum field theories is, to a great extent,
determined by the scaling behavior of the relevant coupling constants
under a change in momentum scale. 
Asymptotically free theories such as QCD lead to vanishing gauge
couplings at short distances, while the opposite is
true for QED. In general the fixed point(s) of the renormalization
group need not be at zero coupling, but can be located at
some finite $G_c$, leading to nontrivial fixed points
or more complex limit cycles \cite{wil75,par76,gro75,zin02}.

These general ideas are realized concretely in the analytic
$2+\epsilon$ expansion for gravity, and reappear later
in essentially the same form in lattice gravity in four dimensions.
In the $2 + \epsilon$ perturbative expansion for gravity
\cite{epsilon,wei79,eps,aid97}
one analytically continues in the spacetime dimension using
dimensional regularization, and applies perturbation theory about
$d=2$, where Newton's constant becomes dimensionless.
A similar method is quite successful in determining the critical properties
of the $O(n)$-symmetric nonlinear sigma model above two dimensions
\cite{sigma}.
In the framework of this expansion the dimensionful bare coupling is written as
$G_0 = \Lambda^{2-d} \, G $, where $\Lambda$
is an ultraviolet cutoff (corresponding on the lattice to a momentum
cutoff comparable to the inverse average lattice spacing, $\Lambda
\sim 1/ l_0 \sim 1 / a $).
There were originally some known technical difficulties with 
this expansion due to the presence of kinematic singularities for the
graviton propagator in two dimension (the Einstein action is
a topological invariant in $d=2$), but these have been overcome
recently.
In addition, one can show that a gauge-choice dependent
renormalization
of the bare cosmological constant $\lambda_0$ can be completely
reabsorbed into an overall rescaling of the metric, with no
physical consequences.
A double expansion in $G$ and $\epsilon= d-2$
then leads in lowest order to a gauge-independent nontrivial 
fixed point in $G$ above two dimensions
\beq
\beta (G) \, \equiv \, { \partial G \over \partial \log \Lambda } \, = \,
(d-2) \, G \, - \, \beta_0 \, G^2 \, + \cdots \;\; ,
\label{eq:beta} 
\eeq
with $\beta_0 > 0 $ for pure gravity.
To lowest order the ultraviolet fixed point is then at
$G_c \, = \, 1 / \beta_0 (d-2) $.
Integrating Eq.~(\ref{eq:beta}) close to the nontrivial fixed point
one obtains for $G > G_c $
\beq
m_0 \, = \, 
\Lambda \, \exp \left ( { - \int^G \, {d G' \over \beta (G') } } \right )
\;\;  \mathrel{\mathop\sim_{G \rightarrow G_c }} \;\;
\Lambda \, | \, G - G_c \, |^{ - 1 / \beta ' (G_c) } \;\;\;\; ,
\label{eq:m_beta}
\eeq
where $m_0$ an integration constant, with dimensions
of a mass or inverse length, expected to be associated with some physical scale.
It is rather natural here to identify this scale with the inverse of
the gravitational correlation length ($\xi=m^{-1}$),
or some equivalent scale associated with the physical large-scale
curvature \cite{npb400,loops}.
Note that the derivative of the beta function at the fixed point defines
the critical exponent $\nu$, which to this order is independent of $\beta_0$,
$\beta ' (G_c) \, = \, - (d-2) \, = \, - 1/ \nu $.

The previous results clearly illustrate how the lattice continuum limit should
be taken.
It corresponds to $\Lambda \rightarrow \infty$,
$G \rightarrow G_c$ with the physical scale $ \xi = 1/m$ held
constant; 
thus for fixed lattice cutoff the continuum limit is approached by tuning $G$ to $G_c$.
In four dimensions the universal critical exponent $\nu$ is defined 
by [see Eq.~(\ref{eq:xi_k})]
\beq
\xi^{-1} (G) \, \equiv \,  m (G) 
\;\; \mathrel{\mathop\sim_{G \rightarrow G_c }} \;\;
A_m \, \Lambda \, | \, G ( \Lambda ) - G_c |^{ \nu } \;\; ,
\label{eq:mass_k}
\eeq
where $\Lambda = 1 /a$ is the inverse lattice spacing,
and the nonperturbative mass scale $m = 1 / \xi $ is defined
as the inverse of the correlation length, with $A_m$ a
nonperturbative but calculable amplitude.
The cutoff independence of the nonperturbative mass scale $m$ implies
\beq
\Lambda \, { d \over d \, \Lambda } \, m ( \Lambda , G (\Lambda )) 
\; = \; 0 \;\; .
\label{eq:callan_lambda}
\eeq
Comparing results in Eqs.~(\ref{eq:m_beta}) and (\ref{eq:mass_k}) one obtains
\beq
\beta ' (G_c) \, = \, - 1/ \nu \;\; ,
\label{eq:nu_eps}
\eeq
so that the universal exponent $\nu$ is directly related to the 
derivative of the Callan-Symanzik $\beta$ function for $G$ in the 
vicinity of the ultraviolet fixed point.
Thus computing $\nu$ is equivalent to computing the universal 
derivative of the beta function at $G_c$.
\footnote{
As a concrete example, in the $2+\epsilon$ expansion for pure gravity using
the background field method one finds at two loop order
$\nu^{-1} = (d-2) \, + \, {3 \over 5} (d-2)^2 + O( (d-2)^3 )$
\cite{aid97}.
Consistency of this expansion generally requires a smooth background
with a small $\lambda_0$ in Eq.~(\ref{eq:ac_cont}).
Nevertheless a renormalization of $\lambda_0$ there is later undone by
the metric rescaling of Eq.~(\ref{eq:metric_scale}), so that the only physical
(and gauge-choice independent) running is in the gravitational coupling
$G$.}

It is easy to see here that the value of $\nu$ determines the
running of the effective coupling $G(\mu)$ in the vicinity of the
fixed point, where $\mu$ is an arbitrary momentum scale. 
The renormalization group tells us
that in general the effective coupling will
grow or decrease with length scale $ r = 1/ \mu$, depending on whether
$G > G_c$ or $G < G_c$, respectively.
This result follows from the fact that the genuinely 
nonperturbative physical mass parameter $m = \xi^{-1} $ 
is itself {\it scale independent}, and
obeys therefore the rather simple Callan-Symanzik renormalization group equation
\beq
\mu \; { d \over d \, \mu } \; m ( \mu, G(\mu)) \; \equiv \;
\mu \; { d \over d \, \mu } \;
\left \{ \; A_m  \, \mu \; | \, G (\mu) - G_c |^{ \nu } \; \right \}
\, = \, 0  \;\;\;\; .
\label{eq:callan} 
\eeq
Here again, by virtue of Eq.~(\ref{eq:mass_k}), the second expression 
on the right-hand-side is appropriate in the vicinity of the 
ultraviolet fixed point at $G_c$.
Nevertheless the above discussion is not necessarily limited to just a region
in the immediate vicinity of $G_c$;
more generally, if one defines the function $F(G)$ via
\beq
\xi^{-1} \; \equiv \; m \; = \; \Lambda \; F ( G( \Lambda) ) \;\; ,
\label{eq:f_function} 
\eeq
then, from the usual definition of the Callan-Symanzik $\beta$ function
$ \beta (G) \, = \, \partial G (\Lambda) / \partial \log \Lambda $,
one obtains
\beq
\beta (G) \; = \; - \, { F (G) \over F' (G) } \;\; ,
\label{eq:callan_gen}
\eeq
which shows that the renormalization group $\beta$-function, and thus
the running of $G (\mu) $ with scale, can be defined also some
distance away from the nontrivial ultraviolet fixed point
[including therefore the higher order corrections in Eq.~(\ref{eq:callan})].
So, more generally, the running of $G (\mu ) $ is obtained by solving
the differential equation
\beq
\mu \; { d \, G (\mu) \over d \, \mu } \, = \, \beta ( G (\mu) ) \; ,
\label{eq:beta_gen} 
\eeq
with $\beta (G) $ obtained from Eq.~(\ref{eq:callan_gen}).
\footnote{
As an example, for $\beta (G) = - {1\over \nu} (G-G_c) - b \,
(G-G_c)^2 $, where $b$ is some numerical constant, one obtains for
the correlation length
$ \xi^{-1} \, \equiv \, m \, = \, a \, A_m \, (G-G_c)^{\nu} 
\left ( 1 - b \, \nu^2 (G-G_c)^2 + {\cal O} (
(G-G_c)^3 ) \right ) $, which relates a sub-leading correction
to $\beta (G)$ to the sub-leading correction in $ m (G)$.} 
It is then clear from the previous discussion that 
the physical mass scale $m = \xi^{-1}$ determines the magnitude of the scaling
corrections, and plays a role similar to the
scaling violation parameter $\Lambda_{\overline{MS}} $ in QCD
(as in gauge theories, this nonperturbative mass scale emerges in spite of
the fact that the fundamental gauge boson remains strictly {\it massless} to all orders in
perturbation theory, and consequently does not violate local
gauge invariance).
Furthermore, as in gauge theories, one expects in gravity that
the magnitude of $\xi$ cannot be 
determined perturbatively, and to pin down its value requires a fully
nonperturbative approach such as the lattice formulation.

Solving explicitly Eq.~(\ref{eq:callan}) for $G(q^2)$, with
$ q $ an arbitrary wavevector scale, one obtains
\beq
G(q^2) \; = \; G_c \left [ \; 1 \, + \, c_0 \, 
\left ( { m^2 \over q^2 } \right )^{1 / 2 \, \nu} \, 
+ \, O ( \left ( { m^2 \over q^2 } \right )^{1 / \nu} ) \; \right ]
\;\; .
\label{eq:run_k} 
\eeq
Here the amplitude of the quantum correction $c_0$ is directly 
related to the constant $A_m$ in Eq.~(\ref{eq:mass_k}) by
\beq
c_0 \; \equiv \; { 1 \over G_c \, A_m^{1/ \nu} } \;\; .
\label{eq:c_zero}
\eeq
One important point is that the magnitude of the quantum correction in 
Eq.~(\ref{eq:run_k}) depends crucially on the magnitude of the 
nonperturbative physical scale $\xi$.
Also, the expression in Eq.~(\ref{eq:run_k}) does not satisfy general
covariance; this in turn can be fixed by performing the replacement
$ q^2 \rightarrow - \Box $ where $\Box ( g_{\mu\nu} )$ is the
covariant D'Alembertian for a given background metric $g_{\mu\nu} (x) $ \cite{eff}.
\footnote{
In the lattice theory of gravity only the smooth
phase with $G>G_c$ exists (in the sense that an instability develops
and spacetime collapses onto itself for $G<G_c$). 
This then implies that the gravitational coupling can only 
{\it increase} with distance \cite{hw84}.
In other words, a gravitational screening phase does not exist in
the lattice theory of quantum gravity.
This situation appears to be true both for the Euclidean theory
in four dimensions and in the Lorentzian version in $3+1$ dimensions \cite{htw12}.}
This then leads, from Eq.~(\ref{eq:run_k}), to
\beq
G(\Box) \; = \; G_c \left [ \; 1 \, + \, c_0 \, 
\left ( { 1 \over - \xi^2 \, \Box^2  } \right )^{1 / 2 \, \nu} \, 
+ \, \dots \; \right ] \;\; .
\label{eq:run_box} 
\eeq
A set of manifestly covariant effective field equations with a
$G(\Box)$ takes the simple form \cite{eff}
\beq
R_{\mu\nu} \, - \, \half \, g_{\mu\nu} \, R \, + \, \lambda \, g_{\mu\nu}
\; = \; 8 \pi \, G  ( \Box )  \, T_{\mu\nu}  \;\; 
\label{eq:run_field}
\eeq
with the nonlocal contribution coming from the quantum correction
in the $G(\Box)$ of Eq.~(\ref{eq:run_box}).
\footnote{
If general covariance is to be maintained, then it is virtually
impossible here to have a running cosmological term in the field
equations with a $\lambda (\Box)$,
by virtue of the simple fact that covariant derivatives of the metric vanish
identically, $ \nabla_\lambda \, g_{\mu\nu} =0$ \cite{lambda}.}
These nonlocal effective field equations can then be solved for a
number of physically relevant metrics.
For the specific case of a static isotropic metric it is possible to obtain 
an exact expression for $G(r)$ in the limit $r \gg 2 M G$ \cite{eff}.
The result, for $\nu = 1/3 $ exactly, reads
\footnote{
One can show that this exact solution only exists 
provided $\nu = 1/(d-1) $ for $d \ge 4$; otherwise 
no consistent solution to the effective nonlocal field equations with
$G(\Box)$ can be found \cite{eff}.}
\beq
G \; \rightarrow \; G(r) \; = \; 
G \, \left ( 1 \, + \, 
{ c_0 \over 3 \, \pi } \, m^3 \, r^3 \, \ln \, { 1 \over  m^2 \, r^2 }  
\, + \, \dots
\right )
\label{eq:run_r}
\eeq
with $m = 1/ \xi$.
This last result is vaguely reminiscent of the Uehling 
(vacuum polarization) correction to the static potential found in QED.
Generally the expressions in Eqs.~(\ref{eq:run_k}), (\ref{eq:run_box}) and (\ref{eq:run_r})
are consistent with a gradual slow increase in $G$ with large distance
$r$, and with a modified Newtonian potential in the same limit.

The remainder of this paper will deal therefore with establishing firm 
values for the nonperturbative amplitudes and exponents defined in the previous
sections, and later determining both qualitatively and quantitatively their
effects on the running of Newton's $G$ and on the long distance behavior of physical
correlation functions, such as the ones defined in the previous sections.



\vskip 40pt

\section{Average Local Curvature}

\label{sec:curv}

\vskip 20pt

Next we come to a discussion of the numerical methods employed in
this work and the analysis of the results.
For the reader who is not interested in such details, a separate section
later summarizes the most important results obtained so far.
As in previous work, the edge lengths are updated by a
Monte Carlo algorithm, generating eventually an ensemble of configurations distributed
according to the action and measure of Eq.~(\ref{eq:z_latt}).
Details of the method as it applies to pure gravity
are discussed in \cite{lesh84,ham00}, and will not be repeated here.

In this work lattices of size $L^4$ with 
$L=4$ (256 sites, 3,840 edges and 6,144 simplices),
$L=8$ (4,096 sites, 61,440 edges and 98,304 simplices),
$L=16$ (65,536 sites, 983,040 edges and 1,572,864 simplices),
$L=32$ (1,048,576 sites, 15,728,640 edges and 25,165,824 simplices),
and
$L=64$ (16,777,216 sites, 251,658,240 edges and 402,653,184 simplices)
have been considered.
These lattices are all constructed by conveniently dividing up hypercubes into
simplice by introducing suitable diagonals \cite{rowi81}, and periodic boundary
conditions are used throughout.
In general for a lattice with $L^4$ sites one has $15 $ edges per vertex
and $24 $ four-simplices per vertex.
For these lattices one should keep in mind
that due to the simplicial nature of the lattice there are many edges
per hypercube with many interaction terms; as a consequence the
statistical fluctuations already for one single hypercube can be comparatively
small, unless one is very close to a critical point. 
The results presented here are
still preliminary, and in the future it should be possible to
repeat such calculations with improved accuracy on even larger lattices.
Also, while the overall statistics on the $32^4$ lattice seems adequate,
the overall statistics on the $64^4$ lattices was yet far too low
to be usable in the present analysis.

On the $32^4$ lattice up to 200,000 consecutive configurations were
generated for each value of $k$, and 9 different values
for the parameter $k$ were chosen. 
On the $16^4$ lattice up to 800,000 consecutive configurations were
generated for each value of $k$, and 32 different values
for $k$ were chosen. 
In addition, results for different values of $k$ can be considered
as completely statistically uncorrelated, since
they originated from unrelated edge length configurations.
On the smaller $8^4$ lattice 200,000 consecutive configurations were
generated for each value of $k$.
On the $4^4$ lattice two million consecutive configurations were
generated for each value of $k$.
To accumulate enough statistics, runs were performed 
around the clock on a 1200 core machine over a period of roughly four months.
As a result, the increase in accuracy is significant compared to
the results presented in previous work done at the time on a 
dedicated 32-node cluster \cite{ham00}.

In this work the topology is restricted to a four-torus (periodic
boundary conditions). 
One could perform similar calculations with other lattices employing different
boundary conditions or topology, but one expects that the universal
long distance scaling properties of the theory to be determined by 
short-distance renormalization effects, which are generally
independent of the boundary conditions at infinity.
A clear example of this is of course the Feynman diagrammatic expansion for
gravity in $2+\epsilon$ dimensions, where boundary conditions
play no role in the renormalization of the couplings. 
In addition, it will be necessary to impose, based on physical
considerations, the constraint
that the correlation length in lattice units be much larger than
the average lattice spacing, and at the same time much 
smaller than the overall linear size of the system,
$ l_0 \; \lsim \; \xi \; \lsim \; L_0 $, 
where $L_0 \sim V^{1/4}$ here is the linear size of the system, and
$l_0 \sim a $ the (average) lattice spacing.

As stated earlier, the bare cosmological constant $\lambda_0$ appearing in the
gravitational action of Eq.~(\ref{eq:z_latt}) is set $1$ since its
value just sets the overall length scale in the problem.
The higher derivative coupling $a_0$ was also set to $0$ (pure Regge-Einstein action).
It is possible to introduce $R^2$-type terms in the action,
nevertheless in this work these terms were not included in
order not to ``contaminate'' the results with the effects of such
higher derivative terms.
These terms were studied extensively in \cite{ham00}, and their effects
is generally to stabilize the theory at the expense of nonunitary
contributions, which cause a visible oscillatory behavior in curvature
correlations at short distances.
The downside of not including any lattice higher derivative terms is that
the theory eventually develops instabilities very close to the critical point
in $G$, which need to be handled properly by an extrapolation or analytic continuation in $G$.
Nevertheless, as has been shown in \cite{ham00} and in the discussion
further below, such an extrapolation or analytic continuation
is fairly unambiguous, given a large enough amount of
high precision numerical data.
Indeed such an instability is in fact {\it expected}
on the basis of the well-known Euclidean conformal mode contribution,
arising from a kinetic energy contribution for the conformal 
mode with the wrong sign \cite{haw79,gib77}.
Its appearance should therefore be regarded as consistent with the
full recovery of a continuum behavior in the vicinity of the ultraviolet
fixed point at $G_c$.

For the measure in Eq.~(\ref{eq:z_latt}) the above choice of
parameters then leads to a well behaved ground state for 
$k < k_c \approx 0.052 $ for $a=0$ \cite{ham00,monte}.
Given this choice of parameters the system then resides in the 
`smooth' phase, with a fractal dimension close to four; on the other hand
for $k > k_c$ the local curvature can become rather large (`rough' phase),
and lattice spacetime collapses into a degenerate configuration
with very long, elongated simplices and thus more akin to a two-dimensional 
lattice \cite{lesh84,hw84,monte,ham00}.

The results obtained for the average curvature ${\cal R}$
[defined in Eq.~(\ref{eq:avr_latt})] as a function
of the bare coupling $k$ are shown in Figures 4 to 8, on lattices of
increasing size with  $4^4$, $8^4$, $16^4$ and $32^4$ sites.
Figures 5 and 7 show the $32^4$ data by itself.
The errors there are quite small, of the order of
a tenth of a percent or less, and are therefore not visible in the graph.
In \cite{ham00} it was found that as $k$ is varied, the average
local curvature is negative for sufficiently small $k$ ('smooth'
phase), and appears to go to zero 
continuously at some finite value $k_c$.
For $k \ge k_c$ the curvature becomes very large, and
the simplices tend to collapse into degenerate configurations
with very small volumes ($ <\!V\!> / <\!l^2\!>^2 \; \sim 0$).
This collapsed phase corresponds to the region of the usual weak field expansion
($G \sim 0$), characterized by unbounded fluctuations
in the conformal mode.

Accurate and reproducible curvature data can only be obtained for $k$
below the instability point $k_u$ since, as already pointed
out in \cite{ham00}, for $k > k_u \approx 0.052$
an instability develops, presumably associated with the unbounded
conformal mode. Its signature is typical of a sharp first order transition,
beyond which the system tunnels into the rough, elongated phase which
is two-dimensional in nature with no physically acceptable continuum limit.
This instability is caused by the appearance of one or more localized singular
configuration, with a spike-like curvature singularity,
and is clearly driven by the Euclidean Einstein term in the action, 
and in particular its unbounded conformal mode contribution.
Nevertheless an important result that emerges from the lattice
calculations is that for sufficiently strong coupling such singular configurations
are suppressed by quantum fluctuations and thus by the nature of the measure,
which imposes nontrivial constraints coming from the generalized triangle
inequalities.
The lattice results suggest therefore that the conformal instability 
is entirely cured for sufficiently strong coupling.
It is characteristic of first order transitions that the free energy
develops an infinitely sharp delta-function singularity at $k_u$, with the metastable
branch developing no nonanalytic contribution at $k_u$. Indeed it
is well known from the theory of first order transitions
that tunneling effects will lead to a purely imaginary
contribution to the free energy, with an essential singularity for
$k > k_u$ \cite{par81}.
In the following we shall therefore clearly distinguish the instability
point $k_u$ from the true critical point at $k_c$. 
Consequently the nonanalytic behavior of the free energy (and its
derivatives which include, for example, the average curvature) has to be
obtained by {\it analytic continuation} of the Euclidean theory into the metastable
branch. 
This procedure  is then formally equivalent
to the construction of the continuum theory exclusively from
its strong coupling (small $k$ or large $G$) expansion, for example starting from
\beq
Z_L (k) \; = \; \sum_{n=0}^{\infty} a_n k^n \;\; ,
\eeq
\beq
{\cal R} (k) \; = \; \sum_{n=0}^{\infty} b_n k^n \;\; ,
\label{eq:series_r}
\eeq
\beq
\chi_{\cal R} (k) \; = \; \sum_{n=0}^{\infty} c_n k^n \;\; .
\label{eq:series_chi}
\eeq
Given a large enough number of terms in this expansion, the nonanalytic behavior in the vicinity
of the true critical point at $k_c$ can then be determined
unambiguously, using for example differential
or Pade approximants \cite{pade,dombgreen} for suitable combinations which are 
expected to be meromorphic in the vicinity of the true critical point.
In the present case, instead of the analytic strong coupling expansion,
one makes use of a set of (in principle, arbitrarily) accurate
data points to which the expected functional form can be fitted.
What is assumed here then is the kind of regularity which is always 
assumed in extrapolating finite series to the boundary of their radius of convergence.
Ultimately it should be kept in mind though that one is really
interested in the pseudo-Riemannian case, and not the Euclidean
one for which such an instability due to the conformal mode is, as stated before,
to be expected. 
Indeed had such an instability {\it not} occurred one
might wonder if the resulting theory still had any relationship to
the original continuum theory: for the lattice theory one expects such 
an instability to develop at some point, since the continuum theory is 
{\it known} to be unstable for weak enough coupling.  
In conclusion, in  the following only data for $k \le k_u$ will be considered;
in fact to add a margin of safety only $k \le 0.051$ will be
considered throughout the rest of the paper.

To extract the critical exponent $\delta$, one fits the computed values
for the average curvature to the form of Eq.~(\ref{eq:r_sing}).
It would seem unreasonable to expect that the computed values
for ${\cal R}$ are accurately described by this function
even for small $k$, away from the critical point at $k_c$.
Instead, the data is fitted to the above
functional form for either $k \ge 0.02 $ or $k \ge 0.03 $.
Then the difference in the fit parameters can be used as one more measure
for the error. 
In addition, it is possible to include a subleading correction of the form
\beq
{\cal R} (k) \;\; \mathrel{\mathop\sim_{ k \rightarrow k_c}} \;\;
- A_{\cal R} \, \left [ \;  k_c - k + B \; (k_c-k)^2 \; \right ]^\delta
\;\;\;\; ,
\eeq
and use the results to further constraint the uncertainties in
the amplitude $A_{\cal R}$, $k_c$ and the exponent $\delta = 4 \nu -1 $.
Using this set of procedures for $ {\cal R} (k) $ one obtains on a
lattice with $L^4$ sites the following set of estimates
\beq 
L=4 \;\;\;\; k_c = 0.07025(20) \;\;\;\;\;  \nu = 0.357(8)
\eeq
\beq
L=8 \;\;\;\; k_c = 0.05811(27) \;\;\;\;\;  \nu = 0.308(16)
\eeq
\beq
L=16 \;\;\;\; k_c = 0.06134(11) \;\;\;\;\;  \nu = 0.322(6)
\eeq
\beq
L=32 \;\;\;\; k_c = 0.06094(10) \;\;\;\;\;  \nu = 0.320(6) \;\; .
\eeq
Then using the same set of procedures for $ \vert {\cal R} (k) \vert^3 $ 
(which assumes $\nu=1/3$ exactly) one obtains on the same $L^4$ lattices
\beq 
L=4 \;\;\;\;  k_c = 0.06485(20)
\eeq
\beq
L=8 \;\;\;\; k_c = 0.06337(27)
\eeq
\beq
L=16 \;\;\;\; k_c = 0.06377(11)
\eeq
\beq
L=32 \;\;\;\; k_c = 0.06387(9) \;\; .
\label{eq:kc_r3}
\eeq
This last result is presumably the most accurate one, since
it is derived form the largest lattice, with the highest
statistics and the smallest errors on the individual data points.
All of these results are displayed in Figures 4 to 8,
and indicate that the exponent $\nu$ (and therefore $\delta$)
is indeed very close to $1/3$.
Specifically, Figures 6,7 and 8 show a graph of the average 
curvature ${\cal R}(k)$ raised to the third power;
one would expect to get a straight line close to the critical point if
the exponent for ${\cal R}(k)$ is exactly $1/3$. 
The numerical results indeed support such an assumption, and
the linearity of the results close to $k_c$ is quite striking.
The computed data is quite close to a straight line over a wide
range of $k$ values, providing further support for the assumption of
an algebraic singularity for ${\cal R}(k)$ itself, with exponent
close to $1/3$.
This last value can be compared to the old estimate
computed in \cite{ham00}, $\nu \approx 0.33 $.


\begin{figure}
\begin{center}
\includegraphics[width=0.7\textwidth]{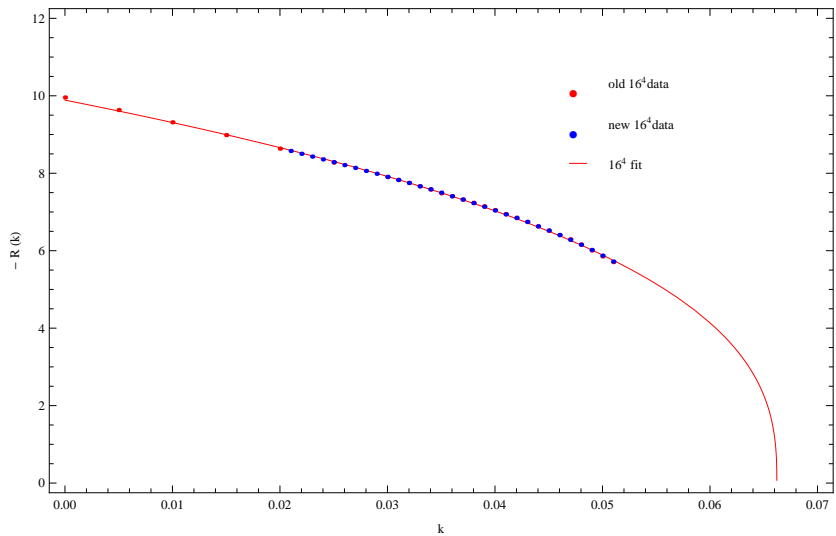}
\end{center}
\caption{
Average local curvature ${\cal R} (k) $ as defined in
Eq.~(\ref{eq:avr_latt}), computed on a lattice with $ 16^4 = 65,536 $ sites.
Statistical errors ($\sim {\cal O}(10^{-4})$)
are much smaller than the size of the symbols.
The continuous line represents a fit of the form $A \; (k_c-k)^{\delta}$
for $k \ge 0.02$, with exponent $\delta = 4 \nu -1$ .
}
\end{figure}


\begin{figure}
\begin{center}
\includegraphics[width=0.7\textwidth]{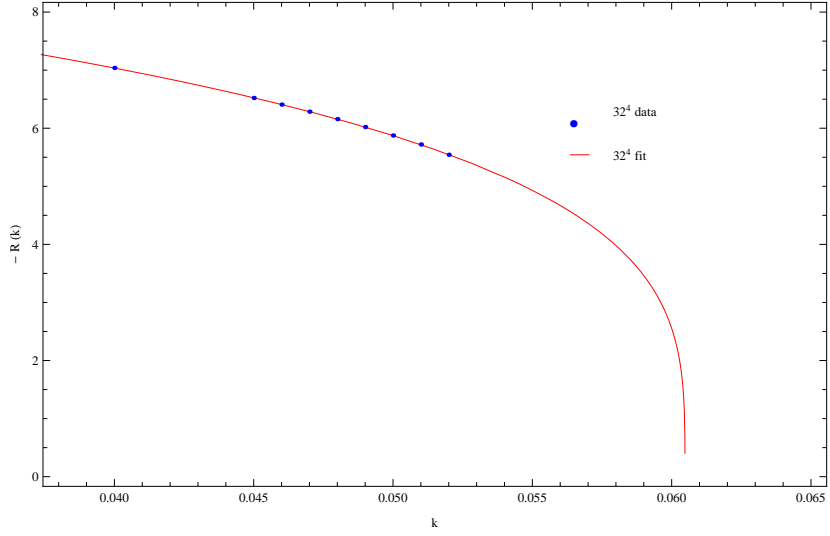}
\end{center}
\caption{
Average local curvature ${\cal R} (k) $ as defined in
Eq.~(\ref{eq:avr_latt}), 
computed on a large lattice with $ 32^4 = 1,048,576 $ sites.
Note the change in horizontal scale as compared to the previous figure.
Statistical errors ($\sim {\cal O}(10^{-4})$)
are much smaller than the size of the symbols.
The continuous line represents a fit of the form $A \;
(k_c-k)^{\delta}$ for $k \ge 0.04$, with exponent $\delta = 4 \nu -1$ .
}
\end{figure}


\begin{figure}
\begin{center}
\includegraphics[width=0.7\textwidth]{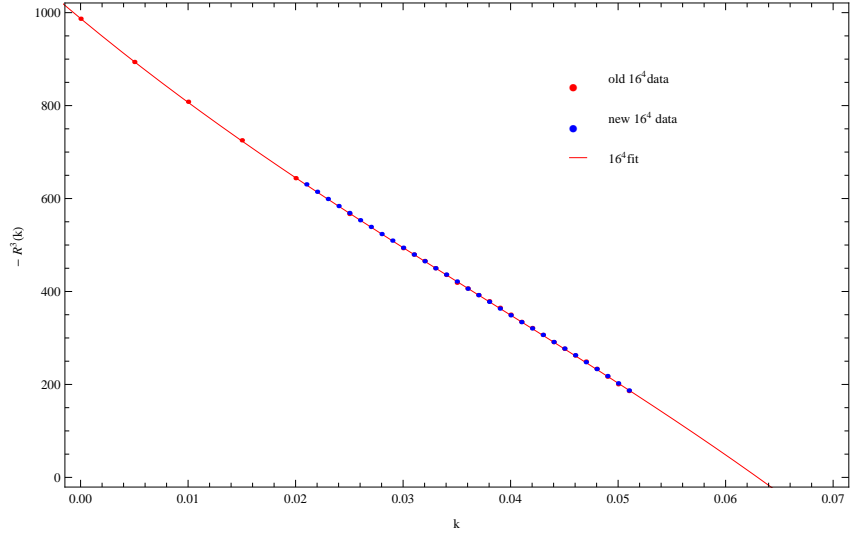}
\end{center}
\caption{
Average local curvature ${\cal R}(k)$ on the $16^4$ lattice, 
raised to the third power.
If $\delta=\nu=1/3$ exactly, then all the data should fall on a
straight line close to $k_c$.
The continuous line here represents a linear fit of the form $A \; (k_c-k)$
for $k \ge 0.02$.
Deviations from linearity of the transformed data are rather small.
}
\end{figure}


\begin{figure}
\begin{center}
\includegraphics[width=0.7\textwidth]{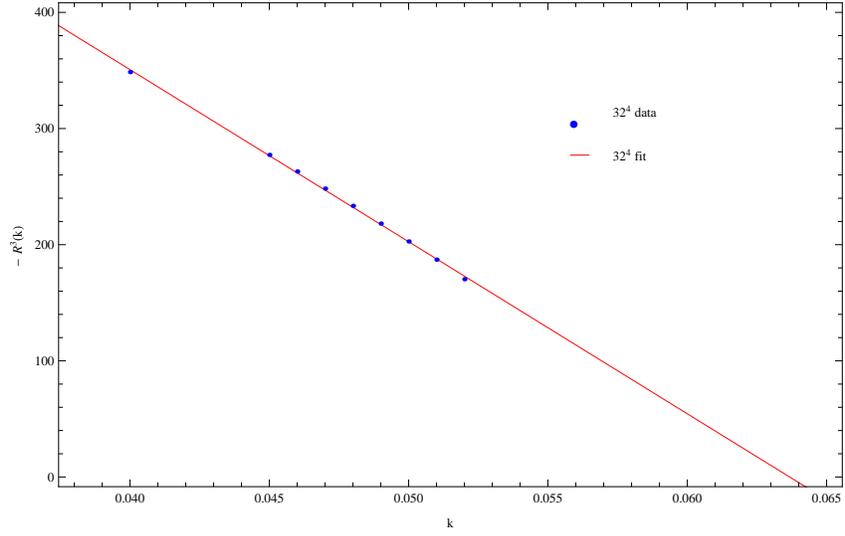}
\end{center}
\caption{
Average local curvature ${\cal R}(k)$ on the larger $32^4$ 
($ 1,048,576 $ sites) lattice, raised to the third power.
Again if $\delta=\nu=1/3$ exactly, then the points should all fall on a straight line.
The continuous line represents a linear fit of the form $A \; (k_c-k)$.
Deviations from linearity of the transformed data are rather small.
}
\end{figure}


\begin{figure}
\begin{center}
\includegraphics[width=0.7\textwidth]{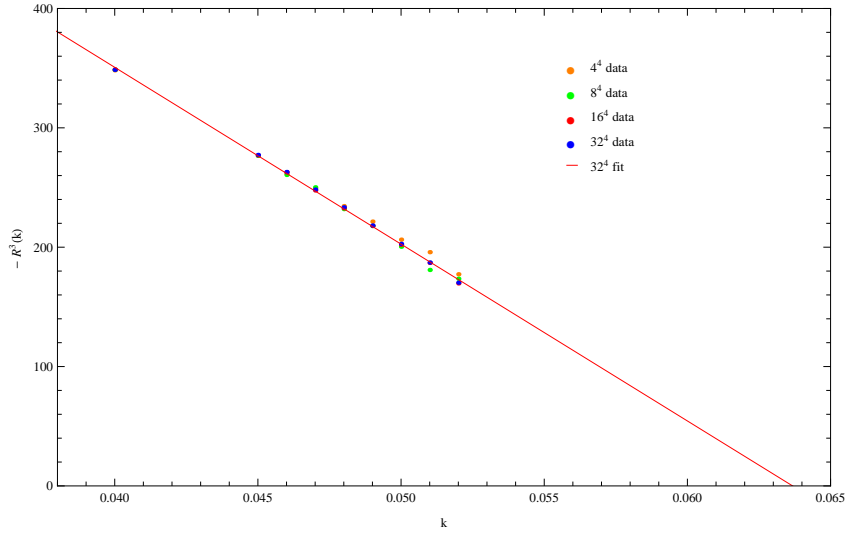}
\end{center}
\caption{
Volume dependence of the average local curvature 
$\vert {\cal R}(k)\vert^3 $ on lattices with $4^4$, $8^4$, $16^4$ and $32^4$ sites.
Again, if $\delta=\nu=1/3$ exactly, then the data should all fall on a
straight line close to $k_c$.
The continuous line represents a linear fit of the form $A \;(k_c-k)$.
The size dependence becomes rather small on the larger lattices, unless
one moves very close to the critical point.
}
\end{figure}


\vskip 40pt

\section{Curvature Fluctutations}

\label{sec:chi}

\vskip 20pt

Figures 9 to 12 show the average curvature fluctuation
$\chi_{\cal R}(k)$ 
defined in Eq.~(\ref{eq:chi_latt}).
At the critical point the curvature fluctuation is expected to diverge, by
definition.
As in the case of the average local curvature 
${\cal R}(k)$ analyzed previously, one can extract the critical exponent
$\delta$ and $k_c$ by fitting the computed values
for the curvature fluctuation to the form given in
Eq.~(\ref{eq:chi_sing}).
And, as for the average curvature itself, 
it would seem unreasonable to expect that the computed values for
$\chi_{\cal R}(k)$ are accurately described by this function
even for small $k$, away from the critical point. 
Instead the data has been fitted to the above
functional form either for $k \ge 0.02 $ or for $k \ge 0.03 $, and
the difference in the fit parameters is then used as a measure
for the error. 
In addition one can include here a subleading correction as well,
of the form
\beq
\chi_{\cal R} (k) \;\; \mathrel{\mathop\sim_{ k \rightarrow k_c}} \;\;
- A_{\chi_{\cal R}}
\left [ \; k_c - k + B (k_c-k)^2 \; \right ]^{-(1-\delta)} \;\;\;\; ,
\eeq
and use the results to further constraint the errors on the
amplitude $A_{\chi_{\cal R}}$, $k_c$ and the exponent $\delta = 4 \nu -1 $.

One finds that the values for $\delta$ and $k_c$ obtained in this fashion
are consistent with the ones obtained from the average curvature
${\cal R}(k)$, but here with somewhat larger errors, since fluctuations
are notoriously more difficult to compute accurately than local averages, and
require therefore significantly higher statistics.
Using these procedures one obtains on the largest lattices with $16^4$ 
and $32^4$ sites
\beq
k_c = 0.05383(102) \;\;\;\;\;  \nu = 0.350(56) \;\;\;\; .
\eeq
Alternatively, one can use for $\chi_{\cal R}(k)$ the best estimate for
$k_c$ obtained earlier from the average curvature.
This then gives
\beq
\nu = 0.321(12) \;\;\;\; ,
\eeq
which is closer to the value obtained from ${\cal R}(k)$.



\begin{figure}
\begin{center}
\includegraphics[width=0.7\textwidth]{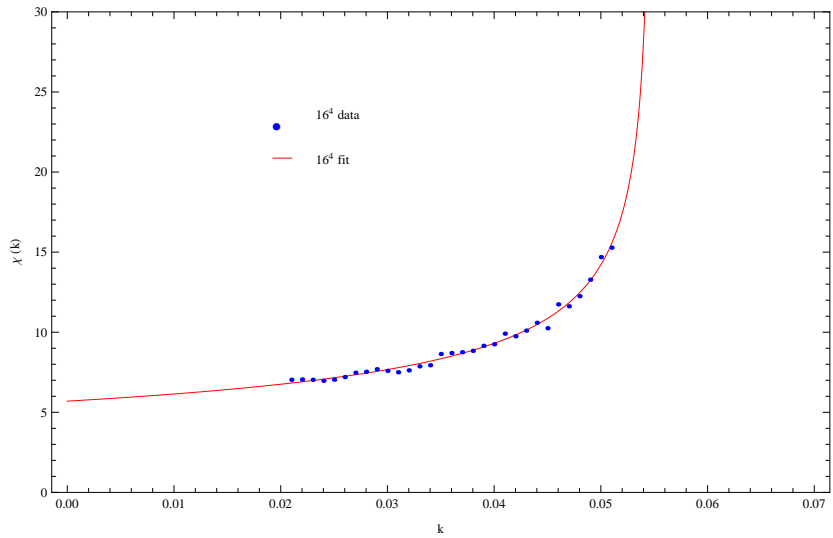}
\end{center}
\caption{
Curvature fluctuation $\chi_{\cal R}(k)$ on lattices with $ 16^4 = 65,536 $ sites.
The continuous line represents a fit of the form
$ \chi_{\cal R} (k) \; = \; A \; (k_c-k)^{ -(1-\delta) } $ for $k \ge
0.02$.
}
\end{figure}


\begin{figure}
\begin{center}
\includegraphics[width=0.7\textwidth]{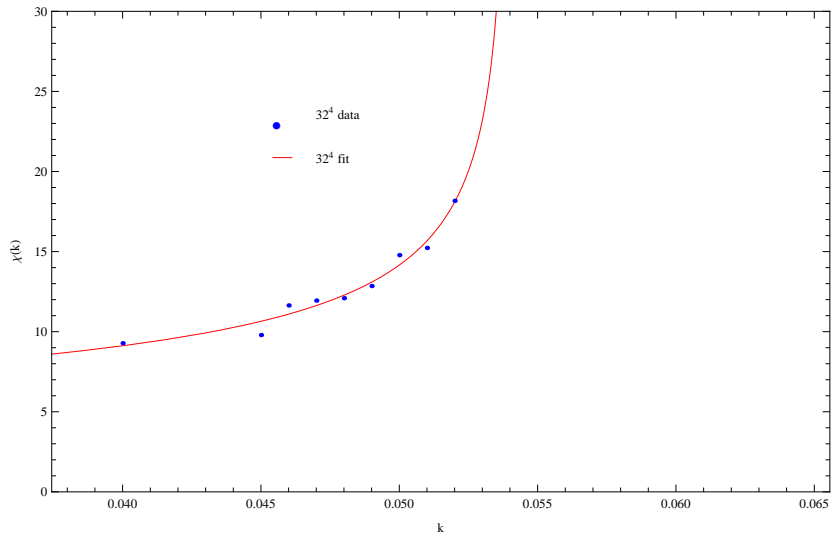}
\end{center}
\caption{
Curvature fluctuation $\chi_{\cal R}(k)$
on the $32^4$ lattice with $ 1,048,576 $ sites.
Note the change in scale from the previous figure.
The line shown is a best fit of the form
$ \chi_{\cal R} (k) \; = \; A \; (k_c-k)^{ -(1-\delta) } $ for 
$k \ge 0.04$.
}
\end{figure}


Figures 11 and 12 show the inverse curvature fluctuation
$\chi_{\cal R}(k)$ on the $16^4$ and $32^4$-site lattices, raised to power $3/2$.
One would expect to get a straight line close to the critical point if
the exponent for $\chi_{\cal R}(k)$ is exactly $-2/3$. 
The computed data is more or less consistent with a linear behavior for
$k \ge 0.03 $, providing further support for an algebraic 
singularity for $\chi_{\cal R}(k)$ itself, with exponent
close to $-2/3$.
Using this last procedure one finds on the largest ($16^4$ and $32^4$)
lattices the improved estimate for the critical point
\beq
k_c = 0.06369(84) \;\;\;\; ,
\eeq
which is consistent with the value obtained earlier
from ${\cal R}^{3}$ (see Figures 6 to 8 and related discussion), and
suggests again that the exponent $\nu$ must be rather close to $1/3$.


\begin{figure}
\begin{center}
\includegraphics[width=0.7\textwidth]{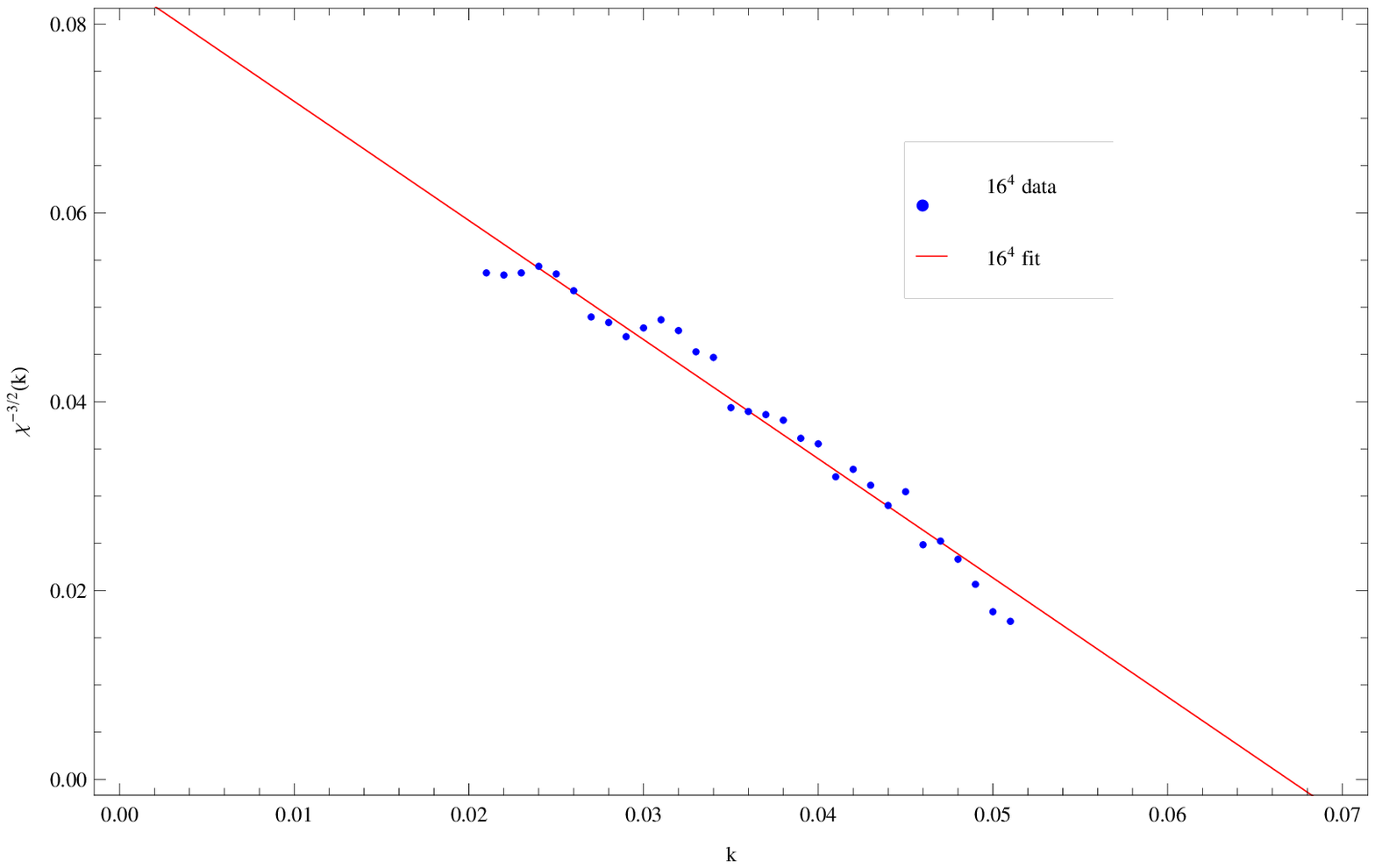}
\end{center}
\caption{
Inverse curvature fluctuation raised
to the power $3/2$, on the $16^4$ lattice;
note that the data is scaled by a factor of $\times 100$.
The straight line represents a linear fit of the form $ A \; (k_c-k) $.
The location of the critical point in $k$ is consistent with the estimate
obtained from the average curvature, but with a somewhat larger error.
}
\end{figure}


\begin{figure}
\begin{center}
\includegraphics[width=0.7\textwidth]{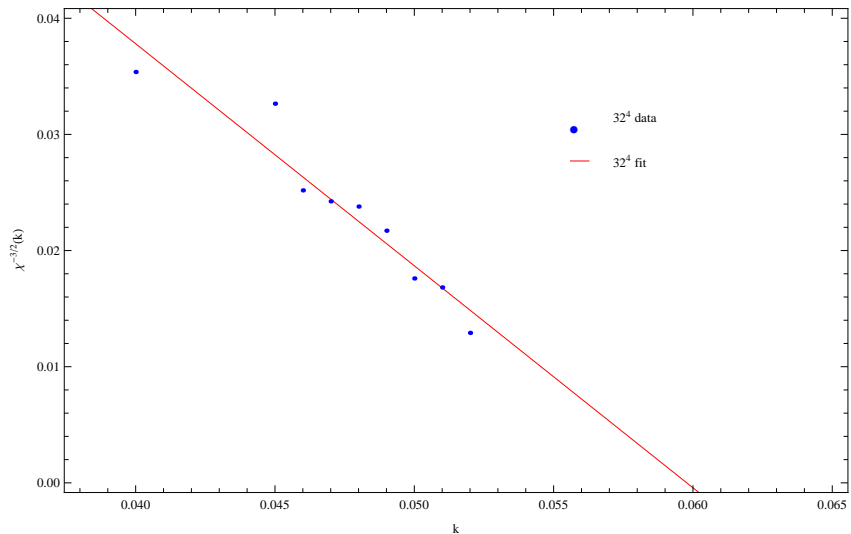}
\end{center}
\caption{
Inverse curvature fluctuation raised
to the power $3/2$, on the $32^4$ lattice;
note that the data is scaled by a factor of $\times 100$.
The straight line represents a linear fit of the form $ A \; (k_c-k) $.
The location of the critical point in $k$ is consistent with the estimate
obtained from the average curvature on the same size lattice, 
here with a larger uncertainty.
}
\end{figure}


In order to check the consistency of the results so far,
it is possible to analyze the previous calculations in a different
way.
From the definition of the average curvature ${\cal R}$ and
curvature fluctuation [Eqs.~(\ref{eq:avr_latt}) and (\ref{eq:chi_latt})],
and the fact that they are both proportional to derivatives of the
free energy $F$ with respect to $k$ [Eqs.~(\ref{eq:avr_z}) and
(\ref{eq:chir_z})], one notices that their ratio is given by
\beq
{ 2 \langle l^2 \rangle \; \chi_{\cal R} (k) \over {\cal R} (k) } \; \sim \;
( \frac{\partial}{\partial k} \ln Z_L ) \, / \,
( \frac{\partial^2}{\partial k^2} \ln Z_L )
\; \sim \; \frac{\partial}{\partial k} \ln 
\left ( \frac{\partial}{\partial k} \ln Z_L \right ) \;\; .
\eeq
The assumption of an algebraic singularity in $k$ for ${\cal R}$
and $\chi_{\cal R}$ (Eqs.~(\ref{eq:r_sing}) and (\ref{eq:chi_sing}))
then implies that the logarithmic derivative as defined above
has a simple pole at $k_c$, with residue $\delta=4\nu-1$
\beq
{ 2 \langle l^2 \rangle \; \chi_{\cal R} (k) \over {\cal R} (k) }
\;\; \mathrel{\mathop\sim_{ k \rightarrow k_c}} \;\;
{\delta \over k - k_c } \;\;\;\; ,
\label{eq:pole} 
\eeq
and the critical amplitudes dropping out entirely for
this particular ratio.
Figures 13 and 14 show the results for the logarithmic derivative
of the average curvature ${\cal R}(k)$, obtained from the data
shown earlier in Figures 4 to 12.
Using this method on the largest $16^4$ and $32^4$ lattices one finds
\beq
k_c = 0.06338(55) \;\;\;\;\;  \nu = 0.3356(84) \;\;\;\; .
\eeq
Note that for the quantity in Eq.~(\ref{eq:pole})
only two parameters are fitted, as opposed to three earlier,
which leads to a slightly improved accuracy.
It is encouraging that the above estimates are in good agreement
with the values obtained previously using the other methods.



\begin{figure}
\begin{center}
\includegraphics[width=0.7\textwidth]{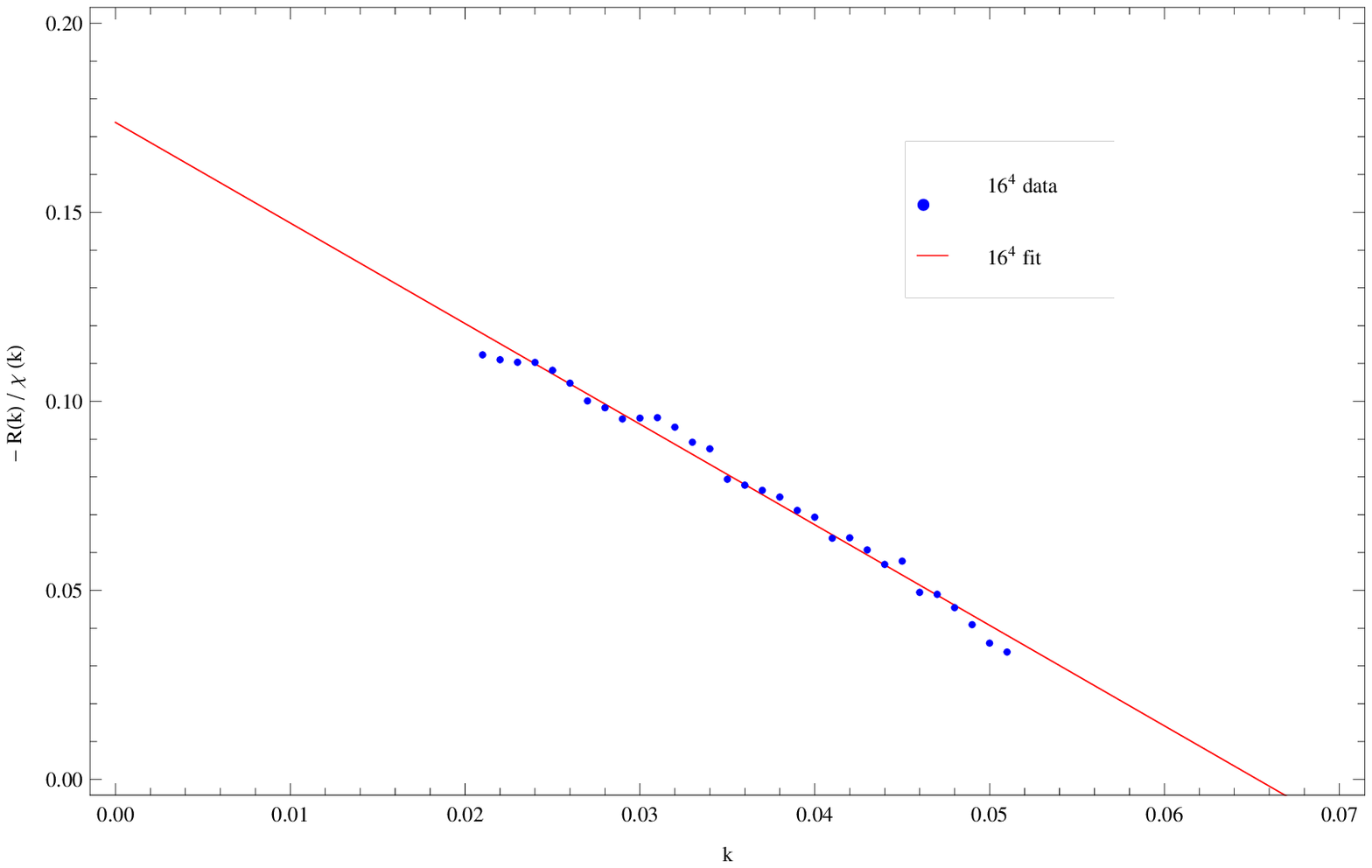}
\end{center}
\caption{
Inverse of the logarithmic derivative of the average
curvature ${\cal R}(k)$ [defined in Eq.~(\ref{eq:pole})]
on the $16^4$ lattice with with $ 65,536 $ sites.
The straight line represents a best fit of the form $ A \; (k_c-k) $ for
$k \ge 0.02$.
The location of the critical point in $k$ is consistent with the 
earlier estimate coming from the average curvature ${\cal R} (k) $
and its fluctuation $\chi_{\cal R} (k) $.
From the slope of the line one then computes directly the exponent $\nu$.
}
\end{figure}


\begin{figure}
\begin{center}
\includegraphics[width=0.7\textwidth]{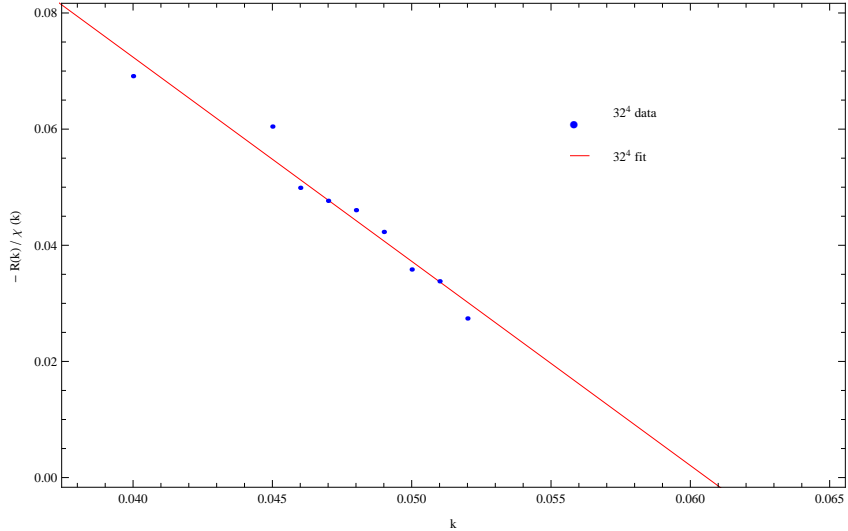}
\end{center}
\caption{
Inverse of the logarithmic derivative of the average
curvature ${\cal R}(k)$ [defined in Eq.~(\ref{eq:pole})]
on the $32^4$ lattice with with $ 1,048,576 $ sites.
The straight line represents a best fit of the form $ A \; (k_c-k) $ for
$k \ge 0.04$.
The location of the critical point in $k$ is consistent with the estimate
coming from the average curvature ${\cal R} (k) $
and its fluctuation $\chi_{\cal R} (k) $.
From the slope of the line one computes directly the exponent $\nu$.
}
\end{figure}


As a further check, it is possible to look at the behavior
of quantities when compared directly to the average local
curvature.
Figure 15 shows a plot of the curvature fluctuation
$\chi_{\cal R} (k)$ versus the curvature ${\cal R} (k) $ (as opposed
to $k$). 
If the average local curvature approaches zero at the critical point 
(where curvature fluctuation diverges), then one would expect these 
curvature fluctuations to diverge precisely at ${\cal R}=0$.
One has from Eqs.~(\ref{eq:r_sing}) and (\ref{eq:chi_sing})
\beq
\chi_{\cal R} ( {\cal R} )
\;\; \mathrel{\mathop\sim_{ k \rightarrow k_c}} \;\;
A \; \vert {\cal R} \vert^{(1-\delta)/\delta}
\; \sim \; A \; \vert {\cal R} \vert^{ (4 \nu -2) / (4 \nu -1) } \;\;\; .
\label{eq:cr_r}
\eeq
An advantage of this particular combination is that it does not require
the knowledge of $k_c$ in order to estimate $\nu$.
Consequently only two parameters are fitted, the overall amplitude
and the exponent in Eq.~(\ref{eq:cr_r}).
Using this method one finds, assuming that the fluctuations
diverge at ${\cal R} = 0 $,
\beq
\nu = 0.3322(71) \;\;\;\; ,
\eeq
which is rather consistent with previous estimates.
Again the error on $\nu$ can be obtained, for example, by reverting
to more elaborate fits of the type
\beq
\chi_{\cal R} ( {\cal R} )
\;\; \mathrel{\mathop\sim_{ {\cal R} \rightarrow 0 }} \;\;
A \; \vert \; {\cal R} + B \, {\cal R}^2 \; \vert^{ 
(4 \nu -2) / (4 \nu -1) } \;\;\; .
\eeq
Note also that for $\nu=1/3$ the exponent simplifies to $-2$, 
and one obtains the simple result (see also Figure 15)
\beq
\chi_{\cal R} ( {\cal R} )
\;\; \mathrel{\mathop\sim_{ {\cal R} \rightarrow 0 }} \;\;
A \; \vert {\cal R} \vert^{-2} \;\;\; .
\eeq
One concludes that the evidence so far supports a vanishing average
local curvature at the critical point, where the curvature fluctuation $\chi_{\cal R}$
and thus the correlation length $\xi$ [in view of
Eqs.~(\ref{eq:chi_sing}) and (\ref{eq:xi_k})] diverge.
These results also show some degree of consistency in the values
for $k_c$ obtained independently from ${\cal R}(k)$ and $\chi_{\cal R} (k)$
(Figures 4 to 15).



\begin{figure}
\begin{center}
\includegraphics[width=0.7\textwidth]{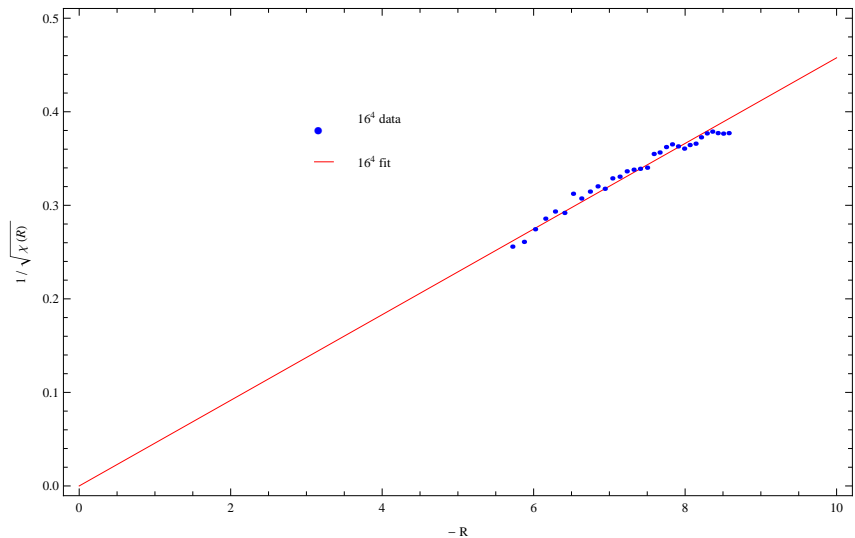}
\end{center}
\caption{
Inverse curvature fluctuation, $1/\sqrt{\chi_{\cal R}}$, versus
the average curvature ${\cal R}$. 
Points shown here are for the largest lattices.
For $\nu=1/3$ exactly, $1/\sqrt{\chi_{\cal R}}$ is expected to be linear in ${\cal R}$
for small ${\cal R}$.
}
\end{figure}



\vskip 40pt

\section{Finite Size Scaling Analysis}

\label{sec:fss}

\vskip 20pt

A further consistency check on the values of the critical exponents
is provided by a systematic finite size scaling (FSS) analysis.
\footnote{
A comprehensive review article can be found in the second of \cite{fis72,barber};
the subject is also covered in numerous books on statistical field theory
\cite{par81,itz91}. 
A systematic field-theoretic derivation of finite-size scaling based
on the renormalization group is given in \cite{qftfss}.
}
Indeed the numerical results presented in the previous sections
have been obtained separately for each lattices of different size.
It would be highly desirable if all those results
could be combined into a single large dataset which then
encompasses all the different lattice sizes, with 
consequently a much higher statistical significance.

Quite in general, the FSS scaling form for a quantity $O$ 
diverging like $ t^{-x_{O}}$ in the infinite volume limit is
\beq
O(L,t) \; = \; L^{x_O / \nu} \; \left [ \tilde{f_O}
\left ( { L \over \xi( \infty ,t) } \right ) \; + \; 
{\cal O} (\xi^{-\omega}, L^{-\omega}) \right ] \;\;\; ,
\label{eq:fss}
\eeq
with $L$ the linear size of the system,
$t$ the reduced temperature or distance from the critical
point, $\tilde{f_O}$ a smooth scaling function,
$\xi(\infty,t)$ the infinite volume correlation length
and $\omega$ a correction to scaling exponent;
but for sufficiently large volumes the correction to 
scaling term involving $\omega$ can be safely neglected.
In the gravity case one has $ t \sim \vert k_c - k \vert $,
and $O (L,t) $ is some physical average such as the local curvature 
${\cal R} (k)$ or its fluctuation $\chi_{\cal R} (k) $, with
the linear size of the system $ L \sim <V>^{1/d} $ .
General properties of the scaling function $\tilde{f_O}(y)$ include
the fact that it is expected to show a peak if the
finite volume value for $O$ is peaked, it
is analytic at $x=0$ since no singularity can develop in a finite
volume, and $\tilde{f_O}(y) \sim \tilde y^{-x_O} $ for large $y$ for a
quantity $O$ which diverges as $t^{-x_O}$ in the infinite volume limit.

The expression in Eq.~(\ref{eq:fss}) is only useful when the 
infinite-volume correlation length $\xi$ is accurately known. 
Nevertheless close to the critical point one can use $ \xi \sim t^{- \nu}$
and then deduce from it the equivalent scaling from
\beq
O(L,t) \; = \; L^{x_O / \nu} \; \left [ \, \tilde{f_O}
\left ( L \; t^{\nu} \right ) \; + \; 
{\cal O} ( L^{-\omega}) \, \right ] \;\;\; ,
\label{eq:fsso}
\eeq
which relies on a knowledge of $t$, and thus of the critical point, instead.


The finite size scaling behavior of the average local curvature, as defined
in Eqs.~(\ref{eq:avr_cont}) and (\ref{eq:avr_latt}) will be discussed
next.
If scaling involving $k$ and $L$ holds according to Eq.~(\ref{eq:fsso}),
with $x_O = 1 - 4 \nu $ the scaling dimension for the curvature,
then all points for different $k$'s and $L$'s should lie on the same universal curve.
From Eq.~(\ref{eq:fsso}), with $t\sim k_c-k$ and $x_O = -\delta = 1 - 4 \nu$,
one has
\beq
{\cal R} (k,L) \; = \; L^{-(4 - 1 / \nu)} \; \left [ \;
\tilde{ {\cal R} } \left ( (k_c-k) \; L^{1/\nu} \right ) \; + \; 
{\cal O} ( L^{-\omega}) \; \right ]
\label{eq:fss_r}
\eeq
where again $ \omega > 0 $ is a correction-to-scaling exponent.
The above argument then suggests that the quantities
\beq
{\cal R} (k,L) \cdot L^{4 - 1 / \nu} \; \sim \; 
\left ( L \over \xi \right )^{4 - 1/ \nu } 
\label{eq:fss_r1}
\eeq
should all lie on a single universal curve when displayed 
as a function of the scaling variable
\beq
x \; \equiv \; (k_c-k) \; L^{1/\nu} \; \sim \; 
\left ( L \over \xi \right )^{1/ \nu } \; .
\label{eq:fss_x}
\eeq
Figure 16 shows a graph of the scaled curvature ${\cal R}(k) \; L^{4-1/\nu}$
for different values of $L=4,8,16,32$, versus the scaled coupling
$ (k_c-k) L^{1/\nu} $. 
The data does indeed support such scaling behavior, and one finds
a best fit for
\beq
k_c = 0.06388(32) \;\;\;\;\;  \nu = 0.3334(4) \;\;\;\; .
\label{eq:kc_fss}
\eeq
Note that the value for $k_c$ found here is in good agreement with
the value given earlier in Eq.~(\ref{eq:kc_r3}).
Thus so far the finite size scaling analysis lead to values
for $k_c$ and $\nu$ which are in good agreement with
what was obtained before, and provides one more stringent
test on the value for $\nu$, which appears to be
again consistent, within errors, with $\nu = 1/3$.



\begin{figure}
\begin{center}
\includegraphics[width=0.7\textwidth]{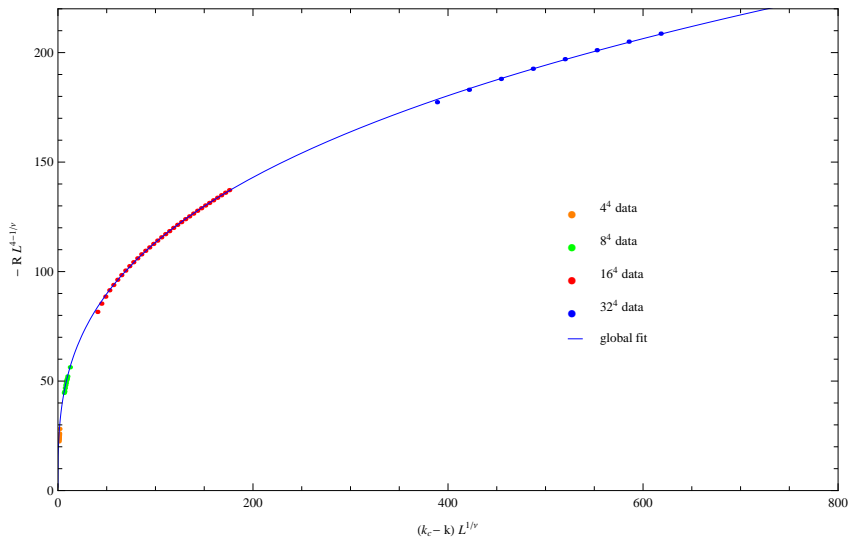}
\end{center}
\caption{
Finite size scaling behavior of the scaled curvature
$ {\cal R} (k,L) \cdot L^{4 - 1 / \nu} $
versus the scaled coupling $ (k_c-k) \cdot L^{1/\nu} $.
Here $L=4,8,16,32$ for the lattice with $L^4$ sites.
Statistical errors are comparable to the size of the dots.
The continuous line represents a best fit to a scaling function
of the form $ a + b \, x^c$, and
finite size scaling predicts that all points should lie 
on the same universal curve. 
The continuous line corresponds to a critical point 
$k_c=0.06388(32)$ and exponent $\nu=0.3334(4)$.
}
\end{figure}


The finite size scaling properties of the curvature fluctuation, defined
in Eqs.~(\ref{eq:chi_cont}) and (\ref{eq:chi_latt}) will be discussed next.
Again, if scaling involving $k$ and $L$ holds according to Eq.~(\ref{eq:fsso}),
with $t \sim k_c-k$ and $x_O = 1 - \delta = 2 - 4 \nu$
then all points should lie on the same universal curve.
From the general form in Eq.~(\ref{eq:fsso}) one expects for
this particular case
\beq
\chi_{\cal R} (k,L) \; = \; L^{2 / \nu - 4} \; \left [ \;
\tilde{ \chi_{\cal R} } \left ( (k_c-k) \; L^{1/\nu} \right ) \; + \; 
{\cal O} ( L^{-\omega}) \; \right ] \;\;\;\; ,
\label{eq:fss_c}
\eeq
where $\omega > 0$ again a correction-to-scaling exponent.
The above arguments then suggests that the quantity
\beq
\chi_{\cal R} (k,L) \cdot L^{4 - 2 / \nu} \; \sim \; 
\left ( L \over \xi \right )^{4 - 2 / \nu } 
\label{eq:fss_c1}
\eeq
should give points all lying on a single universal curve 
when displayed again as a function of the scaling variable 
$x$ in Eq.~(\ref{eq:fss_x}).
Figure 17 shows a graph of the scaled curvature fluctuation
$\chi_{\cal R}(k) / L^{2/\nu-4}$ for different values of $L=4,8,16,32$,
versus the scaled variable $(k_c-k) L^{1/\nu}$.
Using this method one finds approximately
\beq
k_c = 0.06384(40) \;\;\;\;\;  \nu = 0.3389(56) \;\;\;\; .
\label{eq:kc_fss1}
\eeq
Note that the errors in this case are much larger than
for the corresponding average curvature analysis.
Nevertheless the data supports such scaling behavior, and 
suggests again that $\nu$ is close to $1/3$.



\begin{figure}
\begin{center}
\includegraphics[width=0.7\textwidth]{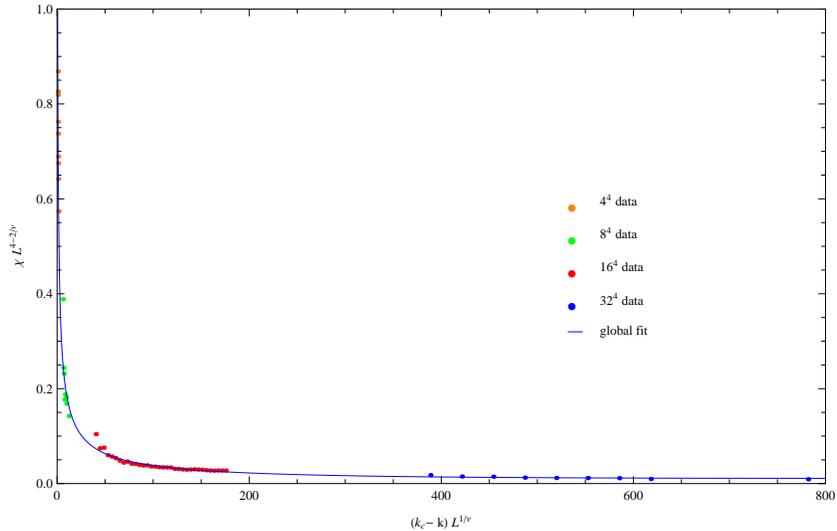}
\end{center}
\caption{
Finite size scaling behavior of the scaled curvature
fluctuation $ \chi_{\cal R} (k,L) \cdot L^{4 - 2 / \nu} $
versus the scaled coupling $ (k_c-k) \cdot L^{1/\nu} $.
Here $L=4,8,16,32$ for a lattice with $L^4$ sites.
The continuous line represents a best fit to a scaling
function of the form $ 1/(a + b \, x^c)$, and
finite size scaling predicts that all points should lie on 
the same universal curve. 
The continuous line corresponds to a critical point $k_c=0.06384(40)$ 
and an exponent $\nu=0.3389(56)$.
}
\end{figure}

The value of $k_c$ itself is expected to have a weak dependence on the
linear size of the system $L_0 \sim V^{1/d}$ .
For a finite system of linear size $L_0$ one anticipates \cite{fis72,barber}
that close to the critical point 
\beq
k_c (L_0) \;\; \mathrel{\mathop\sim_{L_0 \rightarrow \infty}} \;\;
k_c (\infty) \, + \, c \; L_0^{-1 / \nu} \;\;\; .
\label{eq:kc_l0}
\eeq
This is essentially the expression in Eq.~(\ref{eq:xi_k}), with
$\xi \sim L_0 $, and then solved for the finite volume critical point $k_c (L_0)$.
Indeed such a weak size dependence is found when comparing $k_c$ (as obtained
from the algebraic singularity fits discussed previously) on different
lattice sizes. 
Figure 18 shows the size dependence of the critical coupling $k_c$ as
obtained on different size lattices.
In all three cases $k_c (L_0)$ is first obtained from a fit to the average
curvature of the form
${\cal R} (k) \; = \; A \; (k_c-k)^{\delta}$ as in Eq.~(\ref{eq:r_sing}).
Due to the few values of $L$ it is not possible at this point to extract
an estimate for $\nu$ from this particular set of data.
But since $\nu$ is close to $1/3$, it makes sense to use this value
in Eq.~(\ref{eq:kc_l0}), at least as a first approximation.
So if one assumes $\nu=1/3$ exactly and extracts $k_c$ from a linear fit
to $\vert {\cal R} \vert^3 $, then the variations in $k_c$ for
different size lattices are substantially reduced 
(points labeled by smalll circles in Figure 18).
This then gives one additional independent estimate (which now
combines all available lattice sizes, namely $L=4,8,16,32$)
\beq
k_c ( \infty ) \; \simeq \; 0.063862 \;\; 
\label{eq:kc_l}
\eeq
which is in good agreement with the value from the finite
size analysis given in Eq.~(\ref{eq:kc_fss})


\begin{figure}
\begin{center}
\includegraphics[width=0.7\textwidth]{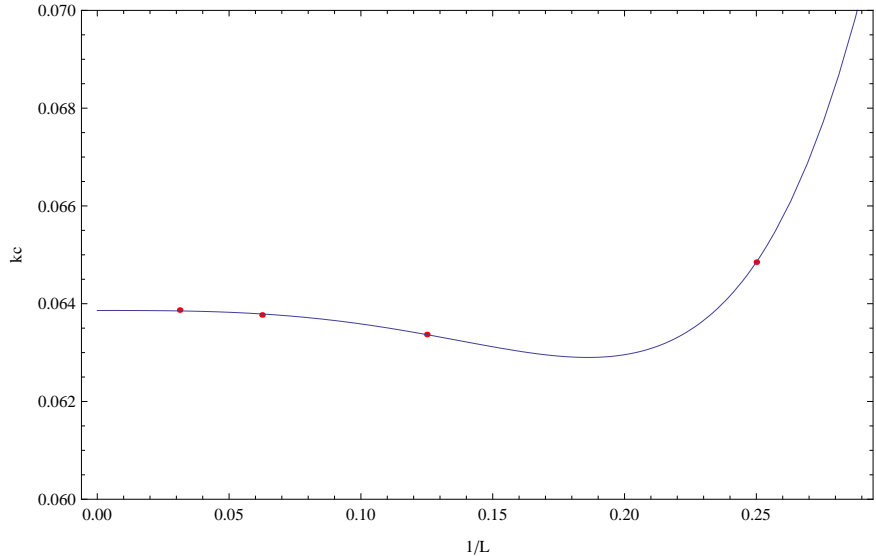}
\end{center}
\caption{
Size dependence of the critical point $k_c (L) $ for different lattices, with $N=L^4$
sites and $L=4,8,16,32$.
The line represents a fit $k_c (L) = k_c (\infty) + A/L^3 + B/L^6 $ and gives
the limiting estimate $ k_c ( \infty ) = 0.0638615 $.
}
\end{figure}

One physicsl quantity of significant interest is the fundamental
gravitational correlation length $\xi(k)$ itself.
It is defined via the exponential decay of physical correlations
[such as the ones given in Eqs.~(\ref{eq:corr_cont}) and (\ref{eq:corr_latt})]
as a function of the geodesic distance between points
[see for example Eq.~(\ref{eq:corr_exp})].
It also appears as the quantity of key significance in the scaling
argument for the free energy [see Eq.~(\ref{eq:z_sing})],
and is expected to diverge in accordance with Eq.~(\ref{eq:xi_k}) 
in the vicinity of the critical point at $k_c$.
The discussion given in the previous sections pointed to
the fact that this quantity
is small and of the order of one average lattice spacing
($\xi \sim l_0 $) in the strong coupling limit (small $k$), and is
expected to increase monotonically towards the critical point at $k_c$
in accordance with Eq.~(\ref{eq:xi_k}).
Indeed all the results presented in the previous two sections
have been analyzed in terms of universal scaling properties
in accordance with the basic assumption of Eq.~(\ref{eq:z_sing}), 
and all the results that follow from it.
From the results presented so far one concludes that the
correlation length exponent $\nu$ defined in Eq.~(\ref{eq:xi_k}) 
is consistent with $\nu =1/3$.

The next step is to fix the correlation length critical amplitude 
$ A_\xi $ as well, which is defined in Eq.~(\ref{eq:xi_k}).
The latter is not obtained in an obvious way from any of
the results presented so far, and requires instead
a direct and separate computation of physical correlations at fixed
geodesic distance, such as the one in Eq.~(\ref{eq:corr_exp}).
These correlations were already computed in \cite{cor94}, and
additional estimates on the correlation length $\xi$ and can be
obtained separately from the size or volume dependence of local averages,
which is expected to behave, for fixed $k \neq k_c$ but close to
the critical point, as
\beq
{\cal R}_{L_0} (k) \;\; \mathrel{\mathop\sim_{ L_0 \, \gg \, \xi }} \;\;
{\cal R}_\infty (k) \, + \, { A \over \sqrt{\xi} \, L_0^{3/2} }
\; e^{- L_0 / \xi } 
\eeq
where here $L_0 \sim V^{1/4}$ is a suitably defined linear size of the system.
Nevertheless, the overall errors for this analysis can be reduced 
significantly if one assumes $\nu=1/3$ exactly (which from the
previous results on $ {\cal R} (k) $ and $ \chi_{\cal R} (k) $
is known to be a very good approximation), and
furthermore if one assumes that the correlation length 
diverges at one and the same critical $k_c$ (also determined
to great accuracy from the previous results for 
${\cal R} (k) $ and $\chi_{\cal R} (k) $).
The latter set of results was largely based on the scaling 
assumption in Eq.~(\ref{eq:z_sing}).
Given these simplifying choices one then obtains
\beq
{\cal R} (k) \;\; \mathrel{\mathop\sim_{ k \, \rightarrow \, k_c }} \;\;
( k_c \, - \, k )^{d \, \nu - 1} \; \sim \; 
\xi^{ 1 / \nu - d}  \; \sim \;  1 / \xi  \;\; ,
\label{eq:r_xi1}
\eeq
and also
\beq
\chi_{\cal R} (k) \;\; \mathrel{\mathop\sim_{ k \, \rightarrow \, k_c }} \;\;
( k_c \, - \, k )^{d \, \nu - 2} \; \sim \; 
\xi^{ d / \nu - 4}  \; \sim \;  \xi^2  \;\; .
\label{eq:chi_xi}
\eeq
Therefore the two combinations $ {\cal R} (k) \cdot \xi (k) $ and
$ \chi_{\cal R} (k) / \xi^2 (k) $ are expected to approach
a constant as $ k \rightarrow k_c $.
Computing these combinations is so far the most accurate 
way of determining the dependence on $k$ of $\xi (k)$,
and in particular for establishing a numerical
value for the key amplitude $ A_\xi $ in Eq.~(\ref{eq:xi_k}).
Via this route one finds close to $k_c$ that
$ {\cal R} \cdot \xi = A_\xi \cdot A_{\cal R} \simeq 19.57 $ and
$ \chi_{\cal R} / \xi^2 = A_\chi / A_\xi^2 \simeq 2.216 $,
which then gives for the correlation length amplitude in 
Eq.~(\ref{eq:xi_k}) the estimate $ A_\xi \simeq 0.80(3)$.
A plot of the correlation length $\xi (k)$ obtained in this way
is shown in Figure 19.
Note that a knowledge of the amplitude $A_\xi$ 
then gives immediately, by the renormalization group
equations in Eqs.~(\ref{eq:run_k}),
(\ref{eq:run_box}) and (\ref{eq:run_r}),
the running of $G$ in the vicinity of the nontrivial
fixed point at $G_c$.

 
\begin{figure}
\begin{center}
\includegraphics[width=0.7\textwidth]{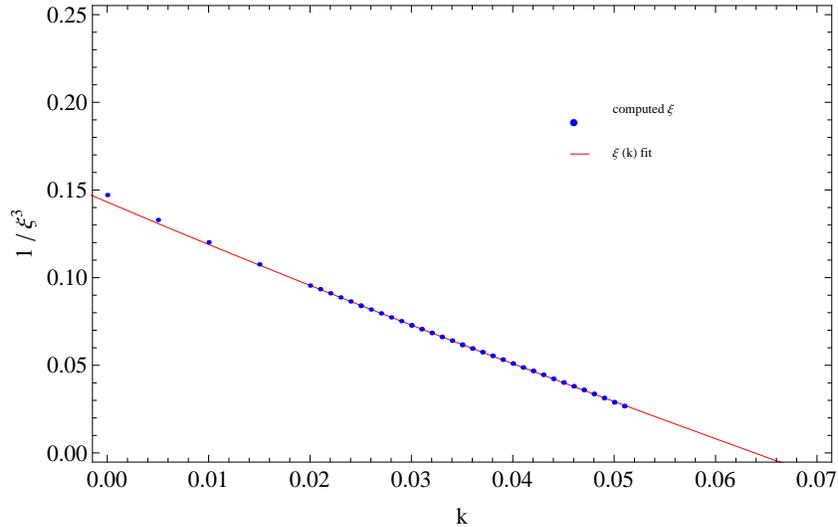}
\end{center}
\caption{
Estimate for the gravitational correlation length $\xi (k)$ versus bare coupling $k$.
For a correlation length exponent $\nu=1/3$ [see Eq.~(\ref{eq:xi_k})], 
$ 1/ \xi (k)^3 $ is expected to be linear in $k$ close to the critical point $k_c$.
}
\end{figure}



\vskip 40pt

\section{Summary of Results}

\label{sec:sum}

\vskip 20pt

Table I summarizes the results obtained for the critical point
$k_c =1/8 \pi G_c$ and for the universal critical exponent $\nu$
obtained so far using a variety of observables and methods.
In view of the detailed discussion of the previous section
one finds from the best data so far (the one with the smallest 
statistical uncertainties, and the least systematic effects)
\beq
k_c = 0.063862(18) \;\;\;\;\;  \nu = 0.334(4) \;\;\;\; ,
\label{eq:kc}
\eeq
which is consistent with the conjecture that $\nu = 1/3$ 
exactly for pure quantum gravity in four dimensions.
In turn this gives for the bare coupling $G$ at the
critical point
\beq
G_c \; = \; { 1 \over 8 \pi k_c } \; = \; 0.623042(25) \;\; .
\label{eq:Gc}
\eeq
In previous work \cite{ham00} the following estimates were given
\beq
k_c = 0.0636(11) \;\;\;\;\;  \nu = 0.33(1) \;\;\;\; ,
\label{eq:kc_old}
\eeq
which have been refined in view of the higher statistics
and larger lattices which are part of the current study.

\begin{table}

\begin{center}
\begin{tabular}{|l|l|l|}
\hline\hline
\\
Observables used to compute $k_c$ and $\nu$ & Critical Point $k_c$ & 
Universal Exponent $\nu$ 
\\
\\ \hline \hline
Average Curvature ${\cal R}$ vs. $k$ & 0.06336(28) & 0.331(4)
\\ \hline
Average Curvature ${\cal R}^{3}$ vs. $k$ & 0.06367(29) & 0.332(2)
\\ \hline
Average Curvature ${\cal R}^{3}$ vs. $k$ & 0.06407(24) & -
\\ \hline
Curvature Fluctuation $\chi_{\cal R}$ vs. $k$ & 0.05383(102) & 0.350(56)
\\ \hline
Curvature Fluctutation $\chi_{\cal R}$ vs. $k$ & - & 0.321(12)
\\ \hline
Curvature Fluctuation $\chi_{\cal R}^{-3/2}$ vs. $k$ & 0.06369(84) & -
\\ \hline
Logarithmic Derivative $ 2 \langle l^2 \rangle \chi_{\cal R}/ {\cal R}
$ vs. $k$ & 0.06338(56) & 0.336(8)
\\ \hline
Curvature Fluctuation $\chi_{\cal R}$ vs. ${\cal R}$ & - & 0.332(7)
\\ \hline \hline
${\cal R}(k,L)$ Finite Size Scaling & 0.06388(11) & 0.333(2)
\\ \hline
$\chi_{\cal R}(k,L)$ Finite Size Scaling & 0.06384(18) & 0.339(6)
\\ \hline
Size Dependence of the Critical Point $k_c (L)$ & 0.063862(30) & -

\\ \hline \hline
\end{tabular}
\end{center}
\label{grav0}

\center{\small {\it
TABLE I. Summary of results for the critical point $k_c$ and the
universal gravitational critical exponent $\nu$, as obtained from the largest 
lattices studies so far.
\medskip}}


\end{table}

\vskip 10pt


Table II provides a comparison between the best lattice estimate
given in Eq.~(\ref{eq:kc}) and the value of the universal 
exponent $\nu$ from other approaches.
These include the calculation of $\nu$ in the $2+\epsilon$ expansion
for gravity \cite{eps} carried out to two loop order \cite{aid97},
a calculation of the same using a truncated renormalization
group approach in four dimensions \cite{reu98,litim}, including
some recent more refined estimates \cite{fal14}.
Further references include a simple geometric argument
based on geometric features of the graviton vacuum polarization
cloud, which gives $\nu = 1 / (d-1)$ for large $d$
\cite{larged},
and the rough estimate for $\nu$ from the lowest nontrivial order strong
coupling expansion for the gravitational Wilson loop \cite{loops},
which gives $\nu =1/2$.
Finally the result of \cite{eff} is mentioned, where
it was found that a solution to the nonlocal effective
field equations of Eq.~(\ref{eq:run_field}) for the static
isotropic metric can only be found provided $\nu = 1 /(d-1)$
exactly for $ d \ge 4 $.
For a plot of the corresponding values for $\nu$ see Figure 20.

\begin{table}

\begin{center}
\begin{tabular}{|l|l|}
\hline\hline
\\
Method used to compute $\nu$ in d=4 & Universal Exponent $\nu$
\\
\\ \hline \hline
Euclidean Lattice Quantum Gravity (this work) &  $\nu^{-1} = 2.997(9)  $
\\ \hline
Perturbative $2+\epsilon$ expansion to one loop \cite{eps} & $\nu^{-1}= 2$ 
\\ \hline
Perturbative $2+\epsilon$ expansion to two loops \cite{aid97} & $\nu^{-1} = 22/5 = 4.40$
\\ \hline
Einstein-Hilbert RG truncation \cite{litim} &  $\nu^{-1} \approx 2.80$
\\ \hline
Recent improved Einstein-Hilbert RG truncation \cite{fal14} &
$\nu^{-1} \approx 3.0 $ 
\\ \hline
Geometric argument \cite{larged} $\rho_{vac \; pol} (r) \sim r^{d-1}$ & $\nu^{-1} = d-1 = 3$
\\ \hline
Lowest order strong coupling (large G) expansion \cite{loops} &
$\nu^{-1} = 2 $
\\ \hline
Nonlocal field equations with $G(\Box)$ for the static metric
\cite{eff} &  $\nu^{-1} = d-1$ for $ d \geq 4$
\\ \hline \hline
\end{tabular}
\end{center}
\label{grav1}

\center{\small {\it
TABLE II. A comparison of estimates for the fundamental scaling
exponent $\nu$, based on a variety of different analytical and
numerical methods.
These include the $2+\epsilon$ expansion for pure gravity carried
out at one and two loops \cite{eps}, an estimate for the leading
exponent in a truncated renormalization group 
expansion \cite{litim,fal14}, a simple geometric argument based on the 
geometric features of the quantum vacuum
polarization cloud for gravity, and finally the only value 
allowed by a consistent 
solution to the nonlocal field equation with a $G(\Box)$
for the static isotropic metric.
\medskip}}


\end{table}
\vskip 10pt

\vskip 20pt

Table III then gives a similar table, where values for
the universal gravitational critical exponent $\nu$ are
given in {\it three} dimensions (for the Euclidean case) or
$2+1$ dimensions (for the Lorentzian case).
Here again it is possible to make a direct
comparison between several approaches, namely
the lattice \cite{gra3d,htw12}, the two-loop $2+\epsilon$ expansion
of \cite{aid97} but with now $\epsilon =1$, the Einstein-Hilbert
truncated renormalization group approach \cite{reu98,litim,fal14}
and the large $d$ estimate of \cite{larged}.
The corresponding values for $\nu$ are shown in Figure 20.


\begin{table}

\begin{center}
\begin{tabular}{|l|l|}

\hline\hline
\\
Method used to compute $\nu$ in $d=3$ & Universal Exponent $\nu$ 
\\
\\ \hline \hline
Euclidean Lattice Quantum Gravity \cite{gra3d} & $ \nu^{-1} = 1.72(5) $
\\ \hline
Exact solution of Lorentzian Gravity (Wheeler-DeWitt Eq.) in 2+1 dim.
\cite{htw12} &  $ \nu^{-1} = 11/6 = 1.8333 $
\\ \hline
Perturbative $2+\epsilon$ expansion to one loop \cite{eps} & $\nu^{-1}= 1$ 
\\ \hline
Perturbative $2+\epsilon$ expansion to two loops \cite{aid97} & $\nu^{-1} = 8/5 = 1.6$
\\ \hline
Einstein-Hilbert RG truncation \cite{litim} &  $\nu^{-1} \approx 1.33 $
\\ \hline
Large $d$ geometric argument \cite{larged} $\rho_{vac \; pol} (r) 
\sim r^{d-1}$ & $\nu^{-1} = d-1 = 2 $
\\ \hline \hline

\end{tabular}
\end{center}
\label{grav2}

\center{\small {\it
TABLE III. A comparison of various estimates for the fundamental scaling
exponent $\nu$ in $2+1$ dimensions, based on a variety of different
analytical and numerical methods.
Included are the $2+\epsilon$ expansion for pure gravity carried
out at one and two loops \cite{eps,aid97}, an estimate for the leading
exponent in a truncated renormalization group expansion \cite{litim},
and a simple geometric argument based on general features of the quantum vacuum
polarization cloud for gravity.
\medskip}}


\end{table}
\vskip 10pt

\vskip 20pt


\begin{figure}
\begin{center}
\includegraphics[width=0.7\textwidth]{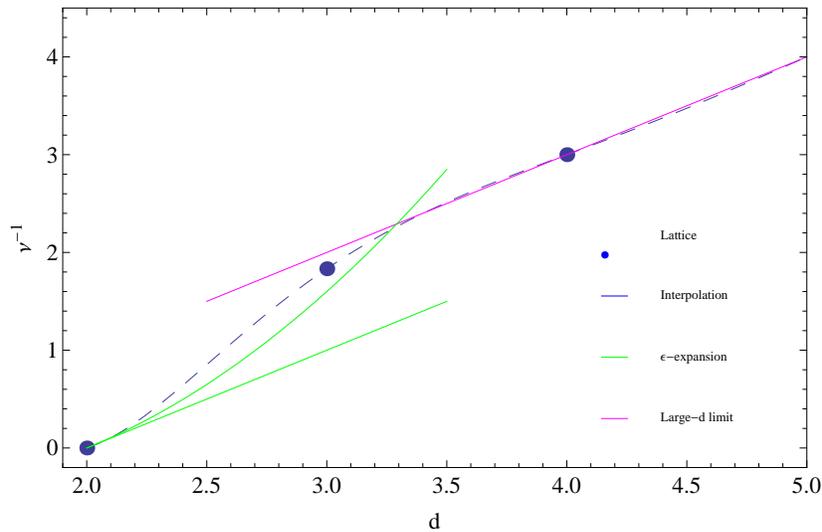}
\end{center}
\caption{
Universal scaling exponent $\nu$ determining the running of $G$ 
[see Eqs.~(\ref{eq:run_k}) and (\ref{eq:run_box})]
as a function of spacetime dimension $d$.
Shown are the results in $2+1$ dimensions obtained from the
exact solution of the lattice Wheeler-DeWitt equation \cite{htw12},
the numerical result in four dimensions (this work and \cite{ham00}), 
the $2+\epsilon$ expansion result to one \cite{eps} and two loops \cite{aid97},
and the large $d$ result $\nu^{-1} \simeq d-1 $ \cite{larged}.
For actual numerical values see Tables II and III.
}
\end{figure}

\vskip 20pt



\vskip 40pt

\section{Implications from a Gravitational Exponent $\nu = 1/3$}

\label{sec:nu}

\vskip 20pt

In this section the consequences of having a definite value
for the critical point $k_c$, as well as a value for the universal
critical exponent $\nu$ [see Eq.~(\ref{eq:kc}], will be discussed.
One notes on the one hand that the value for $k_c$ and thus 
$G_c$ now fixes the value for the ultraviolet cutoff $a$.
At the same time, the specific value for $\nu \simeq 1/3$ gives
predictions for the scaling behavior of local
averages, gravitational correlations and the running of $G$.

First note that the value for the critical point $k_c$ given in Eqs.~(\ref{eq:kc})
and (\ref{eq:Gc}) fixes the lattice spacing $a$, and thus
the value for the ultraviolet cutoff
\beq
G \; \approx \; G_c  \; = 0.623041 \, a^2  \;\; .
\label{eq:gc_phys}
\eeq
From the known laboratory value of Newton's constant $G$, 
$ l_P \equiv \sqrt{ \hbar G / c^3 } $$= 1.616199(97) \times 10^{-33} cm$,
so that one obtains for the fundamental lattice spacing 
$a = 1.2669 \, \sqrt{G_c} \, \equiv \, l_P $, or
\beq
a \; = \; 2.0476 \, \times \, 10^{-33} \, cm \;\; .
\label{eq:a_phys}
\eeq
and from it a definite value for the ultraviolet cutoff $\Lambda \simeq 1/a $.
For the average lattice spacing in units of $a$ one finds
\beq
< \! l^2 \! > \; \equiv \; l_0^2 \; = [ \, 2.398(9) \, a \, ]^2
\label{eq:l0}
\eeq
so that $a$ and $l_0$ are quite comparable in magnitude [this fact can
be traced back to the original overall scale choice $\lambda_0 =1$ 
in Eqs.~(\ref{eq:ac_cont}) and (\ref{eq:ac_latt}), motivated 
by Eq.~(\ref{eq:metric_scale})].

For the average local curvature ${\cal R}(k)$ one has from
Eq.(\ref{eq:r_sing}), using Eq.(\ref{eq:xi_k}) and $\nu =1/3$,
\beq
{ < \int d^4 x \, \sqrt{ g } \, R(x) >
\over < \int d^4 x \, \sqrt{ g } > } \; \sim \; \xi^{1/\nu-d} \; \sim
\; { A'_{\cal R} \over a \; \xi } \;\; .
\label{eq:r_xi_amp} 
\eeq
The dimensionless amplitude $ A'_{\cal R} $ is expected to be $O(1)$ in lattice
units, and is given below in Eq.~(\ref{eq:amp_r_prime}).
This result is based on the fact that the lattice calculations 
allow one to also extract various amplitude coefficients.
For the dimensionless curvature amplitude defined in Eq.~(\ref{eq:r_sing}) one finds
\beq
A_{\cal R} \; = \; 24.46(9) \;\; ,
\label{eq:amp_r}
\eeq
and for the dimensionless curvature fluctuation amplitide defined in 
Eq.~(\ref{eq:chi_sing})
\beq
A_{\chi} \; \equiv \; { 4 \, \nu -1 \over < \! l^2 \! > }  \, A_{\cal R}
\; = \; 1.418(6) \;\; .
\label{eq:amp_chi}
\eeq
Combined with the dimensionless correlation length amplitude defined
in Eq.~(\ref{eq:xi_k})
\beq
A_\xi \; = \; 0.80 (3) \;\;
\label{eq:amp_xi}
\eeq
one finds for the amplitude in Eq.~(\ref{eq:r_xi_amp})
\beq
A'_{\cal R} \; \equiv \; { A_{\cal R} \, A_\xi \over < \! l^2 \! > } 
\; = \; 3.40(13) \;\; .
\label{eq:amp_r_prime}
\eeq
For the curvature fluctuation $ \chi_{\cal R} (k) $ one has 
from Eqs.~(\ref{eq:chi_cont}), (\ref{eq:chi_sing}) and (\ref{eq:chi_xi})
\beq
{ < ( \int d^4 x \, \sqrt{g} \, R )^2 > - < \int d^4 x \, \sqrt{g} \, R >^2
\over < \int d^4 x \, \sqrt{g} > } \; \sim \; 
\xi^{ 2 / \nu - d }  \; \sim \;  A'_{\chi}  \, \xi^2 / a^2 \;\; .
\label{eq:chi_xi2}
\eeq
with dimensionless amplitude
\beq
A'_{\chi} \; \equiv \; { A_{\chi} \, \over A_\xi^2 } \; = \; 
{ 4 \, \nu -1 \over < \! l^2 \! > } \cdot { A_{\cal R} \over A_\xi^2 } 
\; = \; 2.22(9) \;\; .
\label{eq:amp_chi_prime}
\eeq
These results in turn provide useful information for the curvature correlation
function at fixed geodesic distance of Eqs.~(\ref{eq:corr_cont})
and (\ref{eq:corr_latt}).
From Eq.(\ref{eq:corr_nu}) one has for the power appearing in
Eq.~(\ref{eq:corr_pow}) $ 2 n = 2 \, (d - 1 / \nu) = 2(4-3) = 2 $ 
\footnote{
In weak field perturbation theory one finds \cite{modacorr}
$ < \! \sqrt{g} R (x) \sqrt{g} R (y)\! >_c
\; \sim \; < \! \partial^2 h (x) \partial^2 h (y) \! > 
\; \sim \; 1/ \vert x-y \vert^6 $, so the result here is quite
different.
If one defines an anomalous dimension $\eta$ for the graviton propagator
in momentum space, $< \! h \, h \! > \sim 1/k^{2-\eta}$, one finds
$\eta = d-2-2/\nu$ or $\eta =-4$ in four dimensions for $\nu=1/3$,
which deviates significantly from the gaussian or perturbative value.
Such a large deviation is already observed in the $2+\epsilon$ expansion
[see Eq.~(\ref{eq:nu_eps})],
and is not peculiar to lattice quantum gravity.
In gravity such a possibility was already discussed some time ago in \cite{fubini}.}

One then obtains for the curvature-curvature correlation 
function at ``short distances'' $r \ll \xi$ and for $\nu=1/3$
the remarkably simple result
\beq
< \sqrt{g} \; R(x) \; \sqrt{g} \; R(y) \; \delta ( | x - y | -d ) >_c
\;\; \mathrel{\mathop\sim_{d \; \ll \; \xi }} \;\;
{ 1 \over d^{\, 2 \, d - 2/\nu} } \; \sim \; { A_0 \over a^2 \, d^2 }
\;\; .
\label{eq:corr_pow1}
\eeq
Note that in the last term the correct dimensions have been restored,
by inserting suitable powers of the lattice spacing $a$.
It is instructive to compare the above result to the expression for
the local average curvature, Eq.~(\ref{eq:r_xi_amp});
note in particular that both expressions still contain explicitly
the size of the microscopic parallel transport loop $ \sim a \sim l_P$.
Here the dimensionless amplitude $A_0$ is related to the amplitude in 
Eq.~(\ref{eq:amp_chi_prime}) because of Eq.~(\ref{eq:chi_corr}),
and one finds
\beq
A_0 \; \equiv \; { A'_{\chi} \over 2 \pi^2 } 
\; = \; 
{1 \over 2 \pi^2 } \cdot { 1 \over 3 } \cdot { 1 \over < \! l^2 \! > } 
\cdot { A_{\cal R} \over A_\xi^2 } \; = \; [ \, 0.335(20) \, ]^2
\label{eq:corr_amp}
\eeq
so that the dimensionless correlation function
normalization constant is $ N_R \, \equiv \, \sqrt{A_0} = 0.335(20) $.
As expected, all of these amplitudes are close to $O(1)$ in units
of the ultraviolet cutoff (fundamental lattice spacing) $a$.

The exponent $\nu=1/3$ and the amplitude $A_\xi$ now determine
the running of $G$ with scale, see Eq.~(\ref{eq:run_k}), and one
obtains 
\beq
G(q^2) \; = \; G_c \left [ \; 1 \, + \, c_0 \, 
\left ( { m^2 \over q^2 } \right )^{3/2} \, 
+ \, O ( \left ( { m^2 \over q^2 } \right )^{3} ) \; \right ]
\;\; ,
\label{eq:run_xi} 
\eeq
with reference scale $m \equiv 1 / \xi$. 
The coefficient $c_0$ [see Eq.(\ref{eq:c_zero})] determines the amplitude
of the quantum correction, and it is given by
\beq
c_0 \; = \; 8 \pi \, G_c \, A_\xi^{1/\nu} \; = \; A_\xi^3 / k_c \;\; ,
\label{eq:c0}
\eeq
with $ A_\xi = 0.80(3) $ from the numerical solution
[see Eq.~(\ref{eq:amp_xi}); for the definition of the amplitude 
$A_\xi$ see Eq.~(\ref{eq:xi_k})], and also
$\nu=1/3$ and $k_c$ from Eq.~(\ref{eq:kc}).
This then gives for the dimensionless amplitude of the 
leading quantum correction in Eq.~(\ref{eq:run_xi}) 
$c_0 \approx 8.02 $.
This last result can then be translated directly into a covariant
$G (\Box)$ [see Eq.~(\ref{eq:run_box}) with $\nu=1/3$], 
\beq
G(\Box) \; = \; G_c \left [ \; 1 \, + \, c_0 \, 
\left ( { 1 \over - \xi^2 \, \Box^2  } \right )^{3 / 2} \, 
+ \, \dots \; \right ] \;\; .
\label{eq:run_box_xi} 
\eeq
The latter forms the basis for a set of nonlocal effective 
field equations [see Eq.~(\ref{eq:run_field})], and gives
the running of $G(r)$ for the specific choice of the
static isotropic metric [see Eq.~(\ref{eq:run_r})].

In principle it is also possible to estimate the next (sub-leading)
correction to the leading
running of $G$ in Eq.~(\ref{eq:run_xi}), given the knowledge 
of the subleading corrections to $\xi(k)$ or $m(k)$ in 
Eqs.~(\ref{eq:xi_k}) or (\ref{eq:run_xi}).
If one has
\beq
\xi^{-1} (G) \; = \; a^{-1} \, A_m \, ( G \, - \, G_c )^{\nu}
\, \left [ 1 \, - \, b \, \nu^2 \, ( G \, - \, G_c )
\, + \,  {\cal O} ( ( G \, - \, G_c )^2 ) \right ] \;\; ,
\label{eq:xi_k_higher}
\eeq
with a sub-leading correction of amplitude $b$,
then for the running of $G$ one obtains to this order
\beq
{ G(q^2) \over G_c } \; = \; 1 \, + \, c_0 \, 
\left ( { m^2 \over q^2 } \right )^{1 / 2 \nu} \, + \, 
c_1 \, \left ( { m^2 \over q^2 } \right )^{1 / \nu} \, + \, \dots \, ,
\label{eq:run_xi_higher} 
\eeq
with $ c_1 = b \, \nu / ( A_m^{2/\nu} \, G_c ) \approx 2.87 $, given that
$b \approx 0.215$ and $A_m = (k_c/G_c)^{\nu} / A_\xi $, and 
$A_\xi$ given in Eq.~(\ref{eq:amp_xi}).
The domain of validity for the above expression is 
$ q \gg m \equiv 1 / \xi $ or $ r \ll \xi $; 
the strong infrared divergence at $q \simeq 0$ is largely an
artifact of the current expansion, and can be regulated either by
cutting off the momentum integrations at $ q \simeq m = 1 / \xi $,
or by the replacement on the r.h.s. $q^2 \rightarrow q^2 + m^2 $.

Furthermore, the previous results show clearly that the reference
scale for the running of $G$ is set by the correlation length $\xi$, which by
Eqs.~(\ref{eq:r_xi}), (\ref{eq:r_xi1}) and Eqs.~(\ref{eq:r_xi_amp})
appears to be directly related to curvature.
In particular the form of the running of $G$ with scale suggests that
no detectable corrections to classical gravity should arise
until either a) the scale $r$ approaches the very large (cosmological)
scale $\xi$, or b) until one reaches extremely short distances comparable
to the Planck length $r \sim l_p $, at which point higher derivative
terms, light matter corrections and string contributions
come into play.
In other words, the results of Eqs.~(\ref{eq:run_r})
(\ref{eq:run_xi}) or (\ref{eq:run_box_xi}) imply that
classical gravity is largely recovered on atomic, laboratory, solar and
even galactic scales, as long as the relevant distances satisfy $r \ll \xi$.


Therefore one crucial ingredient needed in pinning down the magnitude of
the quantum correction for $G ( q^2 )$ in Eqs.~(\ref{eq:run_xi})
or (\ref{eq:run_box_xi}) is the actual value of the nonperturbative
reference scale $\xi$.
It was argued in \cite{loops} that, in analogy to ordinary
gauge theories, the gravitational Wilson loop provides 
precisely such an insight.
The main points of the argument are rather simple and can
be reproduced in a few lines.
In complete analogy to the gauge theory case, these arguments
basically rely on the concept of universality, the existence
of a universal correlation length at strong coupling,
and the use of the Haar invariant measure to integrate
over large fluctuations of the fundamental local parallel 
transport matrices.
Following \cite{modacorr,modaloop}, in \cite{loops} the vacuum expectation 
value corresponding to the gravitational Wilson loop was defined as
\beq
< W(C)> \; = \; < \tr \left [ \, \omega (C) \; U_1 \; U_2 \; ... \;
  U_n \right ] > \;\; .
\label{eq:wloop_latt}
\eeq
Here the $U$'s are elementary rotation matrices, whose form is 
determined by the affine connection and which therefore
describe the parallel transport of vectors around a loop $C$, see
also Eq.~(\ref{eq:latt_wloop_a}).
Here $\omega_{\mu\nu} (C) $ is a constant unit bivector, characteristic of the overall
geometric orientation of the loop, giving the normal to the loop.
In the continuum the combined rotation matrix ${\bf U}(C)$ is given by the
path-ordered (${\cal P}$) exponential of the integral of the
affine connection $ \Gamma^{\lambda}_{\mu \nu}$, as in
Eq.~(\ref{eq:rot_cont}),
so that the previous expression represents a suitable
regularized and discretized lattice form.
In \cite{loops} it was then shown that quite generally
in lattice gravity for sufficiently strong coupling one 
obtains universally an area law for near planar loops,
\footnote{
A similar result is of course well established in non-Abelian gauge
theories, and by now regarded as standard 
textbook material [see for example, Peskin and Schroeder,
{\sl An Introduction to Quantum Field Theory}, p. 783, Eq. (22.3)
\cite{peskin}].
There $\xi$ represents the gauge field correlation length, defined,
for example, from the exponential decay of connected Euclidean correlations of 
two infinitesimal chromo-magnetic loops separated by a given distance $|x|$.
Following \cite{loops}, we choose to write here the gravitational result
in the same scaling form, involving the invariant gravitational
correlation length $\xi$; an overall, in principle calculable, multiplicative constant
$O(1)$ in the exponent has been set equal to one here.}
\beq
<W(C)> \; \simeq \; \exp \, ( - \, A_C / {\xi }^2 )
\label{eq:wloop_latt1}
\eeq
where $A_C$ is the geometric area of the loop.
This last result relies on a modified first order
formalism for the Regge lattice theory \cite{cas89}, in which
the lattice metric degrees of freedom are separated out into local Lorentz
rotations and tetrads.
Moreover, the result of Eq.~(\ref{eq:wloop_latt}) appears be 
universal since and was shown to hold in all known lattice
formulations of quantum gravity in the strong coupling regime.
In \cite{loops} an expression for the correlation length $\xi$ appearing in 
Eq.~(\ref{eq:wloop_latt1}) was given in the strong coupling limit,
where one finds 
$ \xi = 4 / \sqrt{ k_c \, \vert \log ( k / k_c ) \vert \, + \, O (k^2)}$.
For $k$ close to $k_c$ this gives immediately
$ \xi \, \simeq \, 4 \, \vert k_c - k \vert^{ - 1/2 } $ and thus, to this
order, $\nu = \half $ and also $ A_\xi = 4 $ in Eq.~(\ref{eq:xi_k}).
Nevertheless, the discussion of the previous sections and
the numerical solution of the full lattice theory suggests that 
the correct expression for $\xi$ to be used in
Eq.~(\ref{eq:wloop_latt1}) should be the one in Eq.~(\ref{eq:xi_k}),
with $\nu =1/3 $ [Eq.~(\ref{eq:kc})], $k_c$ given in 
Eq.~(\ref{eq:kc}) and amplitude $A_\xi = 0.80(3) $.

The next step is to make contact between the above results
and a semiclassical description, which requires that one 
connects the nonperturbative result of Eq.~(\ref{eq:wloop_latt1}) to
a suitable semiclassical physical observable.
Indeed, by the use of Stokes's theorem, semiclassically the parallel
transport of a vector round a very large loop depends on the exponential of a 
suitably coarse-grained Riemann tensor over the loop.
In this semiclassical picture one has for the combined
rotation matrix ${\bf U}$
\beq
U^\mu_{\;\; \nu} (C) \; \sim \;
\Bigl [ \;
\exp \, \left \{ \half \,
\int_{S(C)}\, R^{\, \cdot}_{\;\; \cdot \, \lambda\sigma} \, A^{\lambda\sigma}_{C} \;
\right \}
\, \Bigr ]^\mu_{\;\; \nu}  \;\; ,
\label{eq:rot-cont1}
\eeq
where $ A^{\lambda\sigma}_{C}$ is an area bivector associated with the loop in question,
\beq
A^{\lambda\sigma}_{C} = \half \oint_C dx^\lambda \, x^\sigma \; .
\eeq
Then the semiclassical procedure gives for the loop in question
\beq
W(C) \, \simeq \, \tr \left ( \, \omega (C) \, \exp \left \{ \, \half \,
\int_{S(C)}\, R^{\, \cdot}_{\;\; \cdot \, \lambda\sigma} \, A^{\lambda\sigma}_{C} \; 
\right \} \right ) \; .
\label{eq:wloop_curv}
\eeq
Here again $\omega_{\mu\nu} (C) $ is a constant unit bivector,
characteristic of the overall geometric orientation of the 
parallel transport loop.
By carefully comparing coefficients for the two area terms
\cite{loops} one then concludes that the average large-scale curvature is of 
order $ + 1 / {\xi}^2$, at least in the strong coupling limit
considered in the cited references.
Since the scaled cosmological constant can be viewed as a measure of 
the intrinsic curvature of the vacuum, the above argument
then gives a positive cosmological constant for this phase, corresponding to
a manifold which behaves as de Sitter ($\lambda > 0$) 
on large scales \cite{loops}.
These arguments then lead to the suggestion that the macroscopic 
(semiclassical) average curvature is related to $\xi$ by
\beq
\langle \, R \, \rangle_{\rm large \; scales} 
\;\; \sim \; + \, 1 / \xi^2 \;\; ,
\label{eq:xi_r}
\eeq 
at least in the strong coupling (large $G$) limit.
It is important to note here that the result of Eq.~(\ref{eq:xi_r})
applies to parallel transport loops whose linear size $r_C$ 
is much larger than the cutoff, $ r_C \gg l_p , a $;
neverthless in this limit the answer for the macroscopic
curvature in Eq.~(\ref{eq:xi_r}) becomes independent of
the loop size or its area \cite{loops}.
Furthermore, these arguments lead, via the classical field equations,
to the identification of $1 / \xi^2$ with the observed (scaled)
cosmological constant $\lambda_{obs}$, 
\footnote{up to a constant of
proportionality, expected to be of order unity.}
\beq
\third \;  \lambda_{obs} \; \simeq \;  + \, { 1 \over \xi^2 }  \; .
\label{eq:xi_lambda}
\eeq 
In this picture the latter is then regarded as the {\it quantum gravitational condensate},
a measure of the vacuum energy, and thus of the intrinsic curvature of
the vacuum.
\footnote{
Note that, quite generally and independent of the lattice results, it seems
rather difficult to implement a weakly running cosmological constant, if general 
covariance is to be maintained at the level of the effective field
equations. If the running of $\lambda$ is formulated via a $\lambda (\Box)$ then
because of $ \nabla_\lambda \, g_{\mu\nu} =0$ one also has
$ \Box^n \, g_{\mu\nu} =0$, which makes it nearly impossible to have
a nontrivial $\lambda (\Box)$ \cite{lambda}.}
Then a suitable effective action, describing the residual effects of
quantum gravity on very large distance scales, is of the form
\beq
I_{\rm eff} \, [ g_{\mu\nu} ] \; = \; - \, { 1 \over 16 \pi \, G (\mu) }
\int d^4 x \, \sqrt g \; \Bigl ( R \, - \, { 6 \over \xi^2 } \Bigr )
\, + \, I_{\rm matter} \, [ g_{\mu\nu} , \dots ] \;\; ,
\label{eq:ac_eff}
\eeq
with $G(\mu)$ a very slowly varying (on macroscopic scales) Newton's
constant, in accordance with Eqs.~(\ref{eq:run_xi}) or
(\ref{eq:run_box_xi}).

Note that the above results in many ways parallels what is found in
non-Abelian gauge theories, where for example one has
for the color condensate $< F_{\mu\nu}^2 > \, \simeq 1/ \xi^4 $.
Furthermore, this last result can also be obtained largely from purely
dimensional grounds, once the existence of a fundamental
correlation length $\xi$, which for QCD is given by the inverse of the
mass of the lowest spin zero glueball, is established.
Accordingly, for gravity too one would expect, again simply on the basis 
of dimensional arguments, that the large scale curvature (the graviton condensate)
should be related to the fundamental correlation length by
$ \langle \, R \, \rangle \, \simeq \, 1 / \xi^2 $, as in
Eq.~(\ref{eq:r_xi}).



\begin{figure}
\begin{center}
\includegraphics[width=0.7\textwidth]{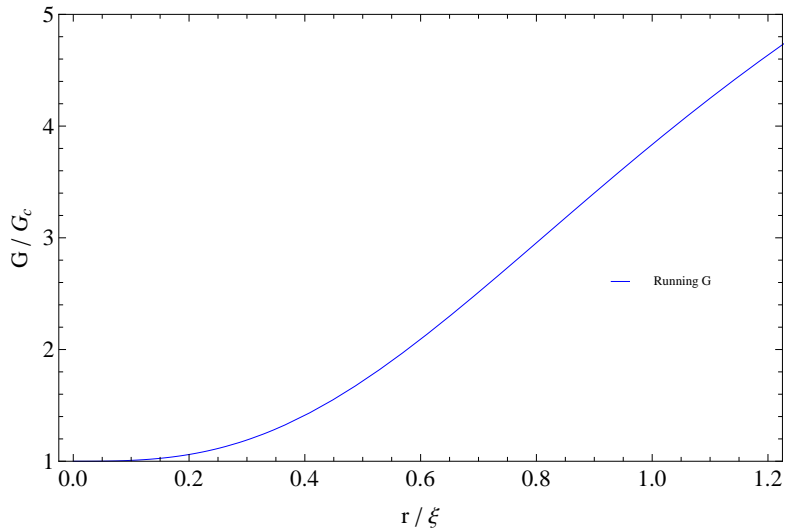}
\end{center}
\caption{
Running gravitational coupling $G(r)$ versus $r$, obtained
from the $G(k)$ in Eq.~(\ref{eq:run_k}) by setting $q \sim 1/r $,
with exponent $\nu=1/3$ and amplitude $a_0 \simeq 8.02(55) $.
The lattice quantum gravity calculations done so far suggest roughly a 5\% effect on
scales of $ 0.187 \times 4890 Mpc \approx 910 Mpc $,
and a 10 \% effect on scales of $ 0.238 \times 4890 Mpc \approx 1160
Mpc $.
}
\end{figure}

The considerations presented so far can to some extent
finally provide a quantitative handle on the physical {\it magnitude} of the 
nonperturbative scale $\xi$.
From the observed value of the cosmological constant one
obtains an estimate for the absolute magnitude of the scale $\xi$
\beq
\xi \; \simeq \; \sqrt{3 / \lambda} \; \approx \; 4890 Mpc \;\; .
\label{eq:xi_mpc}
\eeq
Irrespective of the specific value of $\xi$, this would indicate
that generally the recovery of classical GR results only happens
for distance scales much smaller than the correlation length $\xi$.
In particular, the Newtonian potential acquires a tiny quantum correction
from the running of $G(r)$,
\beq
V(r) \; = \; - \, G(r) \cdot { m_1 \, m_2 \over r } \;\; ,
\label{eq:pot}
\eeq
with $G(r)$ given, for the static isotropic solution, in
Eq.~(\ref{eq:run_r}), and for which quantum effects
become quite negligible on distance scales $r \ll \xi$.
Figure 21 shows the expected qualitative behavior for the
running $G(k)$ over scales slightly smaller or comparable to $\xi$,
with the main uncertainty arising from estimating the physical 
magnitude of $\xi$ itself [Eq.~(\ref{eq:xi_mpc})].
Specifically, from Eq.~(\ref{eq:run_k}) the lattice prediction 
at this point is for roughly a 5\% effect on
scales of $ 0.184 \times 4890 Mpc \approx 900 Mpc $,
and a 10 \% effect on scales of $ 0.232 \times 4890 Mpc \approx 1130
Mpc $.

The above results also suggest that the curvature on very small
scales behaves rather differently from the curvature on very large
scales, due to the quantum fluctuations eventually averaging out.
Indeed when comparing the result of Eqs.~(\ref{eq:r_xi}) 
and (\ref{eq:r_xi1}) to the one in
Eq.~(\ref{eq:xi_r}) one is lead to conclude that the following change
has to take place when going from small (linear size $\sim l_p$) to large 
(linear size $ \gg l_p $) parallel transport loops 
\beq
\langle \, R \, \rangle_{\rm small \; scales} \;\; \sim \; {1 \over l_p \, \xi } 
\;\;\;\;\; \rightarrow \;\;\;\;\;
\langle \, R \, \rangle_{\rm large \; scales} \;\; \sim \; {1 \over \xi^2 } \; .
\label{eq:r_small_large}
\eeq
An intuitive way of understanding the above result is
that on small scales the strong local fluctuations in the 
metric/geometry lead to large values for the average 
rotation of a parallel-transported vector.
But then on larger scales these short distance fluctuations
tend to average out, and the {\it combined} overall rotation is 
much smaller, by a factor $ {\cal O} ( l_p / \xi)$,
\beq
Z_R \; = \; { l_P \over \xi } \;\; .
\label{eq:z_r}
\eeq
The above quantity should then be regarded as an essential and necessary
``renormalization constant'' when comparing curvature on different
length scales, and specifically when going from very small 
(size $\sim l_P $) to large (size $ \gg l_P $) parallel transport loops.
See also the earlier discussion preceding Eq.~(\ref{eq:xi_l}), 
about the issue of comparing correlations of large loops 
versus correlations of small (infinitesimal) loops.

To conclude this section, one can raise the legitimate concern of how 
these results are changed by quantum fluctuations of various matter fields;
so far all the results presented here apply to pure gravity without any
matter fields.
Therefore here and in the rest of the paper what has been followed is the {\it quenched
approximation}, wherein gravitational loop effects (perturbative and
nonperturbative) are fully accounted for, but matter loop corrections
are entirely neglected. 
In the presence of matter fields coupled to gravity (scalars, fermions,
vector bosons, spin-3/2 fields etc.) one would expect, for example, the value for $\nu$
to change due to vacuum polarization loops containing these fields.
A number of arguments can be given though for why these effects
should not be too dramatic, unless the number of light matter fields
is rather large.
Firstly one notices that to leading order in the $2+\epsilon$ expansion
the exponent $\nu$ only depends on the dimensionality of spacetime,
irrespective of the number of matter fields and of their type
\cite{eps}, $\nu \sim 1/(d-2)$.
Also, one can show that in the $2+\epsilon$ expansion for gravity \cite{eps,aid97}
matter loop corrections appear later in the form of factors
$\propto (25-c)$ in the $\beta$-function, where $c$ is the central
charge corresponding to those massless matter fields.
In four dimensions the correction is thought to be even smaller 
$\propto (48-c)$ \cite{nie81}.
Thus unless $c$ is rather large, the matter contribution is quite small
even at next-to-leading order in the $2+\epsilon$ expansion \cite{eps,aid97}.
In addition, in the case of lattice gravity the effects of
a single light scalar field are so small that they are barely 
detectable in the numerical evaluations of the path integral.
In general one would expect significant infrared modifications to gravity
coming from particles that are either massless or very nearly massless.
The evidence so far would therefore suggests that the approximation
in which vacuum polarization effects of light matter fields are
entirely neglected should still be useful, at least as a first step.



\vskip 40pt

\section{Gravitational Scaling Dimensions and Phenomenology}

\label{sec:pheno}

\vskip 20pt

The previous section dealt with the fact that one of the main
implications of quantum gravity is the running of 
the gravitational constant with scale, in accordance
with Eq.~(\ref{eq:run_box}).
There are additional consequences which arise from the fact
that in general gravitational correlations don't follow
free field (gaussian) predictions.
One example is the curvature correlation function of Eqs.~(\ref{eq:corr_cont}),
(\ref{eq:corr_pow}) and (\ref{eq:corr_nu}), for which the final
form at this stage is given in Eq.~(\ref{eq:corr_pow1}).
This section deals therefore with a discussion of the implications
of the result given in this last expression for the curvature
correlation function of Eq.~(\ref{eq:corr_pow1}), specifically
with lattice spacing $a$ from Eq.~(\ref{eq:a_phys}) and
amplitude given in Eq.~(\ref{eq:corr_amp}), $N_{\cal R} \approx 0.335 $.

There is clearly a rather substantial difference
in scale between the curvature appearing in Eq.~(\ref{eq:corr_cont}) and
therefore in Eq.~(\ref{eq:corr_pow1}), and the curvature in
Eq.~(\ref{eq:trace}). 
In the first case the curvature involves
the parallel transport of vectors around infinitesimal loops, whose
size is determined by the ultraviolet cutoff [the lattice spacing, comparable
to the Planck length because of $G \approx G_c$ due to the slow
running of $G$, with $G_c$ given in Eq.~(\ref{eq:Gc})].
In the second case the curvature in question refers instead to the
semiclassical domain, as described by a set of effective 
long distance field equations,
for which the curvature is obtained operationally from the parallel transport
of vectors around macroscopic loops, of linear size much larger
than the Planck length.
Such a concern should be kept in mind when transitioning from
the microscopic result in Eq.~(\ref{eq:corr_pow1}) to the semiclassical
result written down below in Eq.~(\ref{eq:corr_pow2}).

First consider what can be stated purely at the classical level.
One can use the field equations to directly relate the local curvature to the
local matter mass density. 
From Einstein's field equations
\beq
R_{\mu\nu} - \half \, g_{\mu\nu} \, R \; = \; 8 \pi G \, T_{\mu\nu} \;\;
\label{eq:einstein}
\eeq
for a perfect fluid one then obtains for the Ricci scalar, in the limit of
negligible pressure,
\beq
R (x) \; \simeq \; 8 \pi G \; \rho (x) \;\; .
\label{eq:trace}
\eeq
This last result then allows one to relate local fluctuations in the
curvature $ \delta \, R(x)$ to local fluctuations in the matter 
density $\delta \, \rho (x)$,
which could potentially provide a useful connection to the quantum result 
for the correlation function in Eq.~(\ref{eq:corr_pow}).
Of course, in the Newtonian limit the above result simplifies to
Poisson's equation
\beq
\Delta \, h_{00} ( {\bf x}, t ) \; = \; 8 \pi \, G \, \rho ( {\bf x}, t ) \;\; ,
\label{eq:poisson}
\eeq
where $h_{00}= 2 \phi $ and $\rho$ are the macroscopic gravitational field and
the macroscopic mass density, respectively.

In the cosmology literature it is customary to describe matter density 
fluctuations in terms of the density contrast correlation function
\footnote{In the cosmology literature the (dimensionless) galaxy
matter density two-point function is usually referred to as $\xi(r)$,
but here we want to avoid a possible confusion with the gravitational 
correlation length $\xi$.}
\beq 
G(r) \; = \; < \delta \rho (r) \; \delta \rho (0) > \;\; .
\label{eq:galaxy_corr}
\eeq
The latter is related to its Fourier transform $P(q)$ by
\beq
G(r) \; = \; { 1 \over 2 \pi^2 } \int_\mu^\Lambda dq \, q^2 \, P(q) \,
{\sin q r \over q r } \; ,
\eeq
and the above expression has to contain both an infrared regulator ($\mu$) and
an ultraviolet cutoff ($\Lambda$), to make sure the integral converges.
If the power spectrum $P(q)$ is described by a simple power 
law of the form
\beq
P(q) \; = \; { a_0 \over q^s } \;\; ,
\label{eq:power_spec}
\eeq
(where $n = -s $ is commonly referred to as the spectral index),
then one finds in the scaling regime $ 1/\mu \gg r \gg 1/ \Lambda $ for
the density contrast correlation function in real space
\beq
G(r) \; = \; c_s \, a_0 \, \mu \, \Lambda^{2-s} \, \left ( { 1 \over
  \mu \, r } \right )^{3-s} \;\; ,
\label{eq:galaxy_corr1}
\eeq
with $c_s \equiv \Gamma (2-s) \, \sin ( \pi s / 2 ) \, / \, 2 \pi^2 $.
Not unexpectedly, the answer appears to be quite sensitive to the choice for
the ultraviolet and infrared cutoffs.
For the specific value $s=1$ one has $P(q)=a_0 / q $, and this then gives
\beq
G(r) \; = \; \; { a_0 \, \Lambda \over 2 \pi^2 \, \mu }
\cdot { 1 \over r^2 }
\;\;\;\;\;\; (s=1) \;\; ,
\label{eq:gr_a0}
\eeq
which would seem to reproduce the result in Eq.~(\ref{eq:corr_pow1}).
In practice the observational data for such matter density correlations
is often presented in the simple form \cite{peebles}
\beq
G(r) \; = \; \left ( { r_0 \over r } \right )^\gamma \; ,
\label{eq:gr_r0}
\eeq
with exponent $\gamma$ and scale $r_0$ fitted to astrophysical observations.
For an exponent $\gamma$ close to two, one has
by comparing Eq.~(\ref{eq:gr_a0}) to Eq.~(\ref{eq:gr_r0})
$ a_0 \, = \, 2 \pi^2 \, \mu \, r_0^2 / \Lambda $,
which still requires a choice of cutoffs $\mu$ and
$\Lambda$; for the most obvious choice here, namely $\mu \simeq 1 / \xi$ and 
$ \Lambda \simeq 1 / l_P$, one obtains
\beq
a_0 \; = \; { 2 \pi^2 \, l_P \over \xi } \cdot r_0^2 \;\; .
\label{eq:a0}
\eeq
It is rather tempting at this stage to try to connect the observational
result of Eq.~(\ref{eq:gr_r0}) to the quantum
correlation function in Eq.~(\ref{eq:corr_pow1}).
One then expects for the matter density fluctuation correlation also
a power law decay of the form
\footnote{
In weak field perturbation theory one finds
$ < \rho (x) \, \rho (y) >_c 
\; \sim \; < \partial^2 h (x) \, \partial^2 h (y) > 
\; \sim \; 1/ \vert x-y \vert^6 $, so again the result here is quite different.
}
\beq
< \delta \rho ({\bf x},t) \; \delta \rho ({\bf y},t') > \;\;\; 
\mathrel{\mathop\sim_{ \vert {\bf x}-{\bf y} \vert \; \ll \; \xi }} \;\;
{1 \over a^2 (t)} \cdot {1 \over a^2 (t')} \cdot
{1 \over \vert {\bf x}-{\bf y} \vert^2 } \;\; 
\label{eq:corr_pow2}
\eeq
where $a(t)$ here represents the scale factor.
Also, this last correlation function can be made dimensionless by 
suitably dividing it by the square of some average matter 
density $\rho_0 \approx \rho_c = 3 H_0^2 / 8 \pi G $.
By comparing coefficients in Eqs.~(\ref{eq:corr_pow1}) 
and (\ref{eq:gr_r0}) one finds $\gamma=2$, and for the length scale in
Eq.~(\ref{eq:gr_r0})
\beq
r_0 \; = \; { 1 \over 8 \pi \, G \, \rho_0 } 
\cdot { \sqrt{A_0} \over  a } \; ,
\label{eq:r0}
\eeq
with $ \sqrt{A_0} \simeq 0.335 $ the dimensionless amplitude for
the curvature correlation  function of Eq.~(\ref{eq:corr_pow1}), 
and $a$ the lattice spacing given in Eq.~(\ref{eq:a_phys}).

The preceding argument nevertheless still contains a fundamental flaw,
related to the use, at this stage in unmodified form, of the curvature
correlation function result of Eq.~(\ref{eq:corr_pow1}).
As already discussed previously, that form applies to
the correlation of {\it infinitesimal} (Planck length or cutoff size) loops, 
which would not seem to be appropriate for the macroscopic 
(or semiclassical) parallel transport loops, such as
the ones that enter the field equations Eqs.~(\ref{eq:einstein}) 
and (\ref{eq:trace}), and which thus relate locally the macroscopic
$\delta R (x) $ to the $\delta \rho (x)$.
It would then seem desirable to be able to correct for the
fact that the parallel transport loops sampled in
Eq.~(\ref{eq:trace}) are much larger than the infinitesimal
ones sampled in the correlation function in
Eq.~(\ref{eq:corr_pow1}).
As in Eqs.~(\ref{eq:xi_l}), (\ref{eq:r_small_large}) and
(\ref{eq:z_r}), the transition to macroscopic loops (linear size $\gg a$)
can be affected in Eq.~(\ref{eq:corr_pow1}) 
by the replacement of $ a^2 \rightarrow \xi^2 $.
This then gives for large (macroscopic size $\gg a $) parallel transport loops
\beq
< \sqrt{g} \; R(x) \; \sqrt{g} \; R(y) \; \delta ( | x - y | -d ) >_c
\;\; \mathrel{\mathop\sim_{d \; \ll \; \xi }} \;\; 
{A_1 \over \xi^2 \, d^2 } \;\;\; ,
\label{eq:corr_pow3}
\eeq
with the expectation of a comparable amplitude $ A_1 \approx A_0 $.
This last result then leads to the following improved estimate for 
the macroscopic matter density correlation of Eq.~(\ref{eq:galaxy_corr})
\beq
G(r) \; = \; \left ( { 1 \over 8 \pi \, G } \right )^2 \; 
{ 1 \over  \rho_0^2 } \cdot
{ A_1 \over \xi^2 \; r^2 } \; ,
\eeq
so that comparing to Eq.~(\ref{eq:gr_r0}) one finds
for the exponent $ \gamma = 2 $, and for the length scale $r_0$
the improved value
\beq
r_0 \; = \; { 1 \over 8 \pi \, G \, \rho_0 } \cdot 
{ \sqrt{A_1}  \over \xi } \; \approx \; 0.380 \; \xi \;
\label{eq:r0_xi}
\eeq
which seems more in line with observational data.
For the Fourier amplitude $a_0$ in Eq.~(\ref{eq:power_spec}) one has now
\beq
a_0 \; = \; 2 \pi^2 \cdot { l_P \over \xi } \cdot r_0^2
\; \approx \; 2.85 \; l_P \, \xi \; .
\label{eq:a0_xi}
\eeq
Observed galaxy density correlations give indeed for the exponent 
in Eq.~(\ref{eq:gr_r0}) a value close to two, namely
$\gamma \approx 1.8 \pm 0.3 $ for distances in the $0.1Mpc$
to $50Mpc$ range \cite{peebles,bahcall}, and  for the
length scale $r_0 \approx 10 Mpc $.
More recent estimates for the exponent $\gamma$, going up to distance
scales of $100Mpc$, range between $1.79$ and $1.84$
\cite{baugh,longair,tegmark,durkalec,wang,coil}.
Nevertheless at this point the (perhaps rather naive)
identification given in Eqs.~(\ref{eq:r0_xi}) and (\ref{eq:a0_xi}),
while intriguing, is possibly entirely accidental since it
bypasses any concerns about the actual physical origin of the 
galaxy correlation function in Eq.~(\ref{eq:r0}), including the form
and evolution of primordial density perturbation, the detailed nature of
linear relativistic density perturbation theory for a given comoving
background etc.




\vskip 40pt

\section{Conclusions}

\label{sec:concl}

\vskip 20pt

In this work a number of improved estimates have been presented for
gravitational scaling dimensions and amplitudes, obtained from the 
lattice theory of gravity.
Numerical methods combined with modern renormalization group arguments 
and finite size scaling have been shown to provide detailed 
information about rather subtle nonperturbative aspects of the theory.
It has been known for some time that the Euclidean
lattice gravity theory has two phases, only one of which, the gravitational
anti-screening phase for $G > G_c$, is physically acceptable. 
Here we have described in some detail the properties of the
latter smooth phase, and provided quantitative estimates for the critical
point, scaling dimensions and the behavior of physical
correlations for distances large compared to the lattice cutoff.
In many ways the present calculation is still incomplete, in
particular the gravitational Wilson loop and the correlation between
loops has not been studied numerically, and only some general
properties have been inferred. 
Also, more heavy work is needed to accurately
determine the curvature correlation functions versus distance, and
from it the fundamental non-perturbative correlation length and various
amplitudes connected to it.
Furthermore, the derivation of a number of results has relied heavily on
basic renormalization group scaling, with only a handful of explicit checks.
Nevertheless, it would seem from the results presented so far that the
feasibility of these types of calculations should increase
significantly in the near term due to expected rapid advances in hardware
and software tools.

It is encouraging that four different approaches to quantum gravity give
rather comparable results for the scaling dimensions (see the
comparison Tables II and III,
as well as Figure 20), and therefore suggest a unique underlying renormalization
group universality class, associated with the quantum version
of General Relativity.
It is characteristic of the model described here that the growth of $G$ with scale
is described by a nonperturbative correlation length $\xi$, related to
the gravitational vacuum condensate, for which 
a specific quantitative estimate was given earlier.
More generally, the {\it vacuum condensate picture of quantum gravity}
presented in this paper makes in principle a number of specific
and testable predictions, which could either be verified or disproven
in the near future, as new and increasingly accurate satellite 
observations become available.
The main aspects of this picture can be summarized as follows:

\begin{enumerate}

\item[ $\circ$ ]
The vacuum condensate picture of quantum gravity contains from the start
a very limited number of parameters, and is therefore rather strongly constrained.
While it does involve a new nonperturbative scale (the gravitational
vacuum condensate), it is found that this scale simultaneously 
determines the running of $G$ with scale, the value of the scaled cosmological
constant, and the long distance behavior of physical invariant correlations.

\item[ $\circ$ ]
The current theory predicts a slow increase in strength of the
gravitational coupling when very large, cosmological scales are
approached [see Eqs.~(\ref{eq:run_xi}) and (\ref{eq:run_box_xi})].
In this context, the observed scaled cosmological constant $\lambda$
acts as a dynamically induced infrared cutoff, similar to
what happens in non-Abelian gauge theories.
In principle, both the universal power and amplitude for this infrared growth
are calculable with some accuracy from the underlying lattice cutoff
theory.

\item[ $\circ$ ]
For a sufficiently large scale $\xi$ (and therefore small $\lambda$) no observable
deviations from classical General Relativity are expected on
laboratory, solar systems and even galactic scales
[see Eqs.~(\ref{eq:run_r}) and (\ref{eq:pot})].

\item[ $\circ$ ]
The calculations presented here give a number of predictions for the behavior of
invariant curvature correlations as a function of geodesic distance, 
and specifically the powers and
amplitudes involved [see Eqs.~(\ref{eq:corr_pow1}) and (\ref{eq:corr_pow3})].

\item[ $\circ$ ]
The lattice theory appears to exclude at this point the possibility of 
a physically acceptable phase with gravitational screening;
such a (weak coupling) phase in the lattice theory appears to be
inherently unstable,
presumably as a consequence of the conformal mode, and cannot 
lead to a semiclassical regime for gravity.
It leads instead to a pathological degenerate ground state describing
some sort of branched polymer.
Nevertheless for large enough quantum fluctuations (large $G$)
the instability is overcome and a new stable phase emerges.

\item[ $\circ$ ]
In the strong coupling limit (for the Euclidean case) of the lattice
theory the effective, long distance cosmological constant is positive \cite{loops}.
In this same regime it seems nearly impossible from the lattice theory
to get a negative value for this quantity, irrespective of the choice 
of boundary conditions (which in the lattice context play no role in
the argument).
Also, a positive cosmological constant is interpreted here 
as a genuinely nonperturbative gravitational vacuum condensate
[see Eqs.~(\ref{eq:xi_r}) and (\ref{eq:xi_lambda})].
  
\end{enumerate}

If the picture presented in this paper is indeed close to correct, 
then it points to what appears to be a deep analogy 
between the nonperturbative vacuum state of quantum gravity and 
known properties of strongly coupled non-Abelian gauge theories 
(or what could be called the QCD analogy).
Over time this analogy has been helpful in illustrating properties
of quantum gravity, many of which are ultimately based on rather basic
principles of the  renormalization group, connected with the scaling 
properties expected in the vicinity of a nontrivial fixed point.
Indeed in QCD there exists also a nonperturbative mass parameter 
$m = 1/ \xi$ (sometimes referred to as the mass gap) which is 
known to be a renormalization group invariant; that such a mass scale can be 
generated dynamically is known to be a highly nontrivial outcome of the 
renormalization group equations for QCD.
Furthermore, there seems to be a fundamental  relationship between 
the nonperturbative scale $\xi$
(or inverse renormalized mass) and a nonvanishing vacuum
condensate for these theories, both for gravity and QCD,
\beq
\langle \, R \, \rangle \;  \simeq \;  { 1 \over \xi^2 }
\;\;\;\;\;\;\;\;\;\;\;\;
\langle \, F_{\mu\nu}^2 \, \rangle  \;  \simeq \;  { 1 \over \xi^4 }
\; .
\label{eq:vev}
\eeq
An additional relevant example that comes to mind is the fermion 
condensate in gauge theories,
\beq
\langle \, \bar \psi \psi \, \rangle \, \simeq \, { 1 \over \xi^3 } \; ,
\label{eq:vev_fer}
\eeq
a consequence of confinement and chiral symmetry breaking.
Current lattice and phenomenological estimates for QCD cluster around
$ \langle \, { \alpha_S \over \pi } \, F_{\mu\nu}^2 \, \rangle  \;
\simeq \;  (440 \, MeV)^4 $ and 
$ \langle \, \bar \psi \psi \, \rangle \, \simeq \, (290 \, MeV)^3 $
\cite{bro09,dom14,mcn13}.

Modifications to the static potential in gauge theories
are best expressed in terms of the running coupling constant 
$\alpha_S (\mu) $, whose scale dependence is determined by the 
celebrated beta function of $QCD$ with coupling 
$\alpha_S \equiv g^2 / 4 \pi $.
On the one hand, a solution of the renormalization group equations
give for the running of $\alpha_S (\mu )$
\beq
\alpha_S (\mu ) \; = \; 
{ 4 \, \pi \over \beta_0 \ln { \mu^2 / \Lambda_{\overline{MS}}^2  } }
\, + \, \dots \; .
\label{eq:alpha_qcd}
\eeq
On the other hand, the nonperturbative scale $ \Lambda_{\overline{MS}}$ 
appears as an integration constant of the renormalization group equations, and
is therefore - by construction - scale independent,
\beq
\Lambda_{\overline{MS}} \; = \; 
\Lambda \, \exp \left ( 
{ - \int^{\alpha_S (\Lambda)} \, {d \alpha_S' \over 2 \, \beta ( \alpha_S') } }
\right ) \; ,
\label{eq:lambda_qcd}
\eeq
where here $\Lambda$ represents the QCD ultraviolet cutoff.
The physical value of $ \Lambda_{\overline{MS}} $
cannnot be fixed from perturbation theory alone and has to be
determined instead from experiment, $ \Lambda_{\overline{MS}} \simeq 210 MeV$.
In quantum gravity the corresponding statements are given
in Eqs.~(\ref{eq:m_beta}) and (\ref{eq:run_k}).

Wilson loop correlations play an important role in QCD as they
do in quantum gravity.
In non-Abelian gauge theories a confining potential 
is found at strong coupling by examining the behavior of the Wilson
loop, defined for a large closed loop $C$ as
\beq
\langle \, W( C ) \, \rangle \, = \, 
\langle \, \tr {\cal P} \, \exp \Bigl \{ i g \oint_{C} A_{\mu} (x) dx^{\mu} 
\Bigr \} \, \rangle \;\; ,
\label{eq:wloop_sun}
\eeq
with $A_\mu \equiv t_a A_\mu^a $ and the $t_a$'s the group
generators of $SU(N)$ in the fundamental representation. 
In the pure gauge theory at strong coupling, the leading 
contribution to the Wilson loop is known to follow an area 
law for sufficiently large loops.
The analogous quantity for gravity is the gravitational
Wilson loop described, for example, in Eqs.~(\ref{eq:rot_cont}) 
and (\ref{eq:wloop_latt}).
But in contrast to QCD, in gravity the Wilson loop bears {\it no}
relationship to the static potential \cite{modapot}
(the path ordered line integral of the affine connection does not in 
any way describe a gravitational interaction energy).

A central role in this view of quantum gravity is played
by the gravitational correlation length of Eqs.~(\ref{eq:corr_exp}),
(\ref{eq:xi_k}) and (\ref{eq:mass_k}).
Gauge theories also contain a nonperturbative,
dynamically generated quantity $ \xi $, the gauge field 
correlation length, and it is essentially the same [up to a factor $O(1)$] 
as the inverse of  $\Lambda_{\overline{MS}}$.
The same universal quantity $\xi$ also appears in a number of other physical
observables, including the exponential decay of the Euclidean correlation
function for two infinitesimal loop operators separated by a distance $|x|$,
\beq
G_{\rm loop-loop} ( x ) \, = \, 
\langle \, \tr {\cal P} \exp \Bigl \{ i g \oint_{C_\epsilon} A_{\mu} (x') dx'^{\mu} 
\Bigr \} (x)
\, 
\tr {\cal P} \exp \Bigl \{ i g \oint_{C_\epsilon} A_{\mu} (x'') dx''^{\mu} 
\Bigr \} (0)
\, \rangle_c \;\; .
\label{eq:box_sun}
\eeq
Here the $C_\epsilon$'s are two infinitesimal loops centered around $x$ ands $0$
respectively, suitably defined on the lattice as, for example, elementary square loops.
The gravitational analogue of such an (infinitesimal loop)
correlation was given earlier in Eqs.~(\ref{eq:corr_cont}) and (\ref{eq:corr_latt}).
It is also understood that in gauge theories the inverse of the 
correlation length $\xi$  corresponds to the lowest mass excitation 
in the gauge theory, the scalar glueball with mass $m_0 = 1 / \xi$.
If the lightest scalar $0^{++}$ glueball has a mass of approximately
$m = 1750 MeV$ (which then fixes $\xi=1/m$), then $\Lambda_{\overline{MS}}$ 
in QCD is about eight times smaller, which gives rise to what has been
described in QCD as ``precocious'' scaling.
So while the two scales are quite close, they do not necessarily
coincide. And the same could be true in gravity.

Another important difference between gravity and QCD is that fact that
in the former the ultraviolet cutoff still appears explicitly,
hidden in the physical value of Newton's constant $G$.
There exists then a second scale $\xi$ whose magnitude is
not directly related to the value of $G$; instead it reflects
how close the bare $G$ is to the ultraviolet fixed point at $G_c$,
and is therefore fine tuned in this approach (just like a mass squared term
in a scalar field theory).
In QCD on the other hand the scale $\xi$ appears explicitly, wheras
the ultraviolet cutoff is deeply hidden in the renormalization group 
relationship between $\Lambda_{\overline{MS}}$ and the bare coupling
at the cutoff scale $\alpha_S (\Lambda )$.


\vspace{20pt}

{\bf Acknowledgements}

The work of the author was supported in part by the Max 
Planck Gesellschaft zur F\" orderung der Wissenschaften, and
by the University of California.
He wishes to thank prof. Hermann Nicolai and the
Max Planck Institut f\" ur Gravitationsphysik (Albert-Einstein-Institut)
in Potsdam for warm hospitality. 
The author is also grateful for useful discussions with 
James Hartle and Gabriele Veneziano.
The numerical calculations described in this paper were
performed in part on the supercomputers located at the $AEI$.


\newpage

\appendix

\section*{Appendix}

\newsection{Details on the parallel code}
\label{sec:code}

A few details will be given here regarding the performance
of the computer code used on the Datura Supercomputer at AEI.
The latter is a high performance Infiniband cluster, 
primarily used for large scale numerical calculations
in classical General Relativity.
The cluster is based on Intel Xeon X5650 2.66 GHz processor boards which
have two processors per board, each with six cores. 
Thus there are 12 cores per board, 200 nodes and 2400 cores total.
The main communication and storage network is based on 
an Infiniband QDR 324-port 40Gbit/s low latency and high bandwidth switch.
In addition, the Datura cluster is equipped with 4,800 GB of main memory,
corresponding to 2GB per core.

There are presently two main versions of the lattice quantum
gravity code, a scalar (sequential) one and a parallel code.
Both codes run exclusively in double precision (64 bits).
The scalar code takes about 1021 seconds per iteration on a lattice
with $32^4 = 1,048,576$ sites, which includes updates with 
an action containing higher derivative terms.
The scalar code performance on a single core is measured
(by counting raw floating point operations using a
hardware performance monitor) at 3.3 GFlops.

In the case of the parallel code, on a $16^4$ lattice 
(1,572,864 simplices) the edges emanating from 256 sites can all be
updated in parallel, with the work distributed on 256 cores.
A whole lattice is therefore updated in 256 passes, and it takes
about 0.26 seconds per iteration for a whole lattice.

On the $32^4$ lattice (25,165,824 simplices) the edges emanating 
from 256 sites are all updated in parallel, with the work distributed
again on 256 cores.
A whole lattice is therefore updated here in 4096 passes, and
it takes about 4.01 seconds per iteration for a whole lattice.
Thus for this setup the parallel codes is about 255 times faster than the
single core scalar code.
The MPI communication overhead between nodes for this setup accounts 
for only about 1\%.
Using 256 cores the overall code performance on this
machine is around 845 GFlops.

On the $64^4$ lattice (402,653,184 simplices) the edges emanating 
from 256 sites are all updated in parallel, with the work 
distributed on 256 cores.
A whole lattice is therefore updated now in 65,536 passes.
This then gives about 64 seconds per iteration.
When a full complement of 1024 cores are used instead of 256, 
the time for one full lattice iteration goes down to about 16 seconds.
With a full 1024 cores used in parallel the overall code 
performance on this machine is around 3.4 TFlops.
If even more cores are used, the code performance should further
improve provided the network bandwidth is commensurate.



\vfill


\end{document}